\documentclass[manuscript]{aastex}
\usepackage{txfonts}
\usepackage{graphicx}
\usepackage{longtable}
\usepackage{lscape}
\usepackage{natbib}

\shorttitle{Evolution, nucleosynthesis and yields of low mass AGB
stars} \shortauthors{Cristallo et al.}

\begin{document}

\title{Evolution, nucleosynthesis and yields
of low mass AGB stars at different metallicities.}

\author{S. Cristallo\altaffilmark{1} and O.
Straniero\altaffilmark{1}}
\affil{INAF-Osservatorio Astronomico di Collurania, 64100 Teramo,
Italy}
 \and
 \author{R.
Gallino\altaffilmark{2,3} \affil{Dipartimento di Fisica Generale,
Universit\'a di Torino, 10125 Torino, Italy} \affil{Center for
Stellar and Planetary Astrophysics, School of Mathematical
Sciences, Monash University, P.O. Box 28, Victoria 3800,
Australia}} \and \author{L. Piersanti\altaffilmark{1}
\affil{INAF-Osservatorio Astronomico di Collurania, 64100 Teramo,
Italy}} \and \author{I. Dom\'inguez\altaffilmark{4}
\affil{Departamento de F\'isica Te\'orica y del Cosmos ,
Universidad de Granada, 18071 Granada, Spain}} \and \author{M.T.
Lederer\altaffilmark{5} \affil{Institut f\"ur Astronomie,
T\"urkenschanzstra\ss e 17, A-1180 Wien, Austria}}


\date{Received / Accepted}

\begin{abstract} The envelope of thermally pulsing AGB stars
undergoing periodic third dredge-up episodes is enriched in both
light and heavy elements, the ashes of a complex internal
nucleosynthesis involving p, $\alpha$ and n captures over hundreds
of stable and unstable isotopes. In this paper, new models of
low-mass AGB stars (2 M$_{\odot}$), with metallicity ranging
between $Z$=0.0138 (the solar one) and $Z$=0.0001, are presented.
Main features are: i) a full nuclear network (from H to Bi)
coupled to the stellar evolution code, ii) a mass
loss-period-luminosity relation, based on available data for long
period variables, and ii) molecular and atomic opacities for  C-
and/or N-enhanced mixtures, appropriate for the chemical
modifications of the envelope caused by the third dredge up. For
each model a detailed description of the physical and chemical
evolution is presented; moreover, we present a uniform set of
yields, comprehensive of all chemical species (from hydrogen to
bismuth). The main nucleosynthesis site is the thin $^{13}$C
pocket, which forms in the core-envelope transition region after
each third dredge up episode. The formation of this $^{13}$C
pockets is the principal by-product of the introduction of a new
algorithm, which shapes the velocity profile of convective
elements at the inner border of the convective envelope: both the
physical grounds and the calibration of the algorithm are
discussed in detail. We find that the pockets shrink (in mass) as
the star climbs the AGB, so that the first pockets, the largest
ones, leave the major imprint on the overall nucleosynthesis.
Neutrons are released by the $^{13}$C($\alpha$,n)$^{16}$O reaction
during the interpulse phase in radiative conditions, when
temperatures within the pockets attain $T\sim1.0\times10^8$~K,
with typical densities of ($10^6\div 10^7$) neutrons $\times$
cm$^{-3}$. Exceptions are found, as in the case of the first
pocket of the metal-rich models ($Z=0.0138$, $Z=0.006$ and
$Z=0.003$), where the $^{13}$C is only partially burned during the
interpulse: the surviving part is ingested in the convective zone
generated by the subsequent thermal pulse and then burned at
$T\sim1.5\times10^8$~K, thus producing larger neutron densities
(up to $10^{11}$ neutrons $\times$ cm$^{-3}$). An additional
neutron exposure, caused by the $^{22}$Ne$(\alpha$,n)$^{25}$Mg
during the thermal pulses, is marginally activated at large Z, but
becomes an important nucleosynthesis source at low Z, when most of
the $^{22}$Ne is primary. The final surface compositions of the
various models reflect the differences in the initial iron-seed
content and in the physical structure of AGB stars belonging to
different stellar populations. Thus, at large metallicities the
nucleosynthesis of light s elements (Sr,Y,Zr) is favored, whilst,
decreasing the iron content, the overproduction of heavy s
elements (Ba,La,Ce,Nd,Sm) and lead becomes progressively more
important. At low metallicities ($Z$=0.0001) the main product is
lead. The agreement with the observed [hs/ls] index observed in
intrinsic C stars at different [Fe/H] is generally good. For the
solar metallicity model, we found an interesting overproduction of
some radioactive isotopes, like $^{60}$Fe, as a consequence of the
anomalous first $^{13}$C pocket. Finally, light elements (C, F, Ne
and Na) are enhanced at any metallicity.
\end{abstract}

\keywords{stars: AGB and post-AGB --- physical data and processes:
nuclear reactions, nucleosynthesis, abundances}


\section{Introduction}

The fundamental role played by Thermally Pulsing Asymptotic Giant
Branch (TP-AGB) stars in the chemical evolution has been early
recognized \citep{ulri}. These stars are  responsible for the
nucleosynthesis of the main and the strong components of the {\it
s}-process \citep{ga98,tra99} and contribute to the synthesis of
various light elements, such as Li, C, N and F. These stars are
bright red giants ($M_{bol}$ ranging from $-4$ to $-7$). Their
compact CO cores, whose mass ranges between 0.5 and 1 M$_\odot$,
are progressively cooled by plasma neutrino emission. Owing to the
efficient heat conduction provided by degenerate electrons, the
cores are maintained in a quasi isothermal state. Outside, there
is an extended H-rich envelope, largely convective and
progressively eroded by an intense mass loss (from $10^{-8}$ up to
$10^{-4}$ M$_\odot$/yr). The most interesting nucleosynthesis
processes occur in a thin He-rich zone ($10^{-2}-10^{-3}$
M$_\odot$), located between the H and the He burning shells
(hereafter He intershell).

The energy irradiated by these stars is mainly provided by the
hydrogen burning. He burning is off during most of the AGB
lifetime, so that the mass of the He intershell increases as a
consequence of the advancing H burning shell. Then, periodic
thermonuclear runaways ({\it Thermal Pulses}, TPs) are driven by
violent He ignitions occurring when the He buffer attains a
critical value \citep{scha65,we66}. Owing to the sudden release of
energy by the $3\alpha$ reactions, the He intershell becomes
dynamically unstable against convection, while the external layers
expand and cool down.  As a consequence of the convective mixing,
the whole He intershell becomes enriched in C (the main product of
the partial He burning) and heavy elements, as produced by the
s~process (slow neutron captures). After a short period (10 to 200
yr), the thermonuclear runaway ceases, a stationary He burning
takes place and convection disappears within the He intershell.
Meanwhile, at the base of the H-rich envelope, the outgoing
nuclear energy flux and the dropping temperature cause the
radiative gradient to exceed the adiabatic gradient at positions
more and more close to the He intershell. Finally, after a few
hundred years, the convective envelope penetrates the He
intershell zone (Third Dredge Up, TDU). \citet{stra03} showed that
all TP-AGB stars may undergo recurrent third dredge up episodes,
provided their envelope mass is sufficiently large. The minimum
envelope mass for the occurrence of the TDU is of a few tenths of
M$_\odot$, but the precise value depends on the core mass and on
the envelope composition.

The occurrence of many third dredge up episodes in thermally
pulsing low mass AGB stars has two important consequences. First
of all it is responsible for the chemical modification of the
envelope, which becomes progressively enriched in primary C and in
the products of the s~process. This provides a simple explanation
for the observed spectroscopic sequence of AGB stars, from the M
to the C type, through the MS and S stars
\citep{smila,mala,bu95,bu01,ab02}. The carbon excess, mixed within
the envelope by previous third dredge up episodes, causes a
significant increase of the radiative opacity in the envelope, so
that the effective temperature sinks, the stellar radius increases
and the mass loss rate rises. The second important consequence of
the penetration of the convective envelope into the He- and C-rich
He intershell concerns the formation of the so called $^{13}$C
pocket, where the main and the strong components of the classical
s~process are built up \citep[see][]{ga98}. Indeed, when the
convective envelope recedes after a third dredge up episode, it
leaves a small region (of the order of a few $10^{-4}$ M$_\odot$)
characterized by an increasing proton profile embedded in a bath
of C and He (about 20 and 80 \%, by mass fractions, respectively).
During the relatively long period that elapses between two
subsequent thermal pulses (interpulse), this region heats up;
then, $^{13}$C is firstly produced through the
$^{12}$C$($p$,\gamma)^{13}$N($\beta^-\nu$)$^{13}$C and, later on
(at $T\sim 0.9\times10^8$ K), the $^{13}$C$(\alpha,$n)$^{16}$O
provides the slow neutron flux required to activate the s~process
nucleosynthesis \citep{stra95}. This pocket, strongly enriched in
heavy elements, is engulfed into the convective zone generated by
the subsequent thermal pulse. When the temperature at the base of
this convective zone exceeds $2.7\times 10^8$ K,  a small neutron
burst is powered by the $^{22}$Ne($\alpha$,n)$^{25}$Mg reaction.
In low mass AGB stars of nearly solar metallicity, this second
neutron source is marginally activated \citep{stra97}. Several
observations confirm such a theoretical expectation \citep[see
e.g.][]{la95,ab01}. However, this second neutron source plays a
more relevant role at low metallicity, even in low mass AGB stars.

Our knowledge of the physical properties of AGB stars and of their
nucleosynthesis is mainly based on observations and models for
intermediate age stellar populations of the Milky Way. The recent
availability of optical and near IR spectroscopy of AGB stars
belonging to nearby galaxies, dwarf spheroidals as well as the
Magellanic Clouds, allow us to extend our investigation to metal
poor stellar populations with age of the order 1-2 Gyr
\citep{dom04,dela06,rey07,leb08,abi08}. In addition to that, the
growing number of studies on C-enhanced metal-poor stars enriched
in s-elements (CEMPs), as  stimulated by dedicated surveys
\citep{be99,ch03} provide new hints on the nucleosynthesis
occurred in the now extinct AGB-halo population, as due to stars
whose initial masses were in the range 1 to 7 M$_\odot$. The aim
of this paper is to investigate how the theoretical scenario
changes with decreasing metallicity, from the solar value to that
of the metal poor stars belonging to the Galactic halo or to the
nearby galaxies. Extant stellar models have been computed by
adopting limited networks. For example, \citet{kala07} include all
the relevant processes and isotopes from H to Si, plus a few
Fe-group isotopes and an artificial neutron sink to account for
heavy element neutron captures. In this way, they can provide a
reasonable estimate of the light element yields. Our approach is
different. We compute evolutionary models, followed from the
pre-main sequence up to the AGB tip, of stars with initial mass 2
M$_{\odot}$ and different metallicities (namely, $Z$=0.0138,
0.006, 0.003, 0.001 and 0.0001), by adopting a full nuclear
network that includes all the stable and the relevant unstable
isotopes from hydrogen to bismuth. In our previous works
\citep{ga98,bi08}, full nucleosynthesis calculations were obtained
by means of a post-process code: in such a case, the main physical
parameters were obtained by means of stellar models calculated
with a restricted nuclear network involving key isotopes and
reactions \citep{stra97}. Then, an ad hoc (average) $^{13}$C
pocket was imposed. In the present work stellar evolution and
nucleosynthesis are coupled. If the faster post-process
calculation allow us to investigate a large area in the parameter
space, the present, fully coupled, calculations shed light on the
possible feedbacks between chemical and physical evolution. These
new models include several improvements in the input physics and
in the computational algorithms \citep[see][thereinafter
SGC06]{sgc06}: the most important for the AGB evolution are
illustrated in \S~\ref{code}. The calibration of some free
parameters used by the mixing algorithm are discussed in
\S~\ref{beta}. Results are presented in \S~\ref{models},
\S~\ref{varied} and \S~\ref{nsynt}. A final discussion follows in
\S~~\ref{concl}.

\section{The theoretical recipe: radiative opacity, nuclear data, mixing scheme and mass loss}\label{code}

The stellar models here presented have been computed by means of
the FRANEC code \citep[][and references therein]{chi98}. An
updated description of the various input physics adopted in the
AGB computations can be found in SGC06. Owing to the particular
relevance for the present paper, let us describe the radiative
opacity, some key nuclear reaction rates, the mixing algorithm and
the mass loss rate.

As discussed by \citet{marigo}, the increase of the radiative
opacity caused by the C dredged up induces important changes in
the physical structure of the outermost layers of an AGB star,
with important consequences on the effective temperature, the
radius and the mass loss rate. In addition to that, when the CN
cycle takes place in the innermost part of the convective envelope
or immediately below it, an important fraction of the primary C
that is dredged up to the surface may be converted into N. This is
a common phenomenon in massive AGBs (the so called hot bottom
burning, HBB) \citep{su71,ib73}, even if there are several
observational evidence proving that deep-mixing processes can also
occur in low mass AGBs (Cool Bottom Process, CBP) (see
\citealt{no03}). As a consequence of the carbon dredged up, new
molecular species may form in the cool atmosphere of an AGB star,
like CN, HCN or C$_2$. Note that since the amount of primary C
synthesized by the 3$\alpha$ reaction and mixed by
 the third dredge up is of the same order of magnitude at any metallicity,
the resulting overabundance of this element (and eventually N) is
very large at low metallicity. Whilst at solar metallicity the
final C enhancement in the envelope may be a factor of 2 to 4, at
$Z$=0.0001 it rapidly attains a factor of 1000. For this reason,
if the adoption of a proper radiative opacity (including
enhancements of C and N) may only introduce limited quantitative
changes in the overall picture of AGB stars with solar
metallicity, its use at low metallicity is absolutely mandatory,
because it leads to substantial modifications of the whole
theoretical scenario \citep[see][]{cri07}. Therefore, below
$T=10^4$~K we use new opacity tables derived by means of the COMA
code \citep{ari00}, which include all the molecular and atomic
species relevant for AGB atmospheres. At higher temperatures, the
opacity tables have been calculated by means of the the OPAL web
facility (http://www-phys.llnl.gov/Research/OPAL/opal.html).
Table~\ref{tab1} lists the C and the N enhancement factors (with
respect to the solar scaled values) adopted for the opacity tables
at different metallicities. All the elements (except H, He, C and
N) are assumed to be solar scaled (namely [M/Fe]=0). The
enhancement of the $\alpha$ elements (O, Ne, Mg, Si, S, Ca), more
appropriated for low metallicity halo stars, is not considered
here, given that it only adds minor effects on the Rosseland mean
opacity. The full database providing low temperature Rosseland
opacities for a wide range of metallicities with varying carbon
and nitrogen abundances is described in \citet{lea}.

The nuclear network is essentially the same already described in
Section~6 of SGC06: it includes about 500 isotopes (from H to Bi)
and more than 700 nuclear reactions (charged particle reactions,
neutron captures and $\beta$-decays). However, few reaction rates
have been changed or updated. In particular, concerning reactions
among charged nuclei, the $^{14}$N($\alpha$,$\gamma$)$^{18}$F rate
is now taken from \citet{gorres}, whilst for the
$^{22}$Ne($\alpha$,n)$^{25}$Mg rate we follow \citet{jag}. Then,
we have updated the $^{151}$Sm(n,$\gamma$)$^{152}$Sm reaction
rate, following the prescriptions reported by \citet{abbo}, and
the cesium isotopes neutron captures rates, as suggested by
\citet{patro}. Concerning the $^{175}$Lu unthermalized isomeric
state (see Section 7.5 of SGC06), we adopt an isomeric ratio
IR=0.2 for temperatures lower than $2.0 \times 10^8$~K and IR=0.25
for higher temperatures. For a more detailed analysis of this
important branching see \citet{heil08}.

Mixing caused by convection is obtained by means of a
time-dependent mixing scheme in which the degree of mixing between
two mesh points, whose separation is $\Delta$R, is linearly
dependent on the ratio between the time step and the mixing time
scale\footnote{Here the mixing time scale is defined as
$\tau_{mix}=\Delta R/v_c$.}. An exhaustive description of the
mixing algorithm can be found in Section~8 of SGC06 \citep[see
also][]{chi01}. Let us here recall the main concepts. The average
convective velocity ($v_c$) is calculated according to the
mixing-length theory \citep{cg68}, whereas the boundaries of the
convective regions are located as given by the Schwarzschild
criterion. In order to handle the instability of the convective
boundary taking place during the third dredge up episodes, when
the H-rich envelope penetrates the H-depleted core, we assume that
the convective velocity drops to 0 following an exponential
decline, namely:
\begin{equation} \label{param}
v=v_{bce}\exp{\left(-\frac{d}{\beta H_P}\right)} \; ,
\end{equation}
where {\it d} is the distance from the formal convective boundary
(as defined by the Schwarzschild criterion), $v_{bce}$ is the
velocity at the formal convective boundary, $H_P$ is the pressure
scale height and $\beta$ is a free parameter. All the evolutionary
sequences presented in \S~\ref{models} and \ref{varied} of the
present paper have been obtained assuming $\beta=0.1$. In the next
Section we will describe the procedure followed to fix the value
of the $\beta$ parameter. A similar mathematical profile has been
already applied to the calculation of AGB models by \citet{he97}
\citep[see also][]{he20}. However, \citet{he97} used an
exponential decline equation to calculate the diffusion
coefficient, instead of the convective velocity, and applied this
prescription to all the convective boundaries. In this way, they
always obtain overshoot at the top and at the bottom of any
convective zone, while in our models Eq.~\ref{param} works only
when $v_{bce}>0$, as it happens in the case of the bottom of the
convective envelope at the time of the third dredge up. A second
important difference concerns the mixing scheme, which determines
the chemical coupling among mesh points within the same convective
zone. In the diffusive mixing adopted by \citet{he97}, the
variation of the composition during a given time step depends on
the second derivative of the composition with $r$ (the radial
coordinate), whilst in our algorithm this relationship is linear.
These differences directly affect the profile of protons in the
transition zone between the fully-convective envelope and the
radiative core at the time of the third dredge up and the
temperature attained in the He-rich intershell during a thermal
pulse. We will discuss the consequences on the nucleosynthesis in
the next section.

Concerning mass loss, we adopt a Reimers' formula ($\eta =0.4$)
for the pre-AGB evolution, whilst for the AGB, we follow a
procedure similar to the one described by \citet{vw93}, but
revising the mass loss-period and the period-luminosity relations,
taking in to account more recent infrared observations of AGB
stars (see SGC06 for the references). A comparison among our mass
loss rate and those obtained by means of the Reimers' formula and
the Vassiliadis and Wood prescriptions can be found in Section~5
of SGC06. An important remark concerns the dependence of the mass
loss rate on the metallicity. The mass loss-period relation we use
is based on data relative to Galactic AGB stars having nearly
solar iron. In a recent paper, \citet{groe} showed that the same
relation applies also to carbon stars in the Magellanic Clouds,
whose metallicity is lower, on average, than that of the Galactic
C-stars. Similar indications come from Spitzer observations of
mass loss of carbon stars belonging to dwarf Spheroidal Galaxies
\citep[see][]{lazi}. In \S~\ref{varied} and \S~\ref{vario} we
discuss the effect of different mass loss prescriptions on the
evolution and nucleosynthesis of the $Z$=0.0001 model.

\section{Instability of the convective boundary,
dredge up and the formation of the $^{13}$C pocket}\label{beta}

In this Section we describe the models obtained by changing the
value of the $\beta$ parameter introduced in the previous Section
to treat the instability of the convective boundary layer, arising
when the H-rich convective envelope penetrates the He-rich zone.
During a third-dredge-up episode, a chemical discontinuity forms
at the convective boundary, causing an abrupt change of the
radiative opacity. If a bare Schwarzschild criterion is adopted to
limit the mixing, the average convective velocity drops from about
$10^4$ cm/s to 0 in 2 adjacent mesh points. In this condition, any
perturbation causing mixing of material across the boundary layer
leads to an increase of the opacity in the underlying stable zone,
which immediately becomes unstable against convection. As it is
well known, a similar condition occurs at the external border of
the convective core during the central He-burning phase
\citep{caste}. The most important consequence of this instability
is the propagation of the convective instability; if in the case
of the central-He burning such an occurrence implies the growth of
the convective core, in the case of the TDU a deeper penetration
of the convective envelope into the H-exhausted core is expected.
Actually, it should exist a transition region, between the fully
convective envelope and the radiatively stable H-exhausted core,
where the convective velocity smoothly decreases to 0 and,
therefore, a partial mixing takes place. In such a way, after any
dredge-up episode, a zone with a smoothed H profile is left
behind. Since the amount of $^{12}$C is rather large (about 20 \%
by mass fraction) in this transition zone, at H re-ignition those
few protons are captured by the abundant $^{12}$C and form a
$^{13}$C pocket. Later on, the s~process nucleosynthesis is
activated, through the $^{13}$C$(\alpha,$n)$^{16}$O reaction. Note
that a $^{14}$N pocket and a $^{23}$Na pocket form, partially
overlapped to the $^{13}$C pocket (see next Section).

\begin{figure*}[tpb]
\centering
\includegraphics[width=\textwidth]{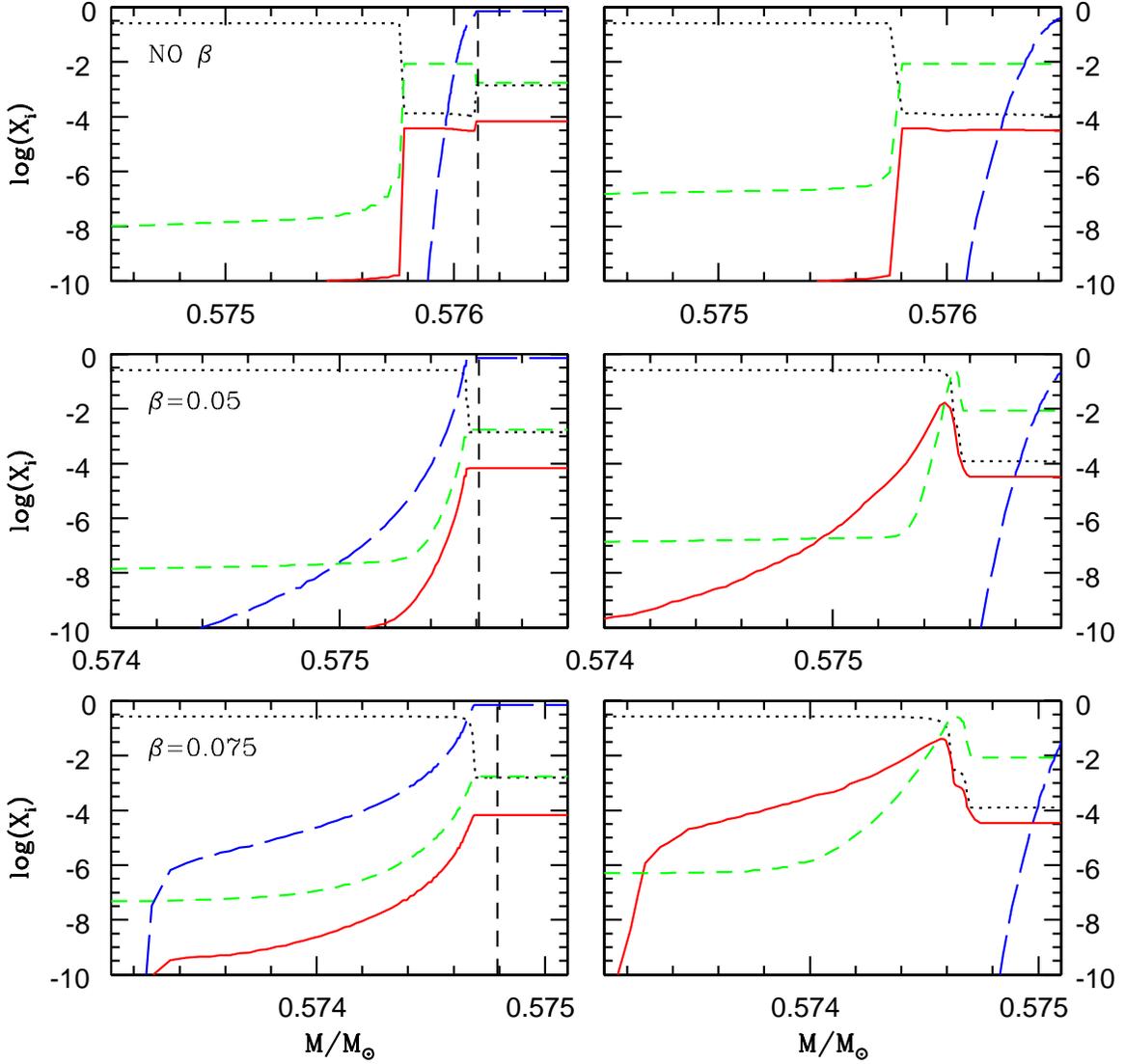}
\caption{Each panel shows the composition of the core-envelope
transition zone of the solar metallicity model (M=2 M$_\odot$) at
the moment of the 3$^{rd}$ TDU, as obtained by adopting low values
of the $\beta$ parameter. We report the mass fractions of $^{12}$C
(dotted line), $^{13}$C (solid line), $^{14}$N (short-dashed line)
and H (long-dashed line) in the transition zone. Left panels refer
to the epoch of the maximum penetration of the convective envelope
into the He intershell during the TDU, whilst right panels show
the same region after the formation of the $^{13}$C pocket. Note
that in the $\beta=0$ case, the convective envelope does not
penetrate the H-exhausted core, so that the showed H profile is
due to the previous shell-H burning, rather than to the operation
of the exponential decline of the convective velocity described by
Eq.~\ref{param}. } \label{fig1a}
\end{figure*}

\begin{figure*}[tpb]
\centering
\includegraphics[width=\textwidth]{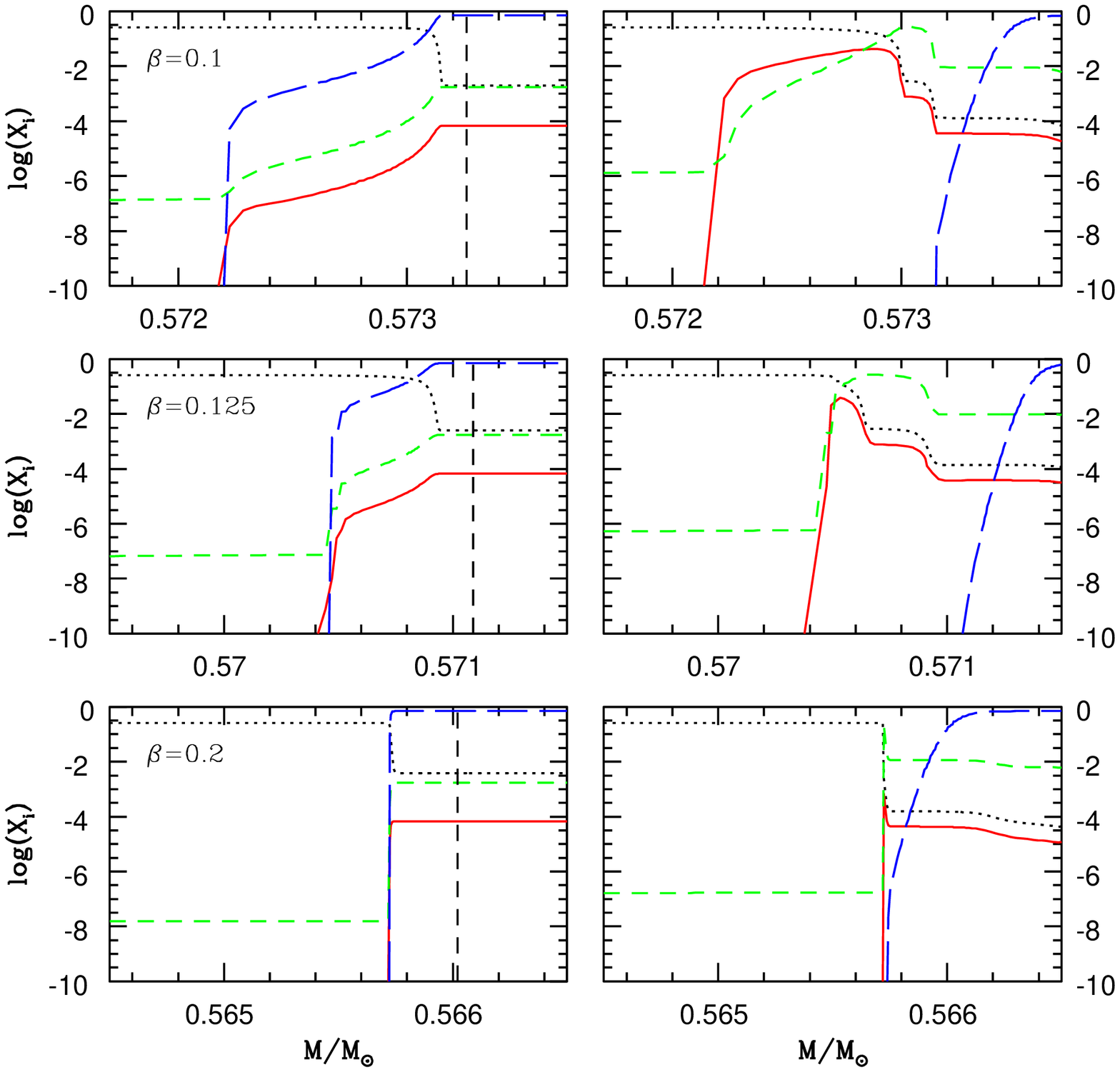}
\caption{Same as in Fig.~\ref{fig1a}, but for larger values of the
$\beta$ parameter.} \label{fig1b}
\end{figure*}

\begin{figure*}[tpb]
\includegraphics[width=\textwidth]{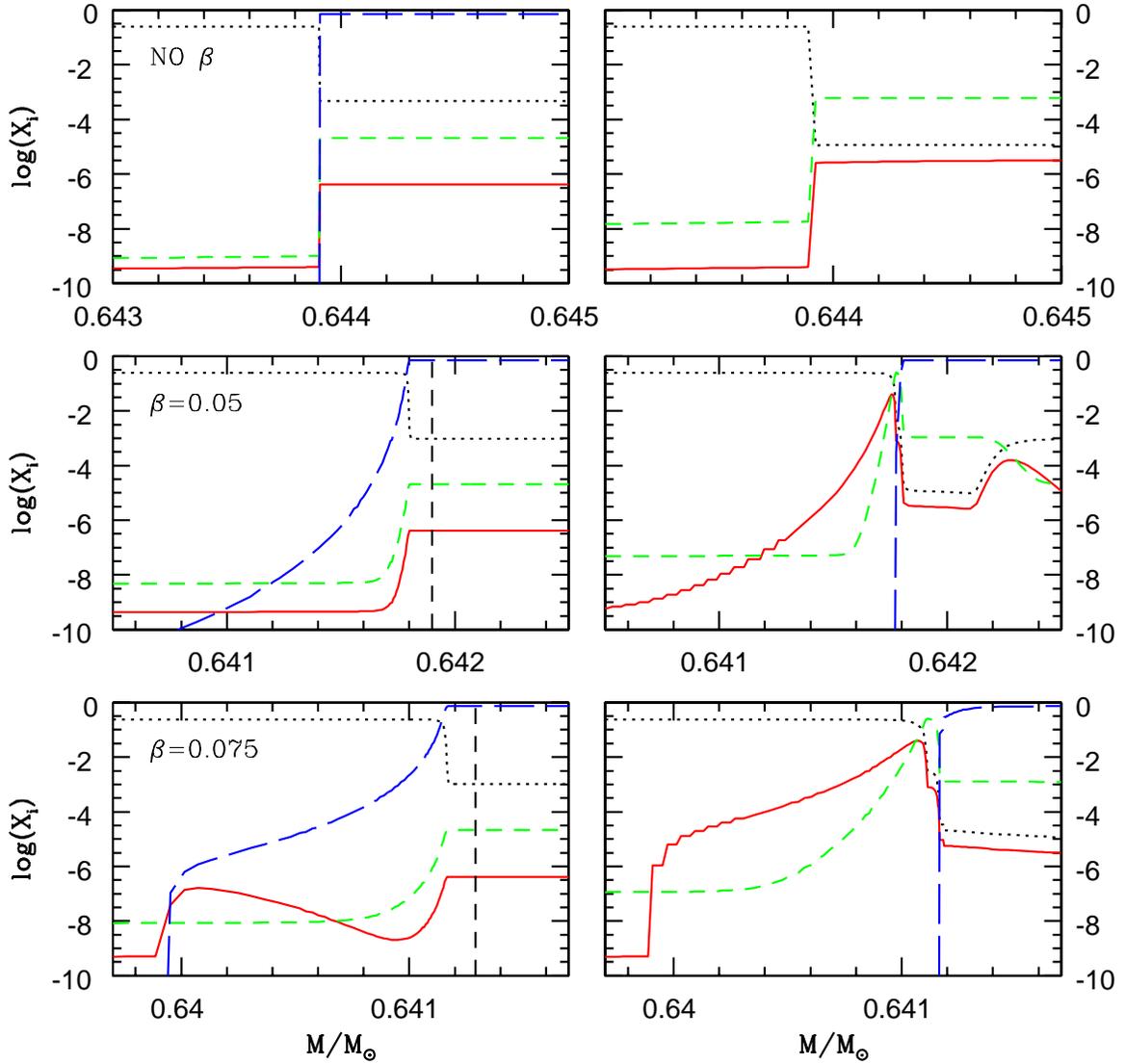}
\caption{Same as in Fig.~\ref{fig1a}, but relative to the 2$^{nd}$
TDU of the model with $Z$=0.0001 and M=2 M$_\odot$.} \label{fig2a}
\end{figure*}

\begin{figure*}[tpb]
\centering
\includegraphics[width=\textwidth]{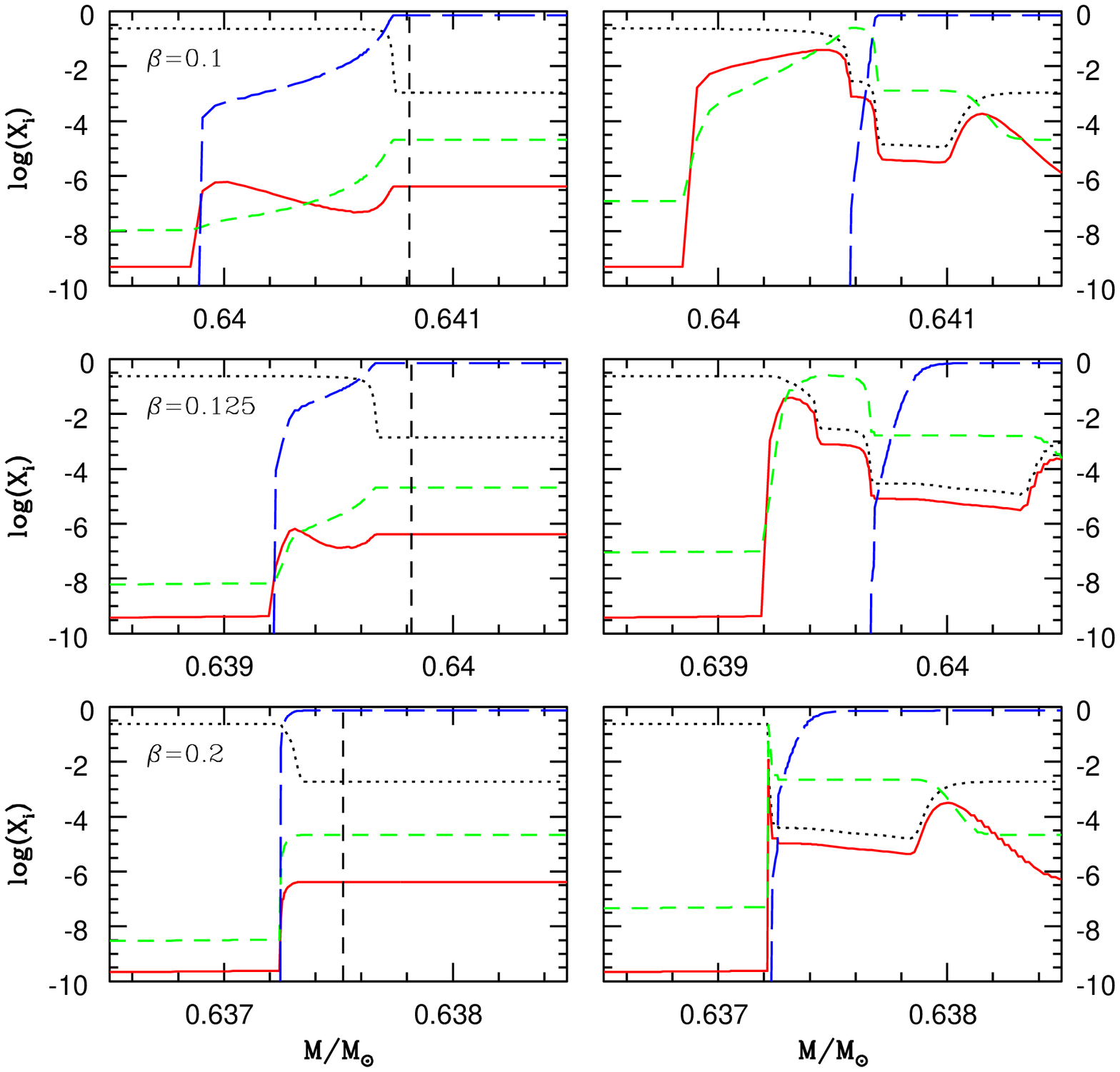}
\caption{Same as in Fig.~\ref{fig2a}, but for larger values of the
$\beta$ parameter.} \label{fig2b}
\end{figure*}

\begin{figure*}[tpb]
\centering
\includegraphics[scale=0.6]{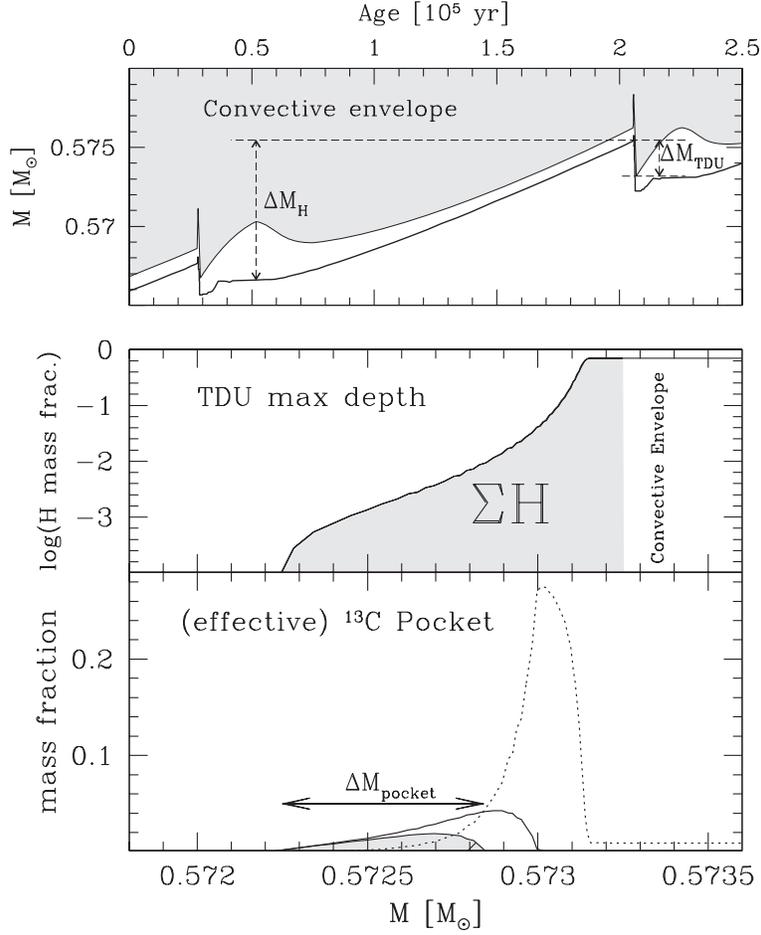}
\caption{{\it Upper panel}: the evolutions of the transition
region between the convective envelope and the H-exhausted core in
between the 4$^{th}$ and the 5$^{th}$ thermal pulse of the 2
M$_\odot$ model with solar metallicity and $\beta=0.1$. The more
external line represents the inner border of the convective
envelope while the lower line shows the location of the layer
where the nuclear energy production is maximum in the H-burning
shell. The 2$^{nd}$ and the 3$^{rd}$ TDU episode are easily
recognized. The quantity $\Delta M_{\rm H}$ and $\Delta M_{TDU}$
are also graphically illustrated. {\it Central panel}: H profile
in the top layer of the H-exhausted core at the epoch of the
maximum penetration of the convective envelope during the 3$^{rd}$
TDU episode. The shaded area represents the quantity $\Sigma$H,
namely the total mass of H left below the formal border of the
convective envelope. {\it Bottom panel}: the same region, but
after the development of the $^{13}$C pocket. The solid and the
dotted lines represent the mass fractions of $^{13}$C and
$^{14}$N, respectively, while the shaded area is
$\Sigma^{13}$C$_{\rm eff}$, namely the mass of the {\it effective}
$^{13}$C in the pocket (see footnote 3 for the definition of {\it
effective} $^{13}$C). The arrow shows the parameter $\Delta
M_{pocket}$, corresponding to the extension (in mass) of the
region where the {\it effective} $^{13}$C is larger than
10$^{-3}$.} \label{figdef}
\end{figure*}

\begin{figure*}[tpb]
\centering
\includegraphics[width=\textwidth]{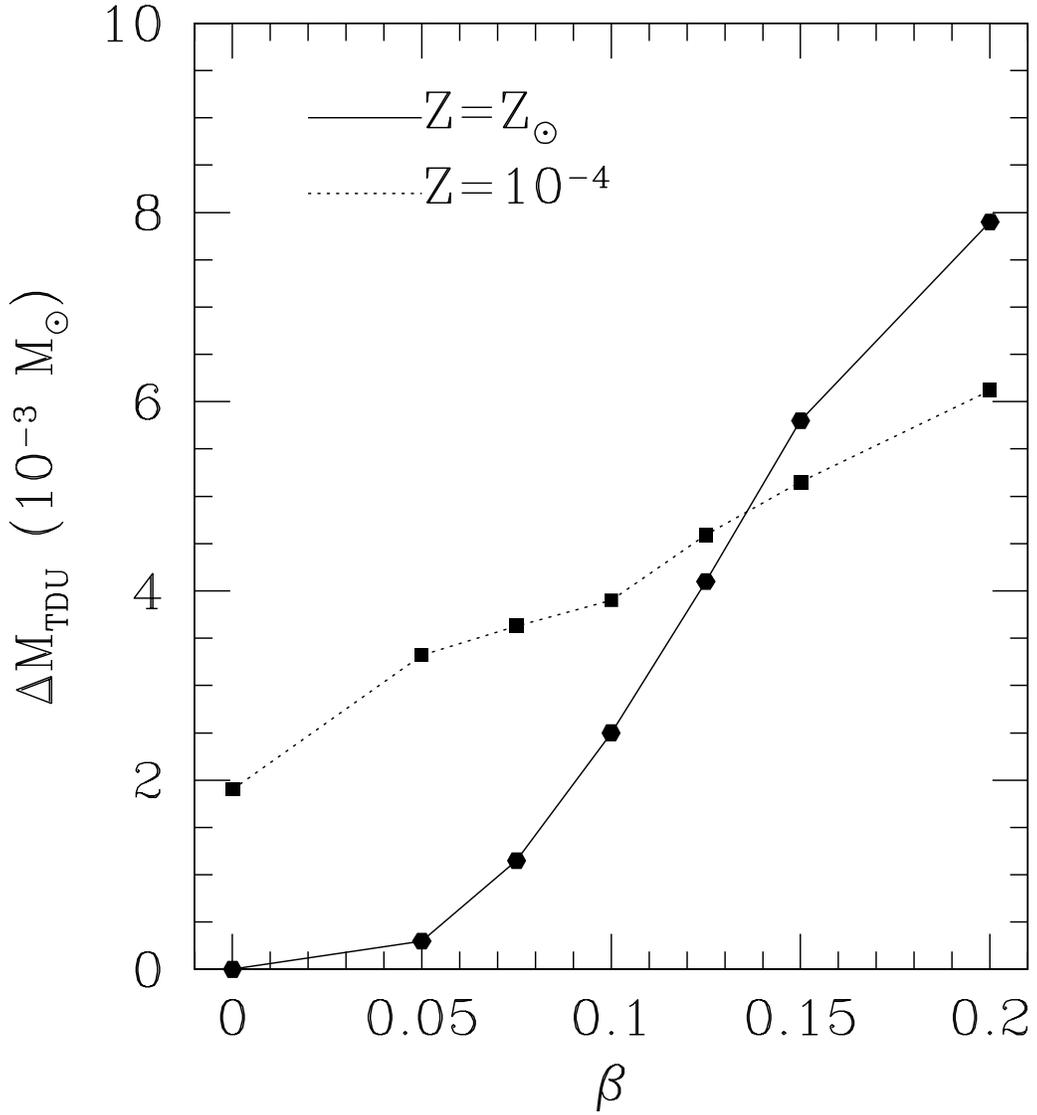}
\caption{The mass portion of the He intershell that is dredged up
versus the $\beta$ parameter after the 3$^{rd}$ pulse followed by
TDU of the solar metallicity model and after the 2$^{nd}$ pulse
followed by TDU of the $Z$=0.0001 model. See text for details.}
\label{fig3}
\end{figure*}

\begin{figure*}[tpb]
\centering
\includegraphics[width=\textwidth]{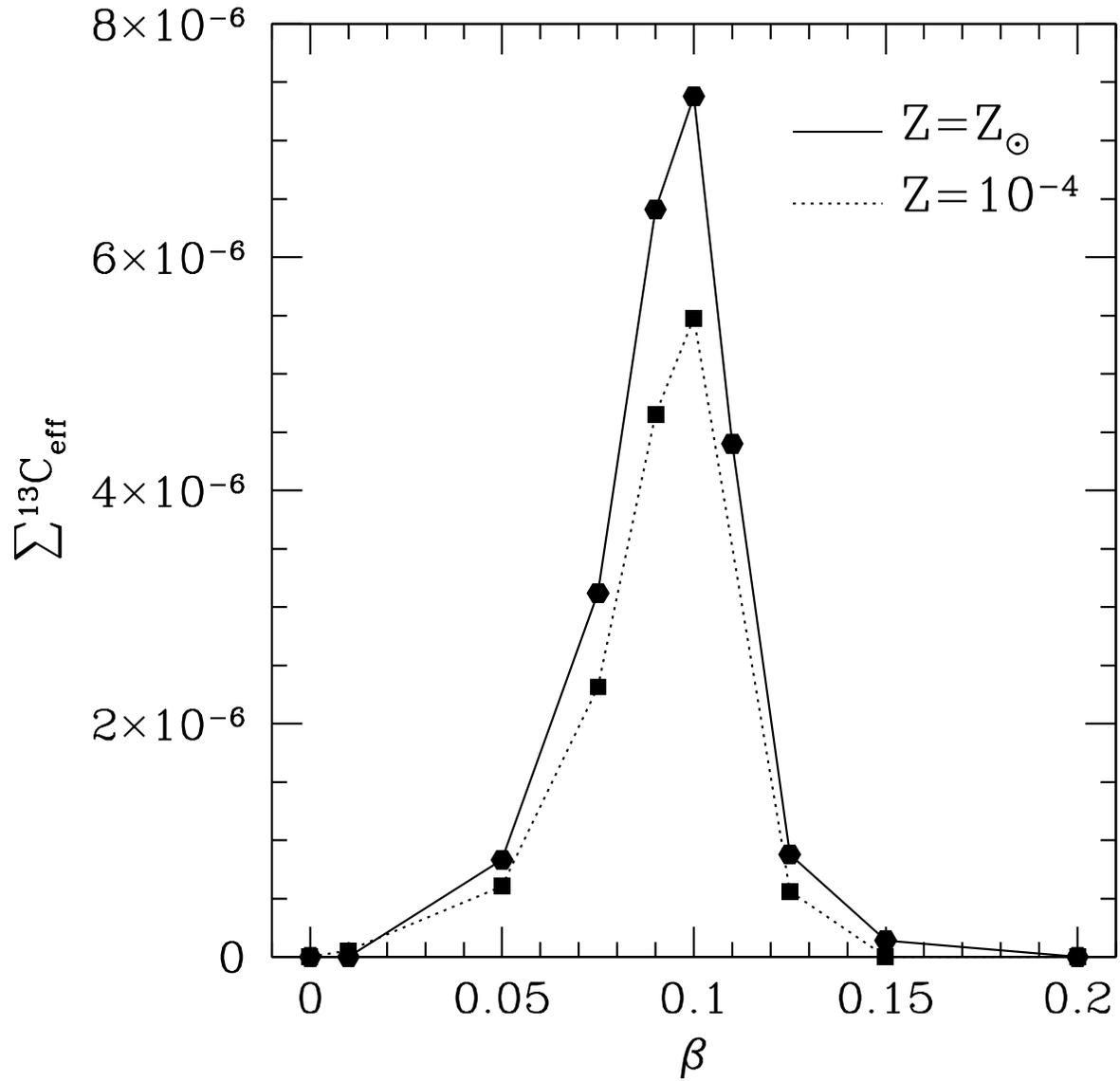}
\caption{The variation with $\beta$ of the effective $^{13}$C mass
in the pocket after the 3$^{rd}$ pulse followed by TDU of the
solar metallicity model and after the 2$^{nd}$ pulse followed by
TDU of the $Z$=0.0001 model. See text for details.} \label{fig4}
\end{figure*}

The mixing algorithm described in the previous Section mimics the
formation of such a transition zone. The free parameter ($\beta$),
used to tune the decline of the average convective velocity at the
base of the convective envelope, determines the extension of such
a transition zone and, in turn, the amount of $^{13}$C available
for the neutron capture nucleosynthesis.
 To calibrate such a parameter, we have repeated
the calculation of the same sequence TP-interpulse-TP for
$0\le\beta\le0.2$. This test has been performed at two
metallicities ($Z$=0.0138 and $Z$=0.0001) in order to test the
sensitivity of our calibration procedure when changing the metal
content. Results are illustrated in Fig.~\ref{fig1a},
Fig.~\ref{fig1b}, Fig.~\ref{fig2a} and Fig.~\ref{fig2b}. Left
panels refer to the epoch of the maximum penetration of the
convective envelope during the third dredge up \footnote{The heavy
vertical-dashed line indicates the location of the formal boundary
of the convective envelope, as defined by means of the
Schwarzschild criterion.}. In the right panels, we show the same
region, but after the H re-ignition, once the convective envelope
recedes and the $^{13}$C pocket is fully developed.  The results
of these test models are quantitatively summarized in
Table~\ref{tab2}, where in columns 1 to 7 we report: the value of
$\beta$, the mass $\Delta M_{TDU}$ of the H-depleted material that
is dredged up, the ratio $\lambda=\Delta M_{TDU}/\Delta M_{\rm H}$
(being $\Delta M_{\rm H}$ the mass of the material that has been
burned by the H-shell during the previous interpulse period), the
mass $\Delta M_{pocket}$ of the zone where the mass fraction of
{\it effective} $^{13}$C is larger than $10^{-3}$, the product
$\beta\times H_P$ at the epoch of the maximum penetration of the
convective envelope, the total mass $\Sigma$H of hydrogen left
below the formal convective boundary at the same epoch and the
total mass $\Sigma ^{13}{\rm C}_{\rm eff}$ of {\it effective}
$^{13}$C within the pocket\footnote{The mass fraction of {\it
effective} $^{13}$C in a given mesh point is defined as
$X_{13}^{eff}=X_{13}-X_{14}\frac{13}{14}$, where $X_{13}$ and
$X_{14}$ are the mass fractions of $^{13}$C and $^{14}$N,
respectively. Owing to its large neutron capture cross section,
$^{14}$N is the most efficient neutron poison, so that the
s-process nucleosynthesis is significantly depressed where it
becomes comparable to the $^{13}$C. For this reason, we define the
mass of the $^{13}$C pocket as the mass of the zone where
$X_{13}^{eff}>10^{-3}$ (see Fig.~\ref{figdef}).}.
Fig.~\ref{figdef} illustrates the definitions of the various
quantities reported in Table~\ref{tab2}. $\Delta M_{TDU}$ and the
masses of the effective $^{13}$C for both metallicities are also
plotted, as a function of $\beta$, in Fig.~\ref{fig3} and
Fig.~\ref{fig4}, respectively. The third dredge up occurs at lower
core masses when a velocity profile is applied at the bottom of
the convective envelope. Moreover, larger $\beta$ values imply
deeper TDU. On the contrary, the total amount of $^{13}$C in the
pocket shows a maximum at $\beta=0.1$. Note that the pocket
practically disappears for the largest $\beta$ value, reflecting
the fact that when enough H is mixed into the transition zone, the
CN cycle goes to the equilibrium and $^{14}$N, rather than
$^{13}$C, is produced. For $\beta=0.1$, the effective $^{13}$C
masses are 7.4 $\times 10^{-6}$ and 5.5 $\times 10^{-6}$
M$_\odot$, at $Z$=0.0138 and $Z$=0.0001, respectively. These
values are close to the standard one adopted by \citet{ga98},
namely 4 $\times 10^{-6}$ M$_\odot$ \footnote{ This amount of
$^{13}$C is often referred as the ST ({\it standard}) case for low
mass AGB nucleosynthesis models.}. \citet{ga98} showed how this
choice of the $^{13}$C-pocket allows the buildup of the main and
the strong components, {\it i.e.} the distribution of s-elements
from the Sr-Y-Zr peak up to the Pb-Bi peak in the solar system,
whilst other works \citep{bu01,ab01,ab02} argue that a certain
spread of the effective $^{13}$C mass is required in order to
reproduce the heavy elements overabundances in a large sample of
Galactic C(N type) stars. More recently, \citet{bi08} confirm the
need for such a spread, extending their analysis to CEMPs stars.
It is worth to note that whilst \citet{ga98} assume a constant
effective $^{13}$C mass, in our models we find a characteristic
evolution of the $^{13}$C pocket (see Fig.~\ref{fig5}). In
particular, a maximum mass of $^{13}$C is attained after very few
thermal pulses (2 or 3); later on, the pocket progressively
shrinks, in mass, until it disappears during the late part of the
AGB evolution (see next section). Nevertheless, the success of the
post-process calculations in reproducing a large amount of
observational data concerning AGB stars
\citep{la95,bu95,ab01,ab02}, their progeny on the post-AGB
\citep{rey07}, Galactic evolution of s~process elements
\citep{tra99,tra01,tra04} and the isotopic composition of C-rich
pre-solar grains (SiC) \citep{zi06a,lu99} suggests to adopt
$\beta$ values allowing the formation of $^{13}$C pocket whose
average size is similar to the ST case. On the other hand, as
shown in Fig.~\ref{fig4}, only values of $\beta$ close to 0.1
provide enough {\it effective} $^{13}$C. For this reason, in all
the computation here presented we have adopted $\beta=0.1$ (but
see \S~\ref{concl} for further remarks).

Let us come back to the differences between our mixing scheme and
the one adopted by \citet{he97}. Basing on a few additional models
computed by using a diffusion algorithm instead of our linear
mixing, we have verified that in this case the extension in mass
of the zone where the proton abundance left by the TDU is
$10^{-3}<X<10^{-2}$ cannot be larger than about $10^{-5}$
M$_\odot$ (for any value of the $\beta$ parameter)\footnote{This
is the range of hydrogen mass fraction required to produce enough
{\it effective} $^{13}$C. If $X>10^{-2}$, $^{14}$N rather than
$^{13}$C would be produced, while for $X<10^{-3}$ too few $^{13}$C
is synthesized.}. Instead, as shown in Fig.~\ref{figdef}, when our
mixing algorithm is adopted, the same region extends for a few
$10^{-4}$ M$_\odot$. Such a difference affects the total mass of
the effective $^{13}$C within the pocket, which may be about 20
times larger in the case of a linear mixing scheme \citep[for
comparisons, see Figure 4 in][]{he20}. As a consequence, the
resulting s-process yields are significantly reduced if a
diffusive mixing is adopted. A further difference concerns the
overshoot at the bottom of the convective zone generated by a
thermal pulse. In our models, the exponential decline described by
Eq.~\ref{param} switches off when, as in this case, the radiative
gradient is equal to the adiabatic gradient at the convective
border, so that $v_{bce}$=0. On the contrary, \citet{he97} force a
certain overshoot below the base of this convective zone, in spite
of the strong negative difference between the radiative and
adiabatic gradient and the huge entropy barrier generated by the
shell-He burning. This implies stronger thermal pulses and larger
temperatures at the base of the He-rich intershell, thus
increasing the efficiency of the s-process nucleosynthesis powered
by the $^{22}$Ne$(\alpha,$n)$^{25}$Mg reaction
\citep[see][]{lu03}.

A reliable (hydrodynamical) model of convection capable to
describe the mixing across the boundaries of the major convective
zones should provides a more realistic description of the AGB
evolution and nucleosynthesis \citep[for a recent attempt,
see][]{mear}. On the other hand, comparisons between the predicted
AGB nucleosynthesis and the abundances observed in AGB stars may
be used to constrain the efficiency of convection. This is the
approach we follow in the present paper.

\section{Reference models at different metallicity}\label{models}

In this Section we present and compare 5 evolutionary sequences of
AGB models having the same initial mass (2 M$_{\odot}$), but
different initial composition, namely: ($Z$;~$Y$)=
(0.0138;~0.269), (0.006;~0.260), (0.003;~0.260), (0.001;~0.245),
(0.0001;~0.245). The mass is representative of low mass AGB stars
and the five metallicities almost span the entire metal
distribution of our Galaxy and of the extragalactic resolved
stellar populations, like the Magellanic Clouds, M31 and several
dwarf spheroidal galaxies. The most metal-rich model corresponds
to the composition of the pre-solar nebula, as derived by means of
an up-to-date standard solar model obtained by adopting the latest
compilation of solar abundance ratios (\citealt{lo03}; see
\citealt{pi07} for more details). The 5 evolutionary tracks, from
the pre-main sequence up to the end of the AGB, are reported in
Fig.~\ref{fig6}. The TP-AGB phase is characterized by large
oscillations in luminosity, associated to the expansions and
contractions powered by thermal pulses, and an evident red
excursion that is the direct consequence of the formation of new
molecular species taking place after the transition from O- to
C-rich atmosphere.

In Fig.~\ref{fig7} we show the evolution, during the TP-AGB phase,
of the positions, in mass coordinates, of the inner border of the
convective envelope, of the location of the maximum energy
production within the H-burning shell and of the location of the
maximum energy production within the He-intershell, top to bottom
line, respectively. The last 2 lines separate 3 main regions
within the star: the innermost zone is the degenerate CO core, the
intermediate one is the He intershell and the more external is the
H-rich envelope. The evolution of the 3 lines mark the most
important events occurring during the TP-AGB phase. Between two
thermal pulses (interpulse phase), the H-burning shell advances in
mass, while the He burning is practically off. Then, when the He
intershell attains a critical mass, the He flash starts. After a
few years, during which the thermonuclear runaway occurs, a
quiescent He burning settles in, while the H burning dies down.
Later on, within about 1000 yr, the convective envelope penetrates
the H-He discontinuity (third dredge up). Note that for a certain
time during the interpulse, the line representing the maximum
energy production within the He-intershell moves suddenly upwards.
At that epoch, the 3$\alpha$ reactions are practically
extinguished in the whole He intershell, but the temperature in
the region occupied by the newly formed $^{13}$C pocket is large
enough ($T \sim 1.0\times 10^8$ K) to ignite the
$^{13}$C($\alpha$,n)$^{16}$O reaction and the consequent s~process
nucleosynthesis. Indeed, the $^{13}$C burning releases about 5-10
MeV per reaction, depending on the energy of the consequent
neutron capture on a seed nucleus. Note that this additional
energy release has a negligible effect on the further
physical evolution.\\
Table~\ref{tab3} reports some properties characterizing the 5
evolutionary sequences, namely (from left to right): the
progressive number of the thermal pulse ($n_{TP}$), the total mass
at the time of the onset of the thermal pulse ($M_{Tot}$), the
corresponding mass of the H-exhausted core ($M_{\rm
H}$)\footnote{The corresponding envelope mass is simply:
$M_{env}=M_{Tot}-M_{\rm H}$.}, the mass of the H-depleted material
dredged up ($\Delta M_{TDU}$), the maximum mass of the convective
zone generated by the TP ($\Delta M_{CZ}$), the mass burnt by the
H-shell during the previous interpulse period ($\Delta M_{\rm
H}$), the overlap factor {\it r} (defined as the fraction of
$\Delta M_{CZ}$ already included in the convective zone generated
by the previous TP), the $\lambda$ factor (defined as the ratio
between $\Delta M_{TDU}$ and $\Delta M_{\rm H}$), the average
number $n_c$ of neutrons captured per initial iron seed nucleus
during the radiative $^{13}$C pocket burning, the duration of the
interpulse period preceding the current TP ($\Delta t_{ip}$), the
maximum temperature attained at the bottom of the convective zone
generated by the TP ($T_{MAX}$), the surface metallicity after the
dredge up ($Z_{surf}$) and the corresponding C/O ratio. Finally,
the total mass of the material cumulatively dredged up during the
whole evolutionary sequence is reported in the last row
($M_{TDU}^{tot}$).

As already discussed in \citet{stra03}, the TDU starts when the
mass of the H-exhausted core exceeds a critical value, whereas it
ceases when the envelope mass, eroded by the mass loss, is reduced
down to a critical value. In the five evolutionary sequences
presented here, the first TDU episode takes place when (from
$Z=0.0138$ to $Z=0.0001$) M$_{\rm H}$ is 0.560, 0.569, 0.576,
0.604, 0.641 M$_\odot$. Note that at the lowest metallicities
($Z=0.001$ and 0.0001) a 2 M$_\odot$ star already develops rather
large core mass during the early-AGB, so that the third dredge up
occurs almost soon after the beginning of the TP-AGB phase. At
those metallicities, stars with smaller mass may experience TDU
episodes at smaller $M_{\rm H}$. The last TDU episode takes place
when $M_{env}$ is 0.534, 0.183, 0.207, 0.292, 0.242 M$_\odot$, for
$Z=0.0138$, 0.006, 0.003, 0.001 and 0.0001,
respectively.\\
As discussed in \S~\ref{beta}, the introduction of a smoothed
profile of the convective velocities at the bottom of the envelope
favors the occurrence of TDU at smaller core masses and enhances
its efficiency, with respect to models where the bare
Schwarzschild criterion is adopted \citep{stra97}. As an example,
in the 2 M$_\odot$ model with $Z=0.02$ of \citet{stra97}, the
first TDU takes place when the core mass is 0.61 M$_\odot$ and has
a reduced efficiency with respect to the model presented here. A
second reason at the base of these differences is the rate of the
$^{14}$N(p,$\gamma$)$^{15}$O reaction: we are currently using the
most recent low-energy laboratory measurement of this rate by
\citet{imbri}, which is a factor of two lower with respect to the
rate proposed by \citet{cf88} and \citet{angulo}. Being the
$^{14}$N(p,$\gamma$)$^{15}$O reaction the bottleneck of the CNO
cycle, a lower rate implies a reduced H-burning efficiency,
stronger TPs and larger TDUs, as already verified by
\citet{stra00} and confirmed by \citet{heau}.

An important quantity affecting the nucleosynthesis is the maximum
temperature attained at the base of the convective zone generated
by a TP. During the AGB evolution, this value progressively
increases, reaches a maximum and, then, slightly decreases toward
the last TP (Fig.~\ref{fig8}, upper panel). In the most metal-rich
model, $T_{MAX}$ remains always below 3.0$\times 10^8$ K: this is
due to the fact the H burning shell is efficient and, therefore,
the TPs are weaker with respect to lower metallicities
\citep[see][]{stra03}. For that reason, in these models the
$^{22}$Ne$(\alpha,$n)$^{25}$Mg reaction is only marginally
activated. On the contrary, in the two most metal-poor models,
$T_{MAX}$ attains higher values (up to 3.2$\times10^8$ K), so that
the $^{22}$Ne$(\alpha,$n)$^{25}$Mg adds a significant contribution
to the s~process nucleosynthesis, in particular for the treatment
of the branchings along the s-path. We will come back on this
issue in the next Section.

Concerning $\Delta M_{TDU}$, it depends non-linearly on the
metallicity, the envelope mass and the core mass \citep[see
formula 3 in ][]{stra03}. Fixing the values of the other two
parameters, it is generally larger at lower metallicity. It is
also larger at larger envelope mass or/and core mass. During the
AGB evolution, the core mass increases, while the envelope mass
decreases, so that $\Delta M_{TDU}$ initially increases, following
the growth of $M_{\rm H}$, but then decreases, when the envelope
erosion by mass loss is stronger. This is clearly showed in
Fig.~\ref{fig8} (lower panel). Looking at the total mass of the
material that is cumulatively dredged up, it has a minimum in the
case of the most metal-rich evolutionary sequence ($\Delta
M^{TDU}_{tot}=3.62\times 10^{-2}$ M$_\odot$), whilst the maximum
is attained at $Z=$0.001 ($\Delta M^{TDU}_{tot}=1.31\times
10^{-1}$ M$_\odot$). In \S~\ref{varied} we will show how the
assumed mass loss rate affects this quantity.

The mass of the convective zone generated by a TP shrinks during
the AGB evolution. In all of the 5 sequences, the last $\Delta
M_{CZ}$ is about a factor of two smaller than the first. It is
also a factor of two smaller at the lowest metallicity with
respect to the largest one. On the contrary, the $\lambda$ factor
is smaller at larger metallicities, whilst the overlap ($r$
factor) is similar in all the five sequences.

The $Z$=0.0138 sequence attains the C-star condition (C/O$>$1)
after 7 dredge up episodes and the final C/O is 1.88. More metal
poor models becomes C-stars sooner and develop larger final C/O
ratios: in the $Z$=0.0001 case, for example, the final C/O is 53.
This is mainly due to the lower initial amount of O, rather than
to the deeper dredge up.

\section{Changing the mass loss and the mixing length efficiency.} \label{varied}

In this Section we will show how the present theoretical scenario
is affected by the assumed treatment of hydrodynamical phenomena,
not explicitly included in the hydrostatic equations, such as mass
loss and convection, whose efficiency in the models relies on some
free parameters. For this purpose, we have calculated 2 additional
evolutionary sequences, both at $Z$=0.0001. In the first one we
have adopted a classical Reimers' formula ($\eta=0.4$), up to the
tip of the AGB, instead of that based on the calibrated mass
loss-period-luminosity relation (see \S~\ref{code}). The second
evolutionary sequence has been obtained by changing the value of
the mixing length parameter ($\alpha=\Lambda/H_P$, where $\Lambda$
is the mixing length) from 2.15, as required by the standard solar
model \citep[see][]{pi07}, to 1.8. The evolutions of $T_{MAX}$ and
$\Delta M_{TDU}$ for these two additional models are compared with
the reference model in Fig.~\ref{fig9}. Either the mass loss rate
and the mixing length parameter are usually calibrated on stars
with solar or nearly solar composition and little is known about
the validity of these calibrations at low metallicity; for a
discussion on the possible calibration of the mixing length at
different $Z$ see \citet{chi95} and \citet{fe06}. The early TP-AGB
evolution of the Reimers' model, which is mainly controlled by the
growth of the H-exhausted core, is similar to the one obtained in
the case of the reference model. However, when the envelope
erosion becomes important, the two sequences depart from each
other. In particular, as a consequence of the higher mass loss
rate, the reference model terminates sooner. As a result, the
Reimers' model experiences a larger number of TPs and third
dredge-up episodes. The total mass cumulatively dredged up is 1.6
$\times 10^{-1}$ M$_\odot$, to be compared with $9.54\times
10^{-2}$ M$_\odot$ in the reference model (see Table~\ref{tab4}).

The main effect of reducing the mixing length parameter is a
smaller extension of the convective regions. As a consequence, the
third dredge up is weaker and the temperature at the bottom of the
convective zone powered by thermal pulses is lower (see
Fig.~\ref{fig9}). In this case the total mass cumulatively dredged
up is just $6.13\times 10^{-2}$ M$_\odot$ (see Table~\ref{tab5}).

The corresponding effects on the nucleosynthesis are discussed in
\S~\ref{vario}.

\section{Nucleosynthesis in the He intershell} \label{nsynt}

Let us start with the solar metallicity model. As noted in
\S~\ref{models}, owing to the low temperature within the
convective zones generated by the various TPs, the
$^{22}$Ne$(\alpha,$n)$^{25}$Mg reaction plays a marginal role as a
neutron source at high metallicities, so that the most important
nucleosynthesis site is the transition region between the core and
the envelope, where the $^{13}$C pocket forms. In Fig.~\ref{fig10}
we report the mass fractions of selected isotopes in this zone,
after the formation of the third pocket. Actually, three different
''pockets'' can be distinguished. The most internal one is the
$^{13}$C pocket (solid line), which is partially overlapped to a
$^{14}$N pocket (long dashed line). Note that the relevant
s~process nucleosynthesis occurs in the left (more internal) tail
of the $^{13}$C pocket, where the amount of $^{14}$N, the major
neutron poison, is low. The third and smallest pocket is made of
$^{23}$Na (short-long-dashed line). The latter forms where the
abundance of $^{22}$Ne (short-dashed line) is comparable to the
$^{12}$C abundance (dotted line), so that the
$^{22}$Ne(p,$\gamma$)$^{23}$Na reaction competes with the
$^{12}$C(p,$\gamma$)$^{13}$N in the proton capture game. As
firstly suggested by \citet{gomo}, this $^{23}$Na pocket may
provide a significant contribution to the synthesis of sodium in
AGB stars.

The most interesting result obtained by coupling stellar structure
evolution and nucleosynthesis concerns the variations of the size
of the 3 sub-pockets when the star climbs the Asymptotic Giant
Branch. Fig.~\ref{fig11a} and Fig.~\ref{fig11b} show, for the
solar metallicity model, the physical conditions in the
core-envelope transition region at the time of the maximum
penetration of the 3$^{rd}$ and of the 11$^{th}$ ({\it i.e.} the
last) TDU, respectively. Solid and dashed lines represent the
hydrogen and the $^{12}$C profiles, respectively. The abundance
curves in both figures have been shifted upward in order to match
the pressure scale axis. The slanting dashed area shows the region
of the fully convective envelope (as defined by the Schwarzschild
criterion), whilst the horizontal and the vertical dashed areas
mark the regions where the $^{13}$C and the $^{23}$Na pockets will
form later (the $^{14}$N pocket, which overlaps with both the
aforementioned pockets, is not reported in the plot for graphical
reasons). Pressure and pressure scale height are also shown. Note
as the pressure jump between the dense core and the loose envelope
is definitely steeper in the case of the last TDU episode. In
fact, when comparing the two physical structures, it comes out
that that the pressure of the H-exhausted core remains of the same
order of magnitude, whilst the base of the envelope at the moment
of the last TDU episode is more expanded with respect to the
3$^{rd}$ one, showing a pressure difference of about five orders
of magnitude. In both cases, external convection penetrates down
to a layer where the pressure is about 10$^{11}$ dyne and $H_P$ is
about 10$^{10}$ cm, but due to the steeper pressure gradient, the
extension of the zone with a smoothed H profile is significantly
reduced in the last TDU episode. In this case, the resulting
pocket is definitely smaller. Table~\ref{tab6} lists the effective
mass of $^{13}$C for all the pockets produced in the $Z$=0.0138
and $Z$=0.0001 models. In the Table, we report (left to right):
the mass of the H-exhausted core ($M_{\rm H}$), the mass of
effective $^{13}$C in the pocket ($\Sigma ^{13}$C$_{\rm eff}$) and
the mass extension of the pocket ($\Delta M$). Note how the last
pocket is more than one order of magnitude smaller than the first
one.

The shrinkage of the $^{13}$C pocket implies a progressive
decrease of the s~process efficiency as the star evolves along the
AGB. For a long time, it has been assumed that the synthesis of
the main component of the s~process would require the partial
superposition of different neutron exposures: \citet{cl61} showed
that an exponential distribution of these exposures could
reproduce the observed $\sigma N_s$ curve \citep[see
also][]{se65}. \citet{ulri} noted that if the s~process takes
place at the base of the convective zone generated by a thermal
pulse, the partial overlap of the recurrent convective zones
provides, in a natural way, an exponential distribution of neutron
exposures, thus reinforcing Clayton's original suggestion.
However, as argued by \citet{ga98} (see also \citealt{stra95};
\citealt{ar99}), when the s~process occurs during the interpulse
in the $^{13}$C pocket, the scenario becomes more complex and
cannot be described by a simple analytical function. Beside its
relatively small thickness, the stratification of the $^{13}$C
(and of the neutron poisons, as $^{14}$N) within the pocket leads
to stratified neutrons irradiations with different intensities. In
addition to that, owing to the variation of the overall amount of
$^{13}$C available in the pocket, as we find in self-consistent
evolutionary models of AGB stars, this stratification changes
pulse by pulse. In particular, the first pockets, the largest
ones, give a major contribution to the overall AGB nucleosynthesis
compared with that of the smallest late pockets. This is
illustrated in Fig.~\ref{fig12}, where we report, for the solar
metallicity model, the production factors, within the He
intershell at the time of the 1$^{st}$, 3$^{rd}$, 6$^{th}$ and
11$^{th}$ TDU episodes, of nuclei whose synthesis is mainly
ascribed to the s~process (namely, nuclei whose s~process
contribution is larger than 80\%) \footnote{The production factor
is defined as $N_j^s/N_j^{\odot}$, where $N_j^s$ is the abundance
by number and $N_j^{\odot}$ is the corresponding solar scaled
abundance.}. The s~process nucleosynthesis occurs in the radiative
$^{13}$C pocket and leaves a thin layer highly enriched with heavy
elements that is located in the middle of the He intershell. When
the convection powered by the thermal pulse takes place, this
material is spread out in the whole He intershell. In spite of the
large dilution caused by the convective mixing \footnote{Note that
the mass of the largest pocket is only $6\times10^{-4}$ M$_\odot$,
while the convective zones attain about $3\times 10^{-2}$
M$_\odot$.}, the s-element overabundances in the He intershell
rapidly increase during the first TPs. A maximum is attained at
the 6$^{th}$ TP. At that time, in the He intershell the abundances
of the s-only nuclei are on the average a factor of 500 larger
than the initial ones (700 for the ls element and 300 for the hs
elements). Later on, the production factors of the various s
elements in the He intershell decrease. This is due to the
shrinkage of the late pockets, which are no more able to
compensate the convective dilution. Nevertheless, quite large
overabundances are found in the He intershell up to the AGB tip,
because of the partial overlap of the recurrent convective zones
generated by the various TPs.

\subsection{Evolution of the surface composition: the main and the strong s~process
components.}\label{sproc}

The variations of the surface composition after selected TDU
episodes of the solar metallicity model are reported in
Fig.~\ref{fig13}. Looking at the elements beyond Fe (Z=26), the
most abundant species are those corresponding to isotopes with
particularly small neutron capture cross sections. In particular,
the peaks corresponding to magic neutron numbers (N=50, 82 and
126) clearly emerge from the bulk production of the s~process. In
general, magic neutron nuclides act as bottlenecks of the
s~process. At solar metallicity, owing to the relatively large
amount of iron seed, compared with the number of neutrons released
by the radiative $^{13}$C burning, the productions of the light,
ls (Sr-Y-Zr), elements and the heavy, hs (Ba-La-Ce-Nd-Sm),
elements, namely the first and the second s-peaks, are favored
with respect to the lead production (the third s-peak). At the end
of the AGB evolution, the surface ratios [hs/ls] and [Pb/hs] are
$-0.21$ and $-0.33$, respectively. This values are attained after
a few TDU episodes, those following the first and largest $^{13}$C
pockets. Later on, the overabundances continue to grow, as a
consequence of the TDU, but these ratios remain constant.

The final surface composition of the five models with different
metallicity are shown in Fig.~\ref{fig14}. Surface abundances
after selected TDUs are shown in Table~\ref{tab7},
Table~\ref{tab8}, Table~\ref{tab9}, Table~\ref{tab10} and
Table~\ref{tab11}. Only key elements are tabulated. Moreover, a
selection of final elemental abundances, isotopic ratios and
spectroscopic indexes are listed in Table~\ref{tab12}\footnote{The
complete set of these Tables is available in the electronic
version of this paper.}. The $^{13}$C($\alpha$,n)$^{16}$O reaction
is primary-like ({\it i.e.} not directly affected by the
metallicity of the pristine material), whilst the iron seeds scale
with the metallicity. Thus, by lowering the metallicity, the
number of neutron per seed nuclei progressively increases: this
trend clearly results from column 9 of Table~\ref{tab3}. Within
the $^{13}$C pockets, in fact, the average number $n_c$ of
neutrons captured per initial iron seed nucleus increases with the
metallicity, starting from about 40 at solar metallicity up to
4000 at $Z=0.0001$\footnote{Note that $n_c$ involves neutron
captures on both light elements and heavy elements. Neutron
captures on light elements act as neutron poisons, thus decreasing
the number of neutrons available for the s-process
nucleosynthesis. The major neutron poison, represented by the
$^{14}$N(n,p)$^{14}$C reaction, has already been considered in the
estimate of $n_c$.}. Therefore, the bulk of the s~process
nucleosynthesis first moves from ls elements to hs elements and,
at low metallicities, it shifts directly to $^{208}$Pb, at the
termination point of the s path. This behavior is well reproduced
by our models (see Fig.~\ref{fig14}). A second important change
occurring at low Z concerns the second neutron source, the
$^{22}$Ne$(\alpha,$n)$^{25}$Mg. In low metallicity models, owing
to the larger temperature attained at the inner border of the
convective zone generated by the thermal pulse, a second neutron
burst takes places, causing interesting changes in the He
intershell composition. At variance with the nucleosynthesis in
the $^{13}$C pocket, since the maximum temperature in the
convective shell generated by a TP increases pulse by pulse (see
Fig.~\ref{fig8}), it is in the late part of the AGB that the major
effects induced by the second neutron burst become important.
Among them, we recall the overproduction of some neutron-rich
isotopes. Owing to the larger neutron density (10$^{11}$ cm$^{-3}$
instead of 10$^7$ obtained in the case of the radiative $^{13}$C
pocket), several interesting branchings along the s~process path
are activated. Thus, neutron-rich isotopes, whose production is
otherwise prevented by $\beta$ decays of lighter isotopes, can be
produced. Table~\ref{tab13} reports some characteristic isotopic
ratios sensitive to the neutron density. The tabulated ratios
relate to the branchings occurring at $^{85}$Kr, $^{86}$Rb,
$^{95}$Zr, $^{133}$Xe and $^{141}$Ce, respectively. As a
comparison, solar ratios are also reported. Note how they increase
when the metallicity decreases. In particular, the $^{87}$Rb is
underproduced at large metallicity with respect to the lighter
$^{85}$Rb, whereas it is overproduced at low Z. This is a direct
consequence of the opening of the $^{85}$Kr and $^{86}$Rb
branchings, taking place when the neutron density exceeds
$10^8-10^9$ neutron/cm$^3$ \citep[see][]{mala}. Note that, since
$^{87}$Rb is a magic neutron nucleus, the overall production of Rb
is significantly enhanced at low Z. As a result, the ratio of Rb
and ls increases at low Z. For example, log(Rb/Sr) is 0.33, 0.42,
0.53, 0.77 and 0.85 at $Z$=0.0138, 0.006, 0.003, 0.001 and 0.0001,
respectively. In Fig.~\ref{ro_n} we report the maximum neutron
densities attained in the models at various metallicities. As
stressed before, the lower the metallicity is, the higher the
neutron density attained during the TP is.

\subsection{Evolution of the surface composition: from C to
Fe.}\label{pippo}

The most striking consequence of the third dredge up is the
surface carbon enhancement. By reducing the metallicity, the
maximum enhancement with respect to iron increases from [C/Fe]=0.6
($Z$=0.0138 model) to [C/Fe]=3 ($Z$=0.0001 model). Obviously, this
implies that the C/O ratio increases with decreasing the
metallicity, passing from 1.87 at solar metallicity to 53 at
$Z$=0.0001 (see Table~\ref{tab12}). Our models do not show
evidences for HBB, therefore at low metallicities we obtain large
C/N ratios and large $^{12}$C/$^{13}$C isotopic ratios. Note that
possible reduction of these ratios due to deep-mixing processes
(as the CBP) is not explicitly considered in our models. Small
enhancements of nitrogen are found, due to the dredge up of the
thin region of incomplete H burning. Almost all the $^{14}$N left
by the H burning in the He intershell is converted into $^{22}$Ne
during the thermal pulse phase, through the chain
$^{14}$N$(\alpha,\gamma)^{18}$F$(\beta^+\nu)^{18}$O$(\alpha,\gamma)^{22}$Ne.
The $^{22}$Ne plays a fundamental role, particularly in metal-poor
stars. It acts as a neutron source, as a poison and, at very low
Z, as a seed capable to counterbalance the scarcity of iron
\citep[see][]{ga06}. Note that at low Z, most of the $^{22}$Ne is
primary: at $Z$=0.0001, in fact, the amount of carbon in the
envelope largely exceeds the pristine C+N+O since the first TDU
episode. As a result, at the end of the AGB the amount of
$^{22}$Ne in the He intershell is comparable with that found in
the same region of the solar metallicity model. Correspondingly,
the surface abundance of neon is significantly enhanced (see
Table~\ref{tab12}). Despite the very low neutron capture cross
section, the large abundance of $^{22}$Ne allows the production of
light isotopes like Na, Mg and Al. Sodium is further enhanced,
because of the already mentioned formation of the $^{23}$Na
pocket, due to protons capture in the most external layer of the
core-envelope transition zone. At solar metallicity, the majority
of the Na overproduction is due to the ingestion of the $^{23}$Na
pocket at the time of the TDU. At $Z$=0.0001, in addition to that,
neutron captures on $^{22}$Ne taking place in the radiative
$^{13}$C pocket and during the thermal pulse account for 13\% and
35\% of the total Na production, respectively \citep{cri06}.

Concerning O, it is only marginally affected by the internal
nucleosynthesis at large Z. In low metallicity models, the oxygen
enhancements results larger ([O/Fe]=1 at $Z$=0.0001) because of
the reduced initial $^{16}$O abundance. We recall that we have
assumed a solar scaled initial composition, so that [O/Fe] is
nearly 0 at the beginning of the AGB for all the models here
presented. Thus, in order to compare theoretical O overabundances
to those measured in C-enhanced stars belonging to the Galactic
halo (CEMP), one has to add an initial overabundance of, at least,
$0.4-0.5$ dex.

Among the light elements, $^{19}$F is produced at all
metallicities, with a significant production at the lowest one.
Fluorine is synthesized by the $^{15}$N$(\alpha,\gamma)^{19}$F
reaction in the convective zone generated by a thermal pulse. The
$^{15}$N production is due to the $^{18}$O(p,$\alpha)^{15}$N
reaction: therefore, the $^{19}$F production requires the
simultaneous presence of $^{18}$O and protons \citep[see,
e.g.,][]{fo92}. In the He intershell, protons are available in the
radiative $^{13}$C pocket and at the beginning of a thermal pulse
\citep[see, e.g.,][]{lu04}, released by the $^{14}$N(n,p)$^{14}$C
reaction: neutrons are therefore
required for a consistent production of $^{15}$N. \\
In the radiative $^{13}$C pocket neutrons are produced by the
$^{13}$C($\alpha$,n)$^{16}$O reaction, while $^{18}$O is
synthesized by means of the
$^{14}$C($\alpha$,$\gamma$)$^{18}$F($\beta^-\nu)^{18}$O chain. \\
Few neutrons are also available at the beginning of a thermal
pulse: they come from the burning of the $^{13}$C left by the
H-burning shell in the upper zone of the He intershell. This
$^{13}$C remains unburned during the interpulse and is engulfed in
to the convective zone generated by the TP. The $^{13}$C burns in
a convective environment producing neutrons, which are almost
totally captured by the (still abundant) $^{14}$N, giving rise to
a certain release of protons. At that time, the temperature at the
bottom of the convective shell is not high enough for a complete
$^{18}$O depletion, therefore $^{15}$N is efficiently produced.
The $^{22}$Ne$(\alpha,$n)$^{25}$Mg reaction (which is the main
source of neutrons during a TP) does not contribute to the
fluorine nucleosynthesis because, when it takes place, the
temperature is so high that no $^{14}$N and too few $^{18}$O
survive in the convective shell. In summary, we have two main
channels for the $^{15}$N production in the He intershell. At
solar metallicity, the two sources equally contribute to the
fluorine production; at low metallicities, the $^{15}$N
accumulated in the radiative $^{13}$C pocket is the main source of
fluorine, because the $^{13}$C left by the full CNO burning is
rather low for the major part of the AGB lifetime (but see
\S~\ref{vario}).

\subsection{Yields}\label{yields}
One of the aims of this paper is to provide a complete and uniform
set of AGB yields, comprehensive of all chemical species, starting
from hydrogen up to the Pb-Bi s~process ending point. Yields of
selected isotopes at different metallicities are reported in
Table~\ref{tab14}: being the number of considered isotopes too
large to provide it in a paper format, a constantly updated
database of AGB yields is available on the web
\footnote{http://www.oa-teramo.inaf.it/osservatorio/personale/cristallo/
data\_online.html.} and in the electronic version of this paper.
According to \citet{ti80}, the yield is:
\begin{equation}
M_y(k)=\int_0^{\tau(M_i)}[X(k)-X^0(k)]\frac{dM}{dt}dt
\end{equation}
where $dM/dt$ is the mass loss rate, while $X(k)$ and
$X^0(k)$ stand for the current and the initial mass fraction of the
isotope $k$, respectively. Yields are given in solar mass units.
We recall that our calculation have been stopped after the last
TDU episode, when the residual envelope mass is of the order of
(0.2$\div0.5$) M$_\odot$. Except for the natural decay of the
eventually surviving unstable isotopes, the envelope composition
is freezed after this moment. Then, in computing the yields, we
assume that the star loses the whole residual envelope through a
single mass loss episode.

\subsection{Short-lived radioactive isotopes.}

In this Section we discuss the synthesis of a few short-lived
radioactive isotopes, namely $^{26}$Al, $^{36}$Cl, $^{41}$Ca,
$^{60}$Fe, $^{107}$Pd and $^{205}$Pb. The present theoretical
predictions for low mass AGB stars may be used to interpret the
evidence for the presence of these radioactive isotopes in the
early solar system (ESS, see \citealt{wa06} for a review on
short-lived radioactive from AGB stars) and in pre-solar dust
grains \citep{zi06b}. Moreover, $^{26}$Al and $^{60}$Fe are
particularly important in the field of $\gamma$-ray astronomy
\citep{diehl}: their detection from selected AGB sources could be
possible in the next future by means of $\gamma$-rays instruments
mounted on high-energy astronomical satellites. Let us start from
these two short-lived isotopes.

The ground state of $^{26}$Al has a terrestrial half-life of
$7.16\times10^5$ yr, which is comparable to the duration of the
whole TP-AGB phase of a low mass star. $^{26}$Al is produced in
the H-burning shell by proton captures on $^{25}$Mg. Within the
H-depleted (and $^{26}$Al-rich) region, we need to distinguish two
zones: the upper one, which extends from the tip of the convective
shell generated by a TP up to the H-shell, and the lower one,
which is engulfed in the convective shell. In the former region
$^{26}$Al survives and is dredged up to the surface, whilst in the
latter one, owing to its large neutron-capture cross section, it
is easily destroyed by neutron captures. This happens in radiative
conditions within the $^{13}$C-pockets and, more importantly,
within the convective TPs. As a matter of fact, when the
temperature at the bottom of the convective shell exceeds
$2.7\times10^8$ K, $^{26}$Al is easily destroyed. In our solar
metallicity model, this condition is attained toward the end of
the TP-AGB phase, so that part of the $^{26}$Al engulfed in the
convective shells is preserved. This is not the case of the low
metallicity models, where the $^{22}$Ne neutron source is
efficient since the first TP (see Table~\ref{tab14}). In that
case, only the $^{26}$Al in the upper part of the He intershell
(not engulfed in the convective shell) survives and is dredged up
to the surface.

The synthesis of $^{60}$Fe during the radiative $^{13}$C burning
is limited, owing to the relatively fast $\beta$ decay of
$^{59}$Fe (half-life 45 days). However, the higher neutron density
generated by the convective $^{22}$Ne burning allows to open the
$^{59}$Fe branching, leading to the production of $^{60}$Fe. This
explains the relatively large $^{60}$Fe yield obtained in low
metallicity models. The yield is higher at $Z$=0.001 than at
$Z$=0.0001, because of the larger total dredged up mass.

A different production channel is active in the metal-rich models
($Z=0.0138$, $Z=0.006$ and $Z=0.003$). The appearance at the
surface of $^{60}$Fe  occurs after the first TDU episode. This
reflects a peculiarity of the first $^{13}$C pocket of these
models: the temperature developed in the pocket during the
interpulse is not large enough to fully consume $^{13}$C before
the onset of the following thermal pulse \citep{cri06}. Then, a
certain amount of unburned $^{13}$C is engulfed into the
convective zone generated by the TP and is rapidly burned at
higher temperature. The resulting neutron densities are rather
large (more than $10^{11}$ cm$^{-3}$). In Fig.~\ref{ro_n}, we mark
with an arrow the neutron peak corresponding to the ingestion of
the first $^{13}$C pocket into the convective shell generated by
the following TP (this fact occurring for the metal-rich models
only). Although the effect of this anomalous convective $^{13}$C
burning on the overall s~process nucleosynthesis is negligible,
some neutron-rich isotopes as, for example, $^{60}$Fe, preserve
the signature of such a peculiar event. Note that the surface
isotopic ratio $^{60}$Fe/$^{56}$Fe we obtain in the solar
metallicity model (namely $4.0\times10^{-5}$) is about 10 times
higher then the value reported by \citet{wa06}, as obtained by
means of a post-process calculation, where the effect of the
peculiar first
$^{13}$C pocket was not considered. \\
Let us finally discuss the production of the short-lived
radioactive isotopes $^{36}$Cl, $^{41}$Ca, $^{107}$Pd and
$^{205}$Pb, whose traces have been found in the ESS and whose
origin can be ascribed to an AGB star. Concerning $^{41}$Ca, an
equilibrium value ($^{41}$Ca$/^{40}$Ca$\sim10^{-2}$) is rapidly
attained whenever a neutron source is activated: this value is
inversely proportional to the ratio of the corresponding neutron
capture cross sections. $^{107}$Pd and $^{205}$Pb are mainly
synthesized during the standard radiative $^{13}$C burning at any
metallicity. The variations of the total yields of these two
short-lived nuclei with the metallicity mainly reflect the
increase of the neutron exposure when the metallicity decreases
(because the number of neutrons per iron seed increases). This is
only partially counterbalanced by the larger dredged up mass.

As a general remark, since the decay rates depend on the
environmental conditions (temperature and electron density), the
use of stellar models, which follow the temporal evolution of
these conditions in detail and in all the layers where radioactive
isotopes are stored after their production, is mandatory. Such a
warning should be seriously considered, particularly, when these
isotopes are engulfed in convective zones where the temperature
and the density vary considerably from the the bottom to the top.
As an example, $^{205}$Pb has a rather long terrestrial half-life
(1.5$\times10^7$ yr), but at the temperature of the He intershell
of an AGB star its half-life is several order of magnitude shorter
\citep{taka}. The faster decays occurring in radiative conditions
during the time elapsed between the thermal pulse and the TDU
(about $10^3$ yr) significantly affects the resulting surface
abundances of $^{205}$Pb \citep[see also][]{mopb}. \\
In Table~\ref{tab15}, we list the final surface abundances (by
mass fraction) of the aforementioned shot-lived radioactive
isotopes (with the corresponding stable isotopes), at different
metallicities.

\subsection{Changing the mass loss and the mixing length efficiency: effects on the
nucleosynthesis.}\label{vario}

In \S~\ref{varied} we have shown how the physical parameters of
the AGB evolutions depends on the choices of the mass loss rate
and the mixing length efficiency. In this Section we discuss the
corresponding effects on the nucleosynthesis. The results are
illustrated in Fig.~\ref{fig15} and Table~\ref{tab16}.

The final overabundances (with respect to iron) of the Reimers
model are generally larger than those found in the reference
model. This is a consequence of the larger duration of the AGB, as
obtained when the mass loss rate is lower, so that the total
dredged up mass is larger. On the average, the abundances of the
s-elements increase by a factor of two, while for the light
elements we found a variation in the range between +0.2 dex
(carbon) and +0.5 dex (magnesium). Concerning the fluorine
production, we note that the delayed end of the AGB phase favors
the second fluorine source even in low $Z$ models (see
\S~\ref{pippo}). The primary $^{13}$C in the ashes of the H
burning becomes, in the late TP-AGB phase of low $Z$ models, as
large as the $^{13}$C found in the reference model with solar
metallicity. As explained before, neutrons released by the burning
of this $^{13}$C at the beginning of each TP provide an additional
channel for the production of $^{15}$N. As a result, a longer AGB
phase, as obtained by reducing the mass loss rate, favors this
fluorine source, at any $Z$. Although the absolute abundances
depend on the mass loss rate, the abundance ratios are less
sensitive to the AGB duration. It occurs because the bulk of the
s~process nucleosynthesis has to be ascribed to the first $^{13}$C
pockets, the largest ones, so that a freeze out of the abundance
ratios takes place after very few TPs in the He intershell
material (see \S~\ref{sproc}). For this reason, the [hs/ls] and
[Pb/hs] obtained in the case of the Reimers model are very similar
to those of the reference model. Interestingly, the final C/N and
$^{12}$C/$^{13}$C isotopic ratios of the Reimers model decrease
with respect to the reference one (where monotonic trends are
found), even if HBB is not at work. This different behavior, which
appears in the late AGB phase, is due to an increase with the core
mass of the temperature at the bottom of the convective envelope
during TDU episodes, which leads to a partial H-burning. During
these phases, therefore, mixing and burning simultaneously occur.
This phenomenon, already found by \citet{gosi} in a model with
initial mass $M$= 3 M$_\odot$ and Z=0.0001, enhances the $^{13}$C
and the $^{14}$N in the envelope more rapidly than $^{12}$C
(which, in any case, increases after each TDU episode). When
applying a velocity profile at the base of the convective
envelope, in fact, protons are mixed to higher temperatures with
respect to the ones attained when using the bare Schwarzschild
criterion. A more detailed analysis of this phenomenon, based on
models with different initial masses, will be presented in a
forthcoming paper.

Finally, we evaluated the effects of varying the mixing efficiency
by reducing the mixing-length parameter (case $\alpha=1.8$). In
that case, the lower cumulative dredged up mass leads to smaller
overabundances. As in the Reimers model, the elemental ratios,
which are sensitive to the metallicity and to the $^{13}$C mass in
the pocket, are less affected by the structural model change.

\section{Conclusion}\label{concl}

This paper reports the first systematic calculation of low mass
AGB models at different metallicities, in which the physical
evolution of the star is coupled to a full nuclear network, from H
to Bi. The major input physics, such as nuclear reaction rates,
radiative opacity, mass loss rate, have been revised, in  order to
provide a reliable set of theoretical stellar yields.

The [hs/ls] and the [Pb/hs] ratios at different $Z$ provide
important hints about the nucleosynthesis occurring at different
metallicities and represent useful tools to verify the goodness of
our theoretical models, when compared with observational data. As
shown in Table~\ref{tab12}, the [hs/ls] is rather low at large
$Z$, because the low number of neutron per seed limits the
production of heavy s elements. When decreasing the metallicity,
this ratio increases, achieving a sort of saturation for
$Z<$~0.001. Below this threshold, the scarcity of seeds
(essentially Fe) is partially compensated by a reduction of the
mass of the $^{13}$C pocket (see below). [Pb/hs] provides a more
sensitive spectroscopic index to test low metallicity models. It
is expected to monotonically increase from high to low $Z$,
indeed. Let us stress that, since the [hs/ls] and the [Pb/Fe] are
practically frozen after a few dredge up episodes, these indexes
are almost independent of the assumed mass loss rate and on the
efficiency of the TDU, which are two of the most uncertain
quantities in AGB modelling. In practice, as noted by \citet{bu01}
\citep[see also][]{ga08,bi08}, they essentially depend on the
effective $^{13}$C mass within the $^{13}$C pockets. In this
context, by comparing the theoretical predictions of these indexes
with spectroscopic ratios at different $Z$, we may have a direct
check of the validity of our choice of the $\beta$ parameter.

In Fig.~\ref{fig16} we have reported the observed [hs/ls] indexes,
as measured in a sample of Galactic and extragalactic C-stars with
different metallicities \citep{ab02,abi08,dela06}. Spectroscopic
and photometric studies indicate that these stars are intrinsic (N
type) C-stars, {\it i.e.} low-mass AGB stars undergoing the third
dredge up. The agreement with our theoretical predictions is
comfortable. In particular, data confirm the increase of [hs/ls],
when the metallicity is reduced below the solar value, up to a
plateau attained at intermediate $Z$. Only the most metal-poor
C(N) star of the sample, namely ALW-C7, a carbon star of the
Carina dwarf galaxy, shows a too large [hs/ls] compared with the
low $Z$ plateau. However, as suggested by \citet{abi08}, this star
shows a particularly low abundance of Zr compared with other light
s elements. Moreover, the fitting process in determining the ls
abundances in ALW-C7 resulted more difficult with respect to
ALW-C6, making the determination of the [hs/ls] even more
uncertain (C. Abia, personal communication).

\begin{figure*}[tpb]
\centering
\includegraphics[width=\textwidth]{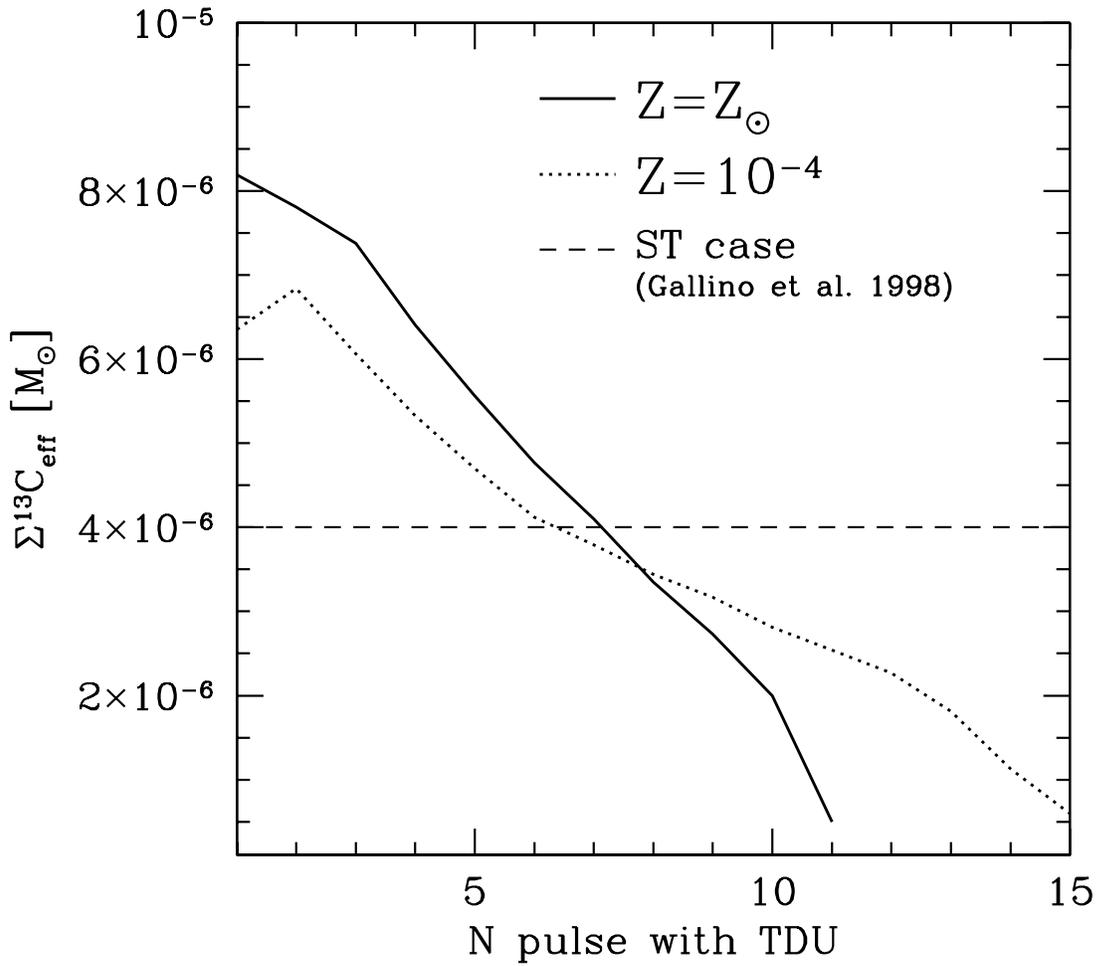}
\caption{Evolution of the effective $^{13}$C mass in the pockets
vs. the number of pulses followed by TDU, for models at
$Z$=$Z_\odot$ and $Z=0.0001$. Note the decrease of $^{13}$C in the
pockets along the AGB evolution. For comparison, the ST case of
\citet{ga98} is reported.} \label{fig5}
\end{figure*}

\begin{figure*}[tpb]
\centering
\includegraphics[width=\textwidth]{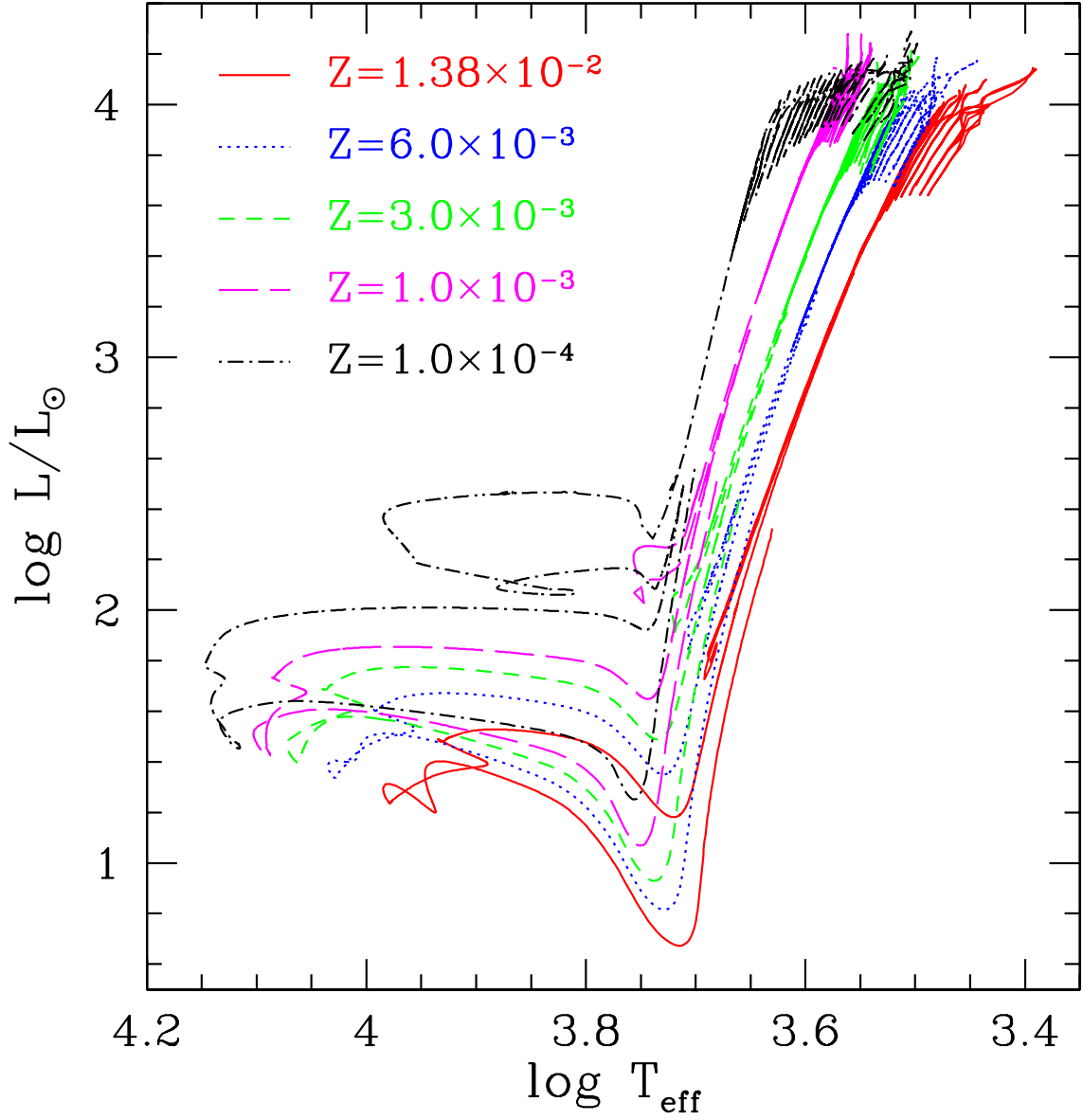}
\caption{Evolutionary tracks in the HR diagram of the five 2
M$_\odot$ models.} \label{fig6}
\end{figure*}

\begin{figure*}[tpb]
\centering
\includegraphics[width=\textwidth]{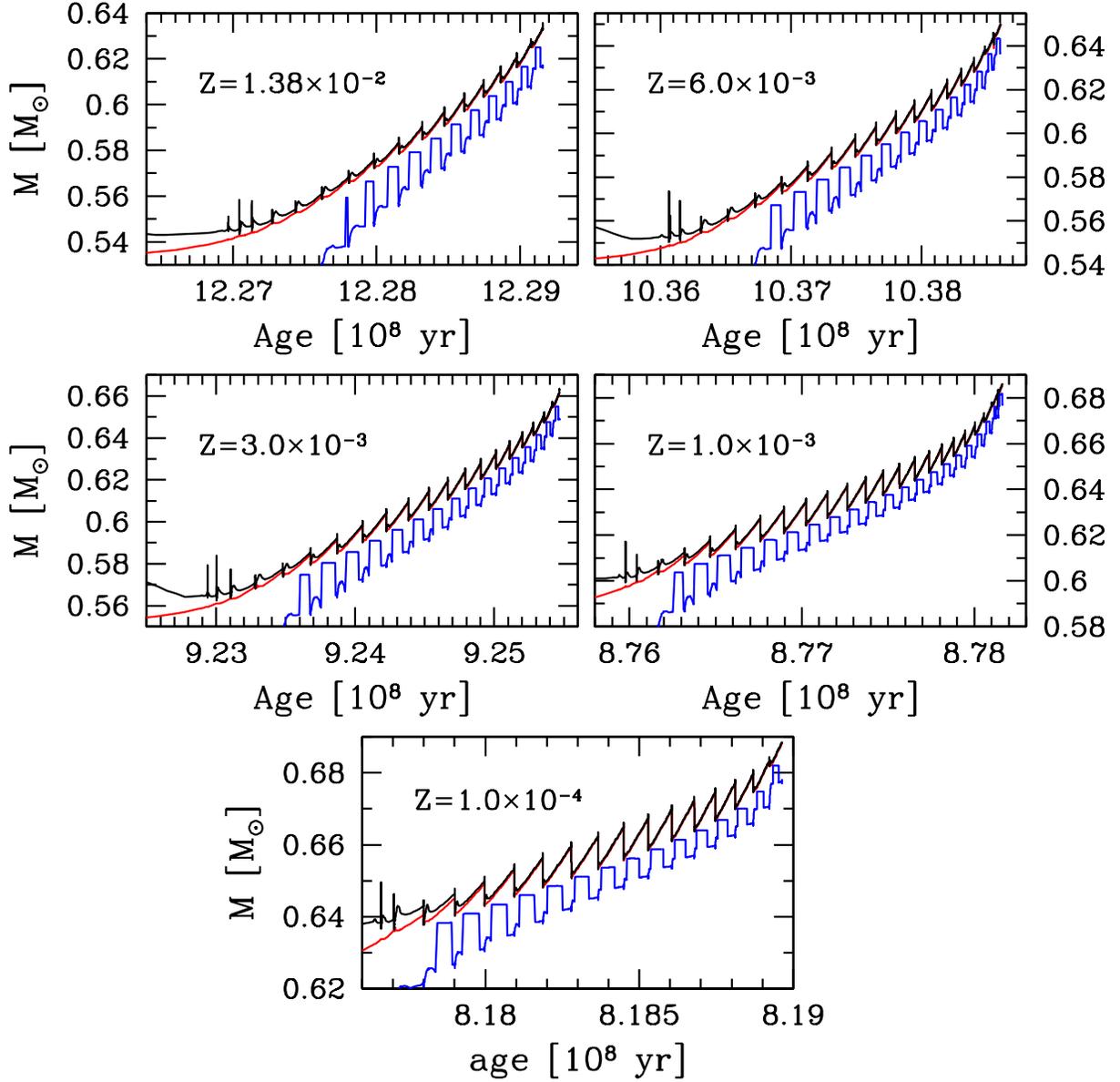}
\caption{Evolution of the positions, in mass coordinates, of the
inner border of the convective envelope, of the location of the
maximum energy production within the H-burning shell and of the
location of the maximum energy production within the He-burning
shell, top to bottom line, respectively. Blue spikes correspond to
the activation of the $^{13}$C($\alpha$,n)$^{16}$O reaction (see
text for details). Each panel represents one of the five 2
M$_\odot$ models, as labelled in the plots.} \label{fig7}
\end{figure*}

\begin{figure*}[tpb]
\centering
\includegraphics[width=\textwidth]{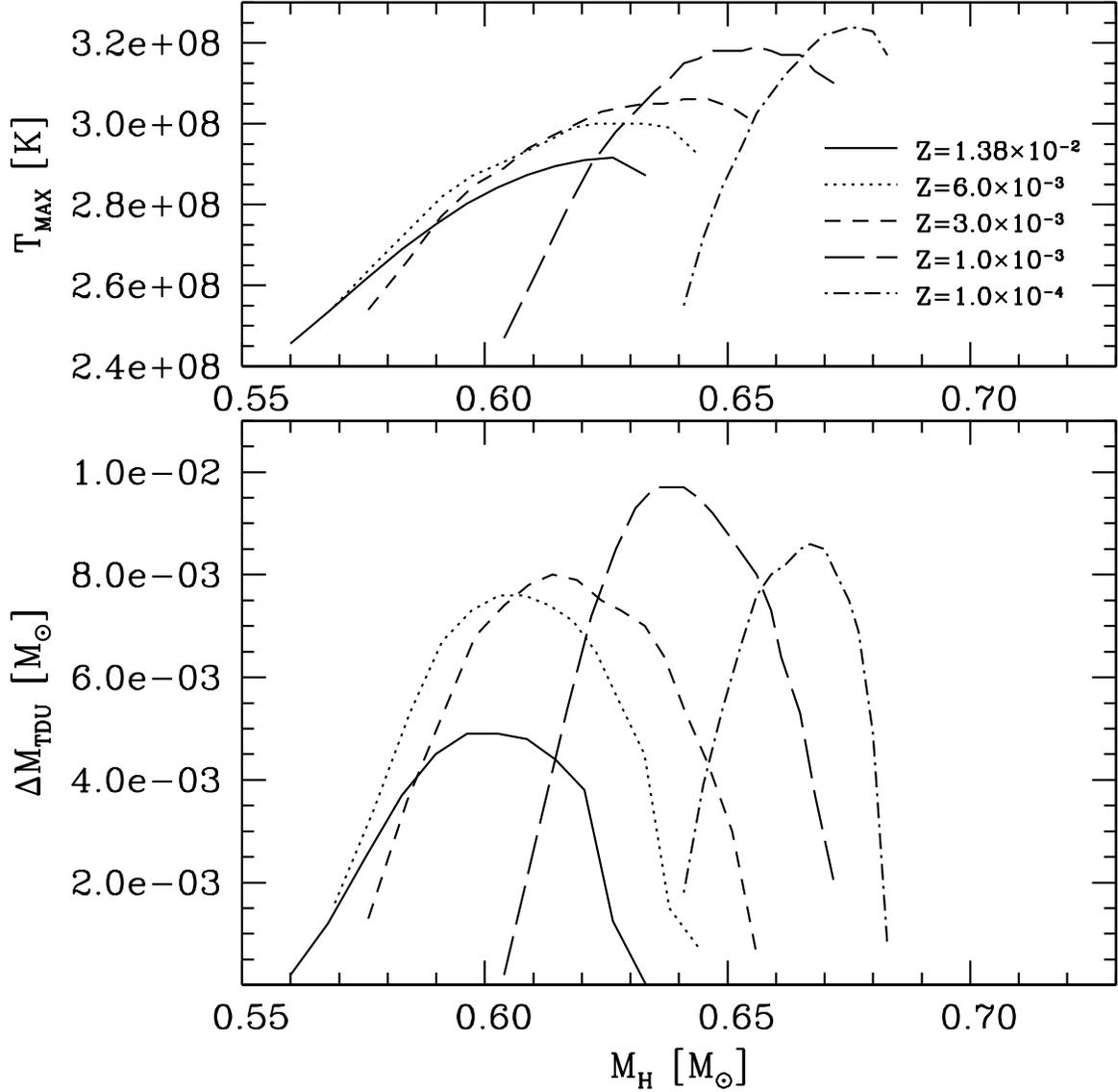}
\caption{Upper panel: maximum temperature attained at the bottom
of the convective zone generated by a thermal pulse versus the
mass of the H-exhausted core (M$_H$). Lower panel: the mass of
H-depleted material that is dredged up after each thermal pulse
versus M$_H$. The five lines of each panels refer to the five
different models, as labelled in the plots.} \label{fig8}
\end{figure*}

\begin{figure*}[tpb]
\centering
\includegraphics[width=\textwidth]{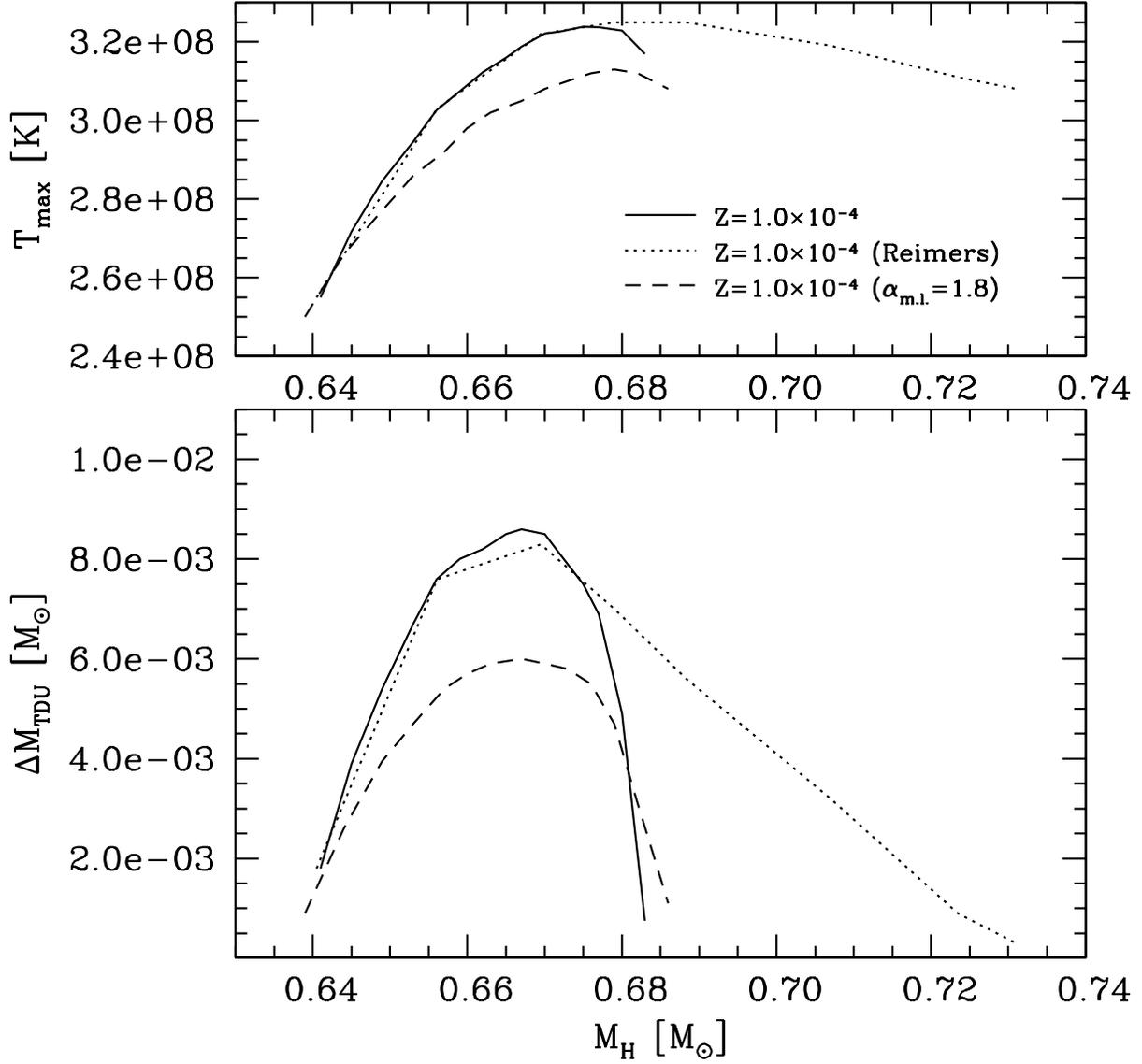}
\caption{The same as in Fig.~\ref{fig8} for the reference
$Z$=0.0001 model (solid lines) and for the two additional models
with varied mass loss (dotted lines) and mixing length (dashed
lines).} \label{fig9}
\end{figure*}

\begin{figure*}[tpb]
\centering
\includegraphics[width=\textwidth]{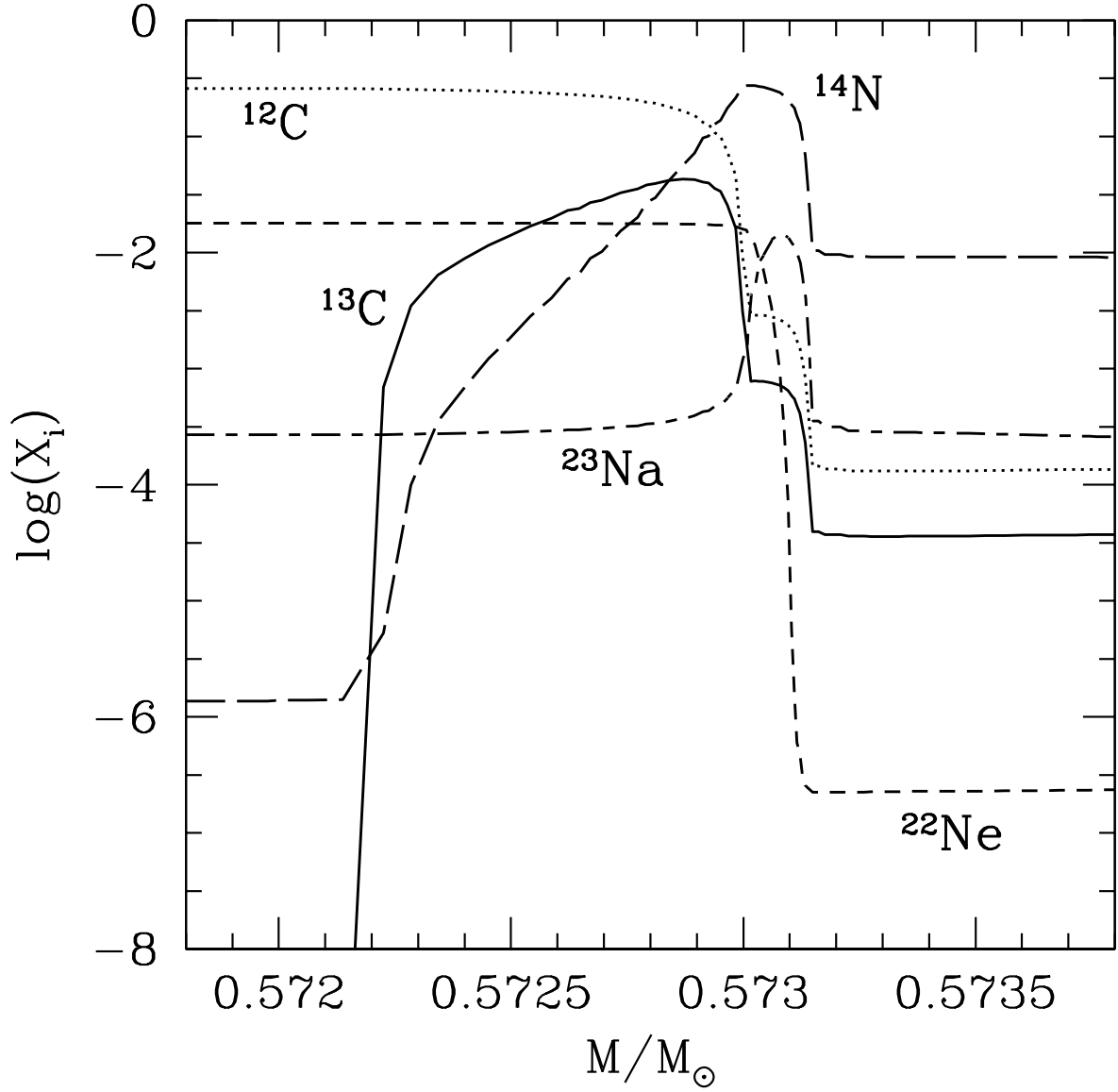}
\caption{Mass fractions of selected isotopes in the third $^{13}$C
pocket of the solar metallicity model. The $^{14}$N pocket and the
$^{23}$Na pocket clearly appear in the figure. See text for
details.} \label{fig10}
\end{figure*}

\begin{figure*}[tpb]
\includegraphics[width=\textwidth]{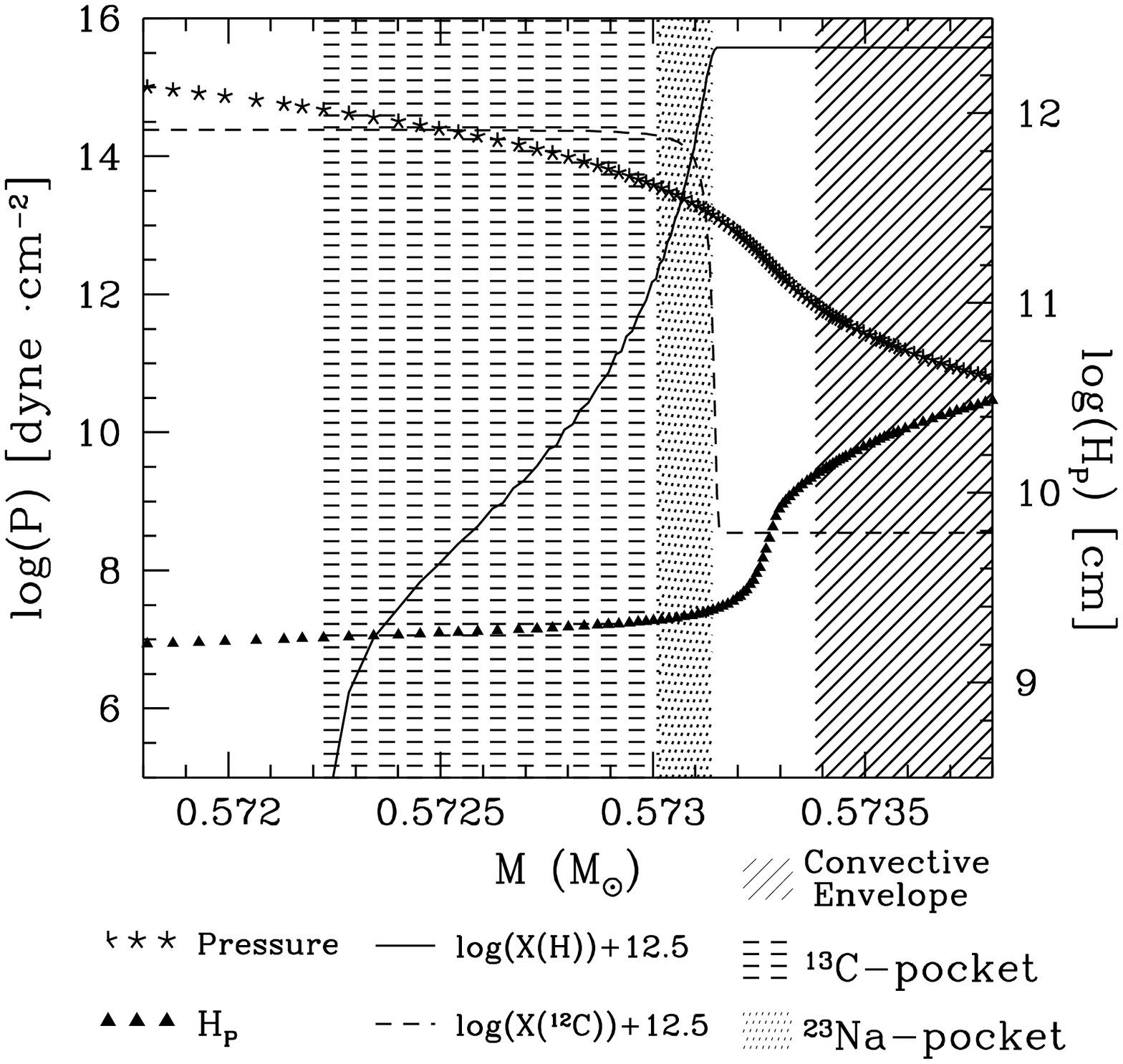}
\caption{Variations of the pressure and of the pressure scale
height ($H_P$) in the core-envelope transition region at the time
of the maximum penetration of the 3$^{rd}$ TDU of the solar
metallicity model. Solid and dashed lines represents the H and the
$^{12}$C profiles, respectively (abundances have been shifted
upward in order to match the pressure scale axis). Slanting dashed
area shows the envelope region, whilst the horizontal and the
vertical dashed areas mark the $^{13}$C and the $^{23}$Na pockets
(the $^{14}$N pocket is not reported in the plot).\label{fig11a}}
\end{figure*}

\begin{figure*}[tpb]
\centering
\includegraphics[width=\textwidth]{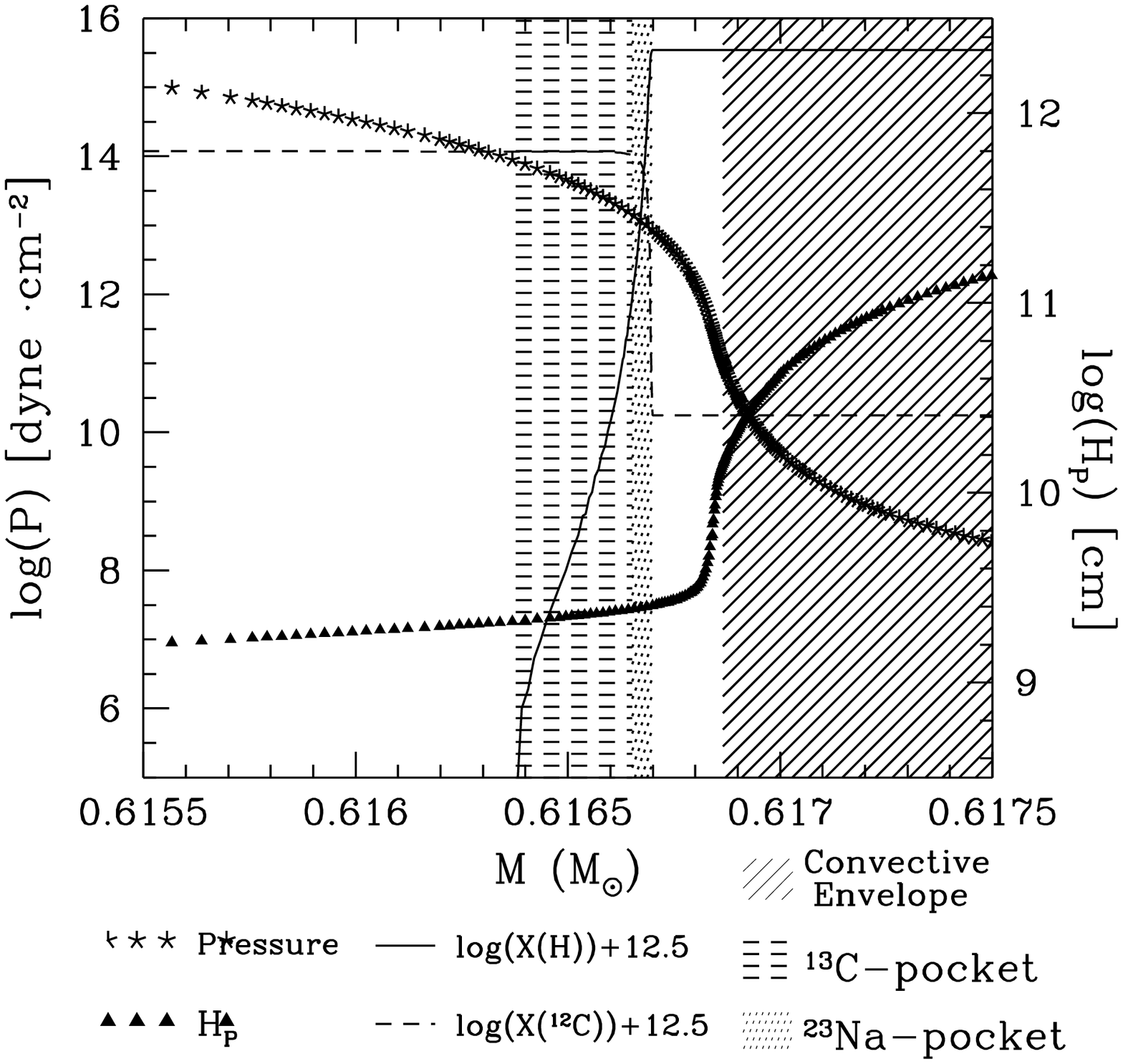}
\caption{As in Fig.~\ref{fig11a}, but relative to the 11$^{th}$
TDU of the solar metallicity model.} \label{fig11b}
\end{figure*}

\begin{figure*}[tpb]
\centering
\includegraphics[width=\textwidth]{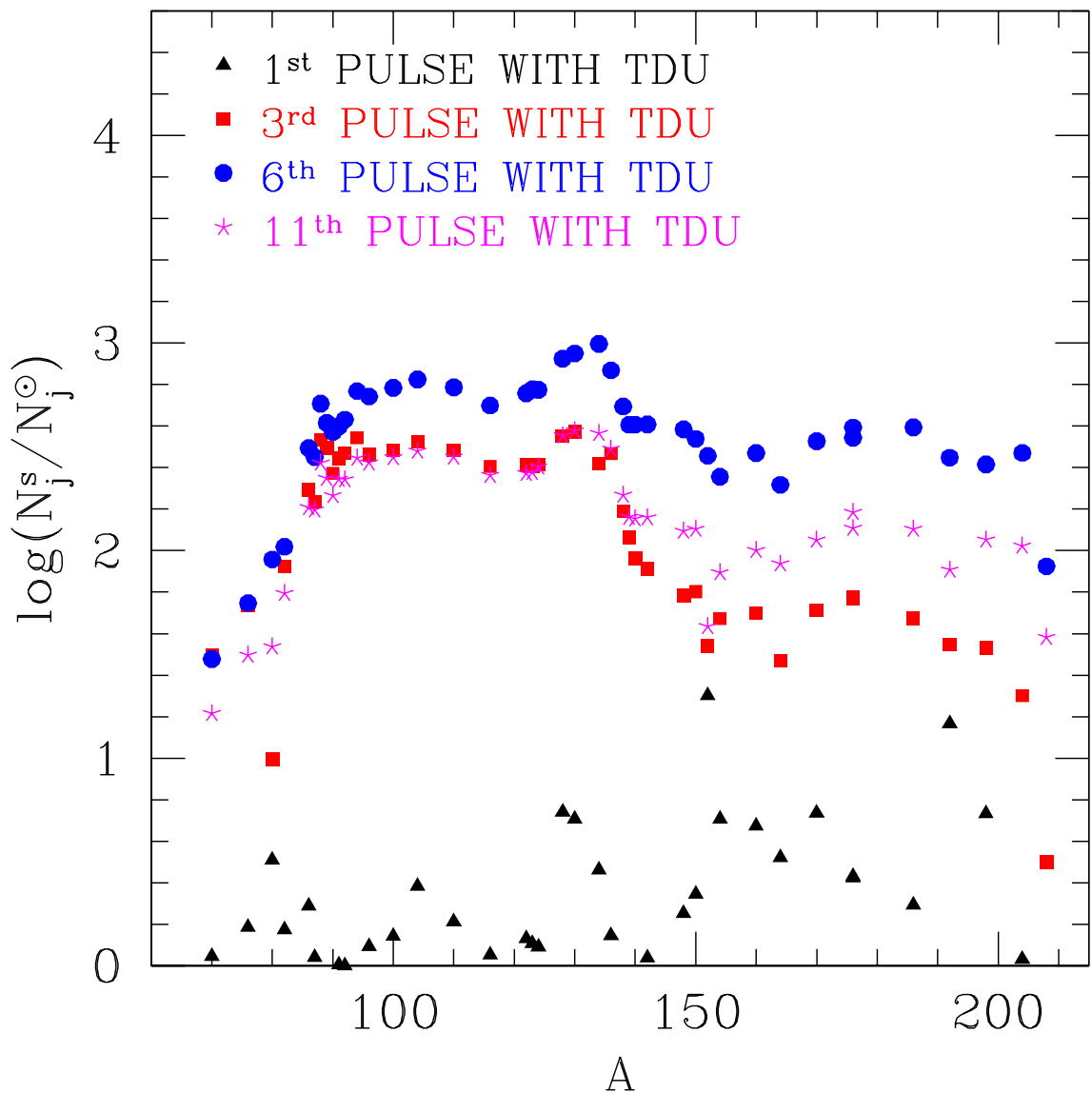}
\caption{Production factors of nuclei whose synthesis is mainly
ascribed to the s~process (namely, nuclei whose s~process
contribution is larger than 80\%) within the He intershell at the
quenching of the convective shells generated by the TPs before the
1$^{st}$, 3$^{rd}$, 6$^{th}$ and 11$^{th}$ TDU episodes of the
solar metallicity model. The abundances in the He intershell are
the result of the mixing of three zones: the deeper layer saw the
cumulative mixing of the overlapped convective zones generated by
the previous TPs, the central layer, which corresponds to the last
$^{13}$C pocket and is heavily enriched in s-elements, and the
upper layer, which corresponds to the portion of the He intershell
containing the ashes produced by the H-burning during the last
interpulse period, with a low s-enhancement. The decreases of the
production factors in the late AGB is due to the shrinkage of the
$^{13}$C pockets.} \label{fig12}
\end{figure*}

\begin{figure*}[tpb]
\centering
\includegraphics[width=\textwidth]{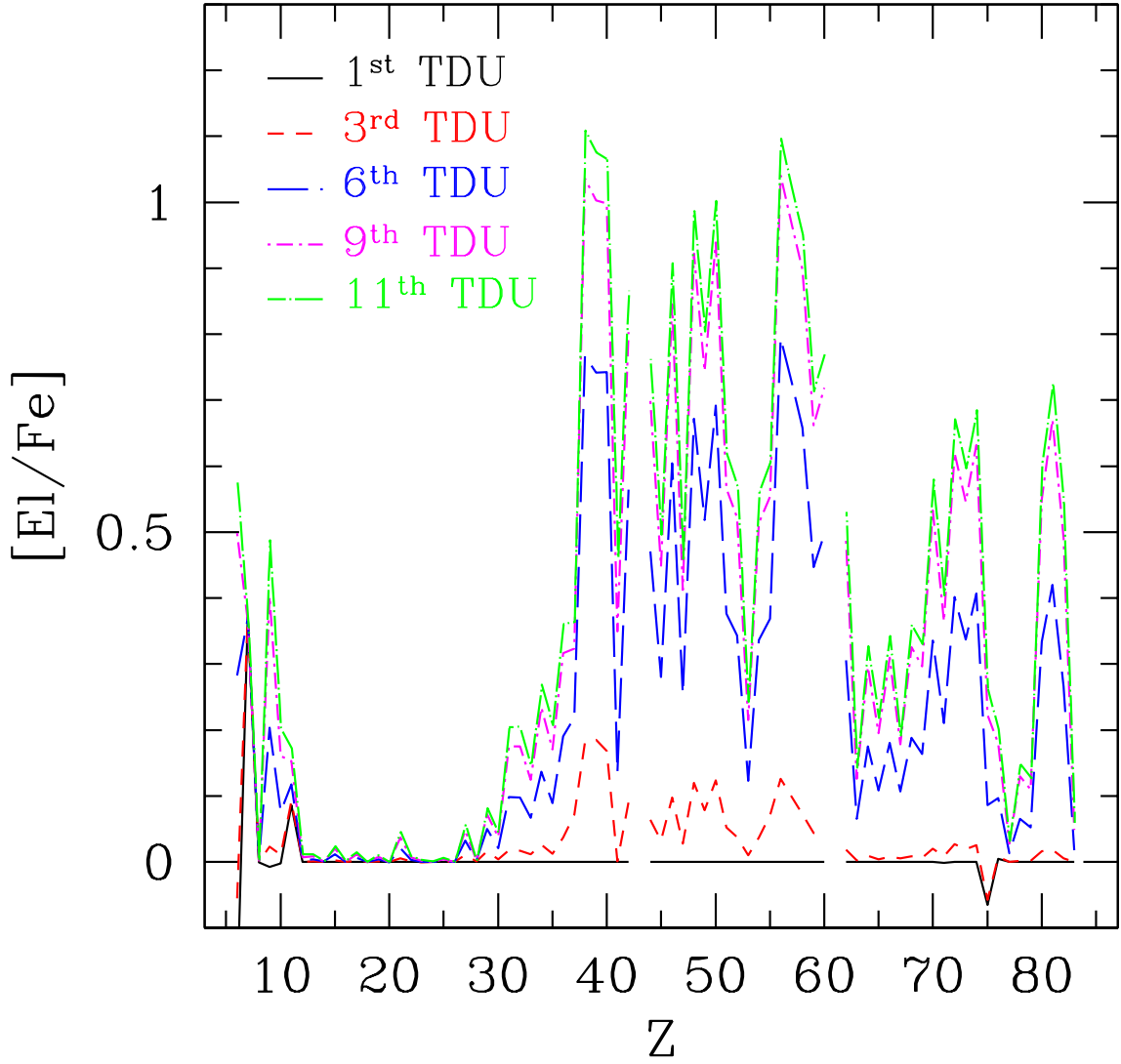}
\caption{Elemental surface composition of the solar metallicity
model as it results after selected third dredge up episodes (see
labels inside the plot).} \label{fig13}
\end{figure*}

\begin{figure*}[tpb]
\centering
\includegraphics[width=\textwidth]{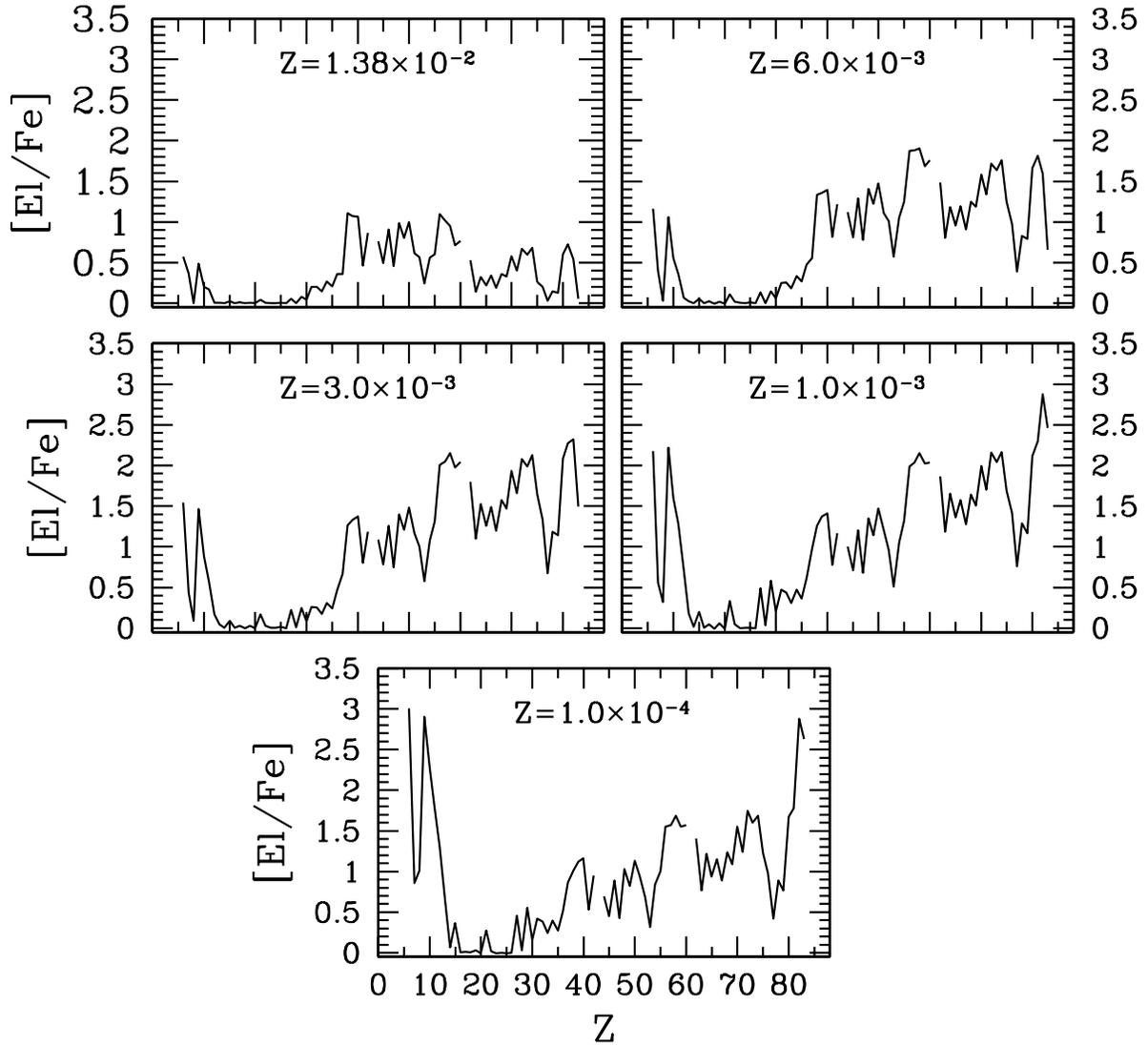}
\caption{Final elemental surface composition of the five
evolutionary sequences with different $Z$, as labelled in the
plots.} \label{fig14}
\end{figure*}

\begin{figure*}[tpb]
\centering
\includegraphics[width=\textwidth]{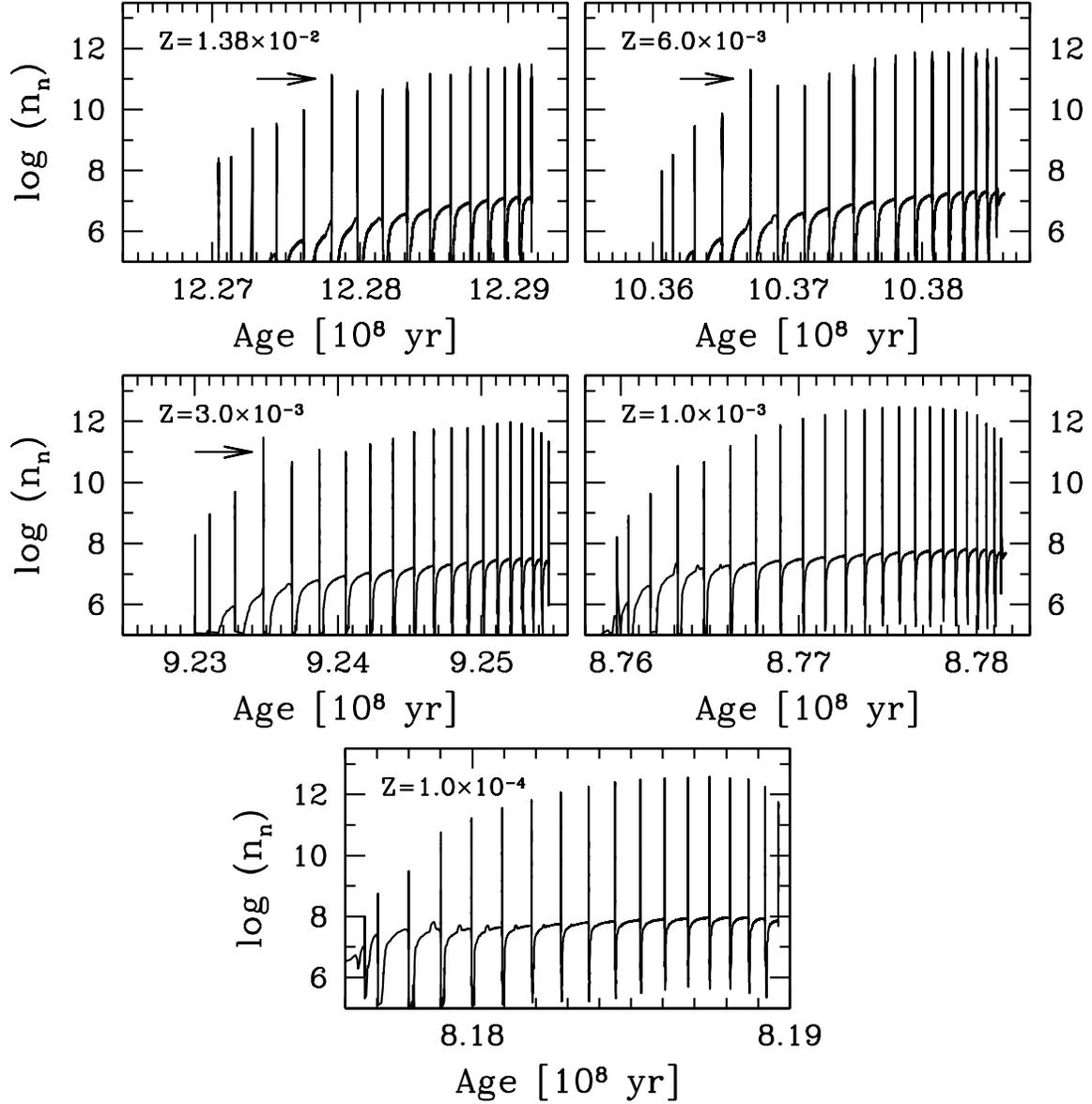}
\caption{Maximum neutron densities attained in models with
$M=2$M$_\odot$ and different metallicities. In the metal-rich
models, arrows mark the ingestion of the first $^{13}$C pockets
into the convective shells generated by the following TPs. See
text for details.} \label{ro_n}
\end{figure*}

\begin{figure*}[tpb]
\centering
\includegraphics[width=\textwidth]{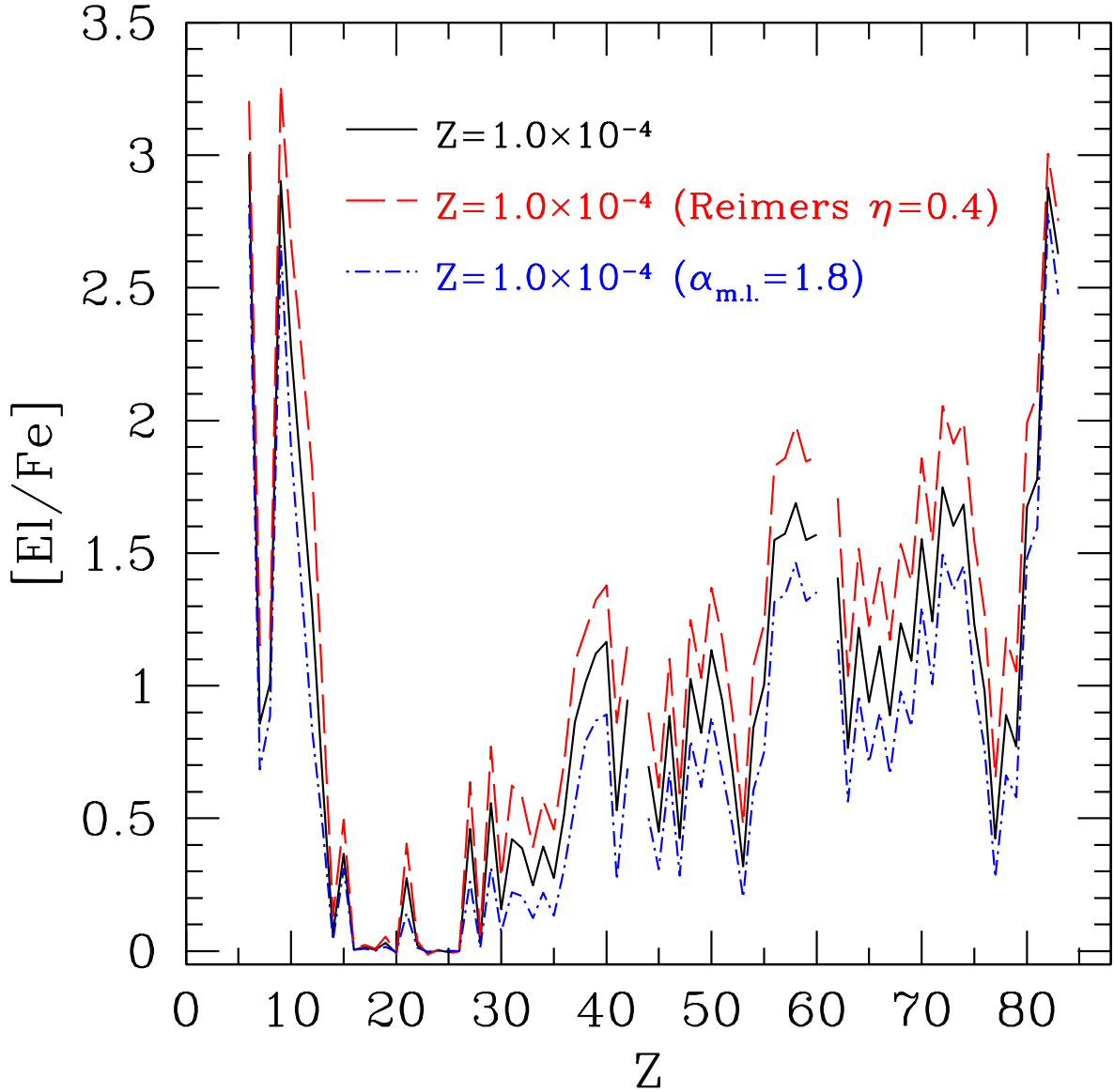}
\caption{Final elemental surface composition of the $Z$=0.0001
reference model is compared with those of the two additional
models obtained by adopting a different mass loss rate (Reimers
$\eta=0.4$) or by decreasing the mixing-length parameter
($\alpha_{m.l.}=1.8$). See text for details.} \label{fig15}
\end{figure*}

\begin{figure*}[tpb]
\centering
\includegraphics[width=\textwidth]{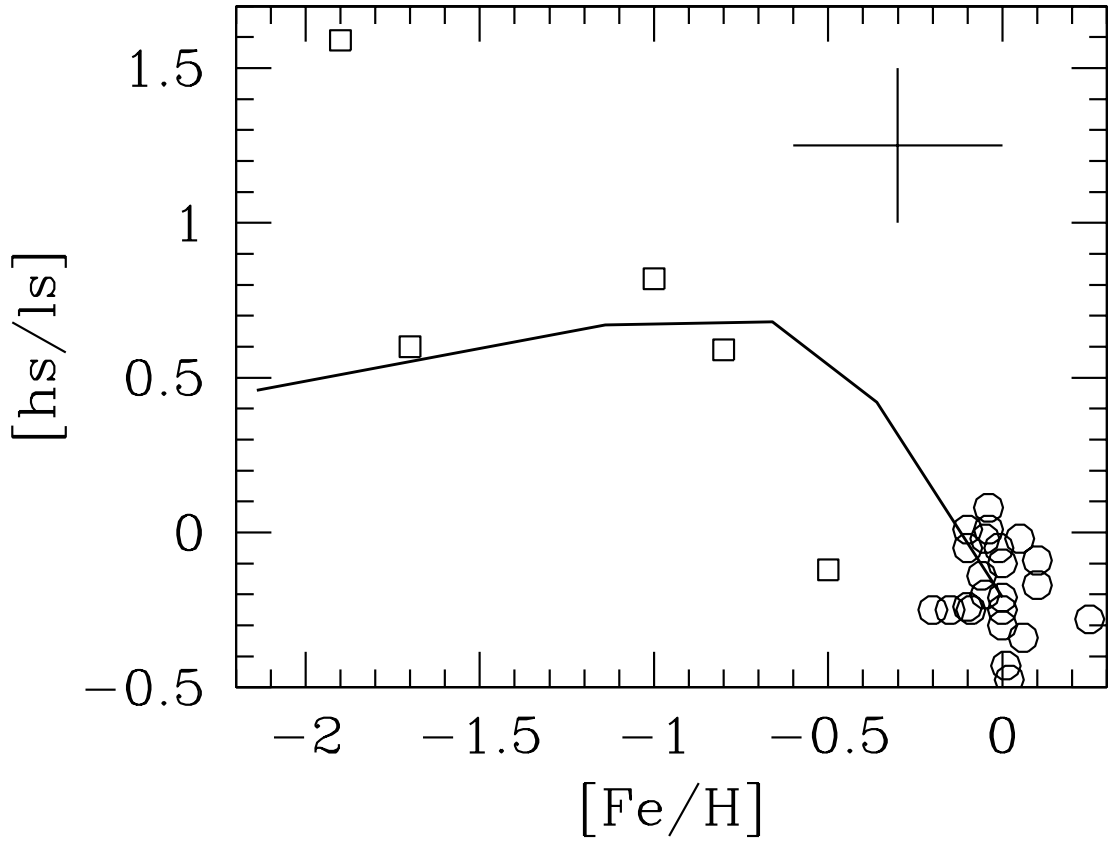}
\caption{The theoretical [hs/ls] index (solid curve) compared with
the same ratios as observed in intrinsic Galactic (circles,
\citealt{ab02}) and extragalactic (squares,
\citealt{dela06,abi08}) N type C stars. Typical error bars are
also reported.} \label{fig16}
\end{figure*}

\begin{figure*}[tpb]
\centering
\includegraphics[width=\textwidth]{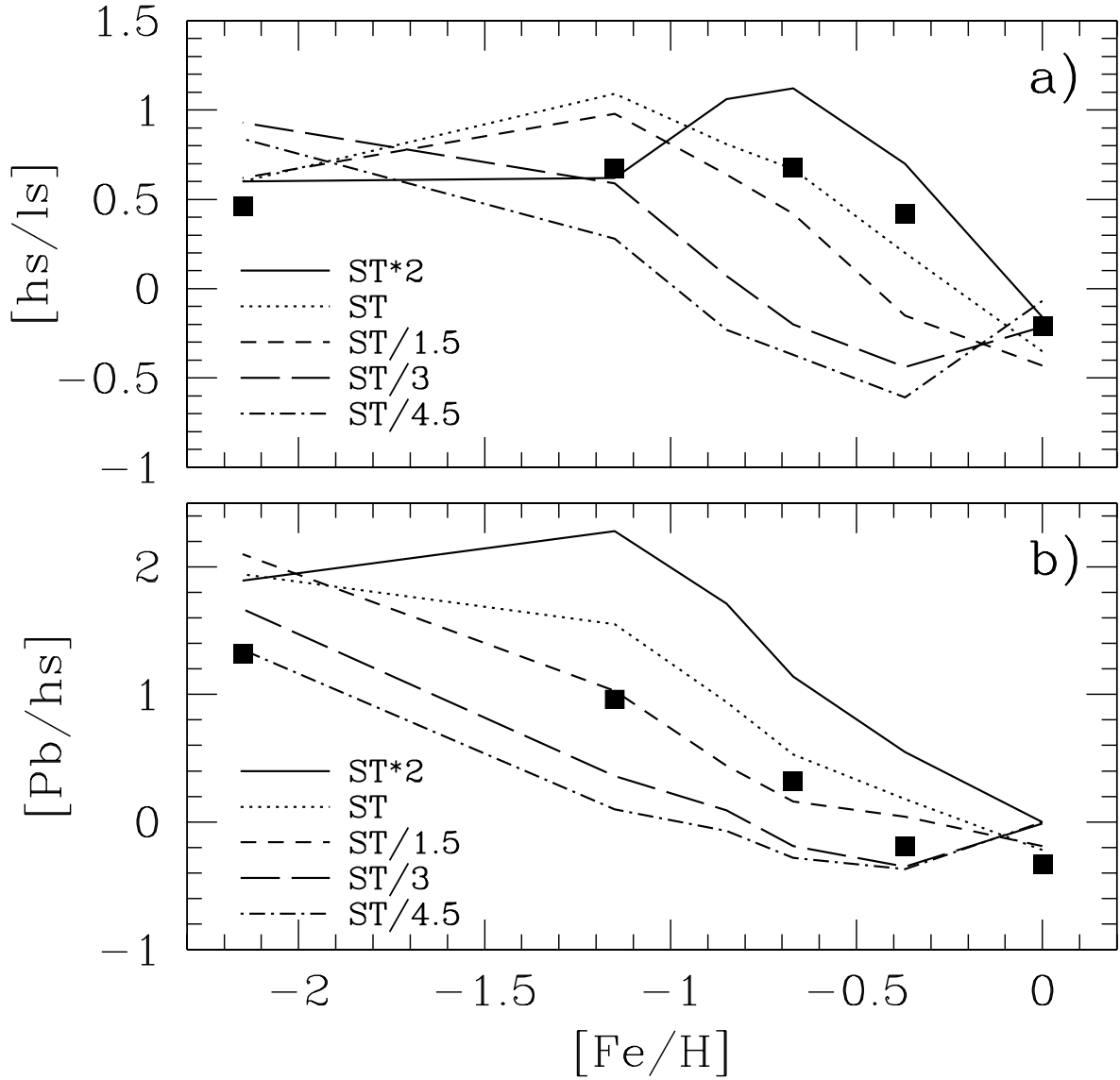}
\caption{Spectroscopic indexes predicted by our new models (filled
squares) compared with post-process calculations
\citep{ga08,bi08}. In panel a) we report the [hs/ls] index, whilst
in panel b) we plot the [Pb/hs] index. See text for details.}
\label{fig17}
\end{figure*}

A second interesting check for the new theoretical scenario
concerns the comparison with nucleosynthesis models based on
post-processes calculations \citep{ga08,bi08}. In the post-process
calculations, the relevant stellar parameters are derived
according to older stellar evolutionary models \citep{stra97} and,
where models were not available, by using the interpolation
formulas provided by \citet{stra03}. Since these models have been
computed by assuming a Reimers mass loss rate, the duration of the
AGB and, in turn, the total mass of H-depleted and s-enriched
material dredged up is larger than that found in the present
computations. Although this difference affects the predicted
overabundance of a single element, it has negligible effects on
the [hs/ls] or the [Pb/hs] indexes. A second important difference
of the post process calculations concerns the mass of the $^{13}$C
pocket, which is fixed to a constant value for the full AGB
evolution; according to \citet{ga98}, the value of the $^{13}$C
mass is a free parameter of the nucleosynthesis model and the ST
(standard) case corresponds to $\Sigma
^{13}$C$_\mathrm{eff}=4\times10^{-6}$ M$_\odot$ (see the
discussion in \S~\ref{beta}). The third important difference
concerns the chemical profiles of $^{13}$C and $^{14}$N within the
pocket (for the post process calculation see Fig. 1 of
\citealt{ga98}). In Fig.~\ref{fig17}, the various lines represent
the results of the post-process calculations \citep{ga08,bi08},
whilst the filled squares are our new predictions. In the plot, ST
refers to the standard case, whilst the cases labelled with ST*k
or ST/k correspond to $^{13}$C pockets whose mass is k and 1/k
times the standard one, respectively. At large metallicities, our
predictions for [hs/ls] and [Pb/hs] are in good agreement with
those of the ST case (dotted lines). When the metallicity is
reduced, however, the two spectroscopic indexes resulting from the
new calculation move progressively toward the lines corresponding
to smaller $^{13}$C masses. This behavior is likely due to the
smaller $^{13}$C pockets we found at low $Z$ (see Fig.~\ref{fig5}
and Tab.~\ref{tab6}), whose effect is to limit the maximum neutron
exposure.

Let us conclude with a comment on the question of the spread of
the $^{13}$C pocket. As discussed by \citet{bi08}, a certain
spread is required to explain the variety of [hs/ls] observed in
the available sample of carbon-enhanced metal-poor and s-rich
stars (CEMPs), whose overabundance of heavy elements is supposed
to be originating from mass transfer from an ancient AGB companion
(extrinsic C-stars). Similar conclusions arise from the abundance
analysis of Barium stars, which may be considered the CEMPs
homologous at intermediate $Z$  \citep{hu08,hu082}. By limiting
our consideration to the Galactic (N type) carbon stars, those
belonging to the disk of the Milky Way and having a nearly-solar
metallicity, for which a quite large sample of [hs/ls] is now
available, we note that the dispersion around the average value,
namely $<$[hs/ls]$>=-0.19\pm0.3$, is of the same order of
magnitude of the error bar of a single measurement ($\pm 0.25$).
In addition to that, it should be noted that the prediction of our
solar metallicity model, namely [hs/ls]=$-0.21$, is in excellent
agreement with the measured average value. It should be recalled
that the present models refer to only one AGB mass (2 M$_\odot$).
The natural variation of the stellar parameters among stars with
similar metallicity, as for example the mass or the initial He, C,
N or O, may account for a certain spread in the observed [hs/ls]
and [Pb/hs] ratios. In the next future, we will extend the
theoretical database to investigate such a possibility.

\acknowledgements {SC, OS and RG are supported by the Italian
MIUR-PRIN 2006 Project "Final Phases of Stellar Evolution,
Nucleosynthesis in Supernovae, AGB stars, Planetary Nebulae". ID
is supported by the spanish MEC project AYA2005-08013-C03-03 and
by the andalusian FQM-292. MTL has been supported by the Austrian
Academy of Sciences (DOC programme) and acknowledges funding by
the Austrian Research Fund FWF (project P-18171). We gratefully
thank Carlos Abia for many enlightening discussions.}

\bibliographystyle{99}

\newpage

\begin{deluxetable}{cccccc}
\tablecolumns{6} \tablewidth{0pc} \tablecaption{The C- and
N-enhancement factors for the H-rich opacity tables at various
metallicities. f=1 corresponds to [C/Fe]=0 and [N/Fe]=0. For each
metallicity, a routine for the calculation of the opacity
coefficients interpolates on a 25 elements matrix, as originated
by the the possible combination of the five C-enhancement and five
N-enhancement factors.} \tablehead{\colhead{Z} &
\multicolumn{5}{c}{enhancement factor (f)}} \startdata
0.0138 & 1 & 1.5 & 1.8  & 2.2 & 4\\
0.0060 & 1 & 2   & 5    & 10  & 50\\
0.0030 & 1 & 2   & 5    & 10  & 50\\
0.0010 & 1 & 5 & 10 & 50  & 200\\
0.0001 & 1 & 10 & 100 & 500  & 2000\\
\enddata
 \label{tab1}
\end{deluxetable}

\begin{deluxetable}{ccccccc}
\tablecolumns{7} \tablewidth{0pc} \tablecaption{Variation with
$\beta$ of the $^{13}$C pocket.} \tablehead{ \colhead{$\beta$} &
\colhead{$\Delta M_{TDU}$ [M$_{\odot}$]} & \colhead{$\lambda$} &
\colhead{$\Delta M_{pocket}$ [M$_{\odot}$]} & \colhead{$\beta \times H_P$
[cm]} & \colhead{$\Sigma{\rm H}$ [M$_{\odot}$]} &
\colhead{$\Sigma{^{13}{\rm C}_{\rm eff}}$ [M$_{\odot}$]}}
\startdata
\multicolumn{7}{c}{\textbf{$Z$=0.0138 (3$^{rd}$ Pulse followed by TDU)}}\\
0    &  0.00E+00   & \nodata & \nodata & \nodata & \nodata & \nodata \\
0.01 &  0.00E+00   & \nodata & \nodata & \nodata & \nodata & \nodata \\
0.05 &  3.00E-04   &  0.04   &  9.40E-05 & 5.62E+08 & 4.37E-05 & 8.32E-07 \\
0.075&  1.15E-03   &  0.14   &  3.47E-04 & 8.82E+08 & 8.49E-05 & 3.12E-06 \\
0.09 &  1.70E-03   &  0.20   &  7.66E-04 & 1.06E+09 & 9.56E-05 & 6.41E-06 \\
0.10 &  2.50E-03   &  0.28   &  5.54E-04 & 1.22E+09 & 1.18E-04 & 7.38E-06 \\
0.11 &  2.90E-03   &  0.32   &  2.54E-04 & 1.36E+09 & 1.29E-04 & 4.40E-06 \\
\enddata
\label{tab2}
\end{deluxetable}

\begin{deluxetable}{ccccccccccccc}
\rotate \tablecolumns{13} \tablewidth{0pc} \tablecaption{Evolution
of selected physical parameters of the 5 reference models.}
\tablehead{ \colhead{$n_{TP}^a$} & \colhead{$M_{Tot}^{b}$} &
\colhead{$M_{\rm H}^{b}$} & \colhead{$\Delta M_{TDU}^{b}$} &
\colhead{$\Delta M_{CZ}^{b}$} & \colhead{$\Delta M_{\rm H}^{b}$} &
\colhead{$r$} & \colhead{$\lambda$} & \colhead{$n_c$}
&\colhead{$\Delta t_{ip}^{c}$} & \colhead{$T_{MAX}^{d}$} &
\colhead{Z$_{surf}$} & \colhead{C/O}}\startdata
\cutinhead{\textbf{$Z$=0.0138 \ Y=0.269}}
-2 & 1.91E+00 & 5.48E-01 & 0.00E+00 & 3.25E-02 & 4.20E-03 & 8.99E-01 & 0.00E+00 & \nodata & 14.0 & 2.20 & 1.39E-02 & 0.31 \\
-1 & 1.90E+00 & 5.53E-01 & 0.00E+00 & 3.07E-02 & 4.80E-03 & 8.66E-01 & 0.00E+00 & \nodata & 16.5 & 2.27 & 1.39E-02 & 0.31 \\
1  & 1.89E+00 & 5.60E-01 & 2.00E-04 & 3.12E-02 & 7.00E-03 & 7.90E-01 & 2.90E-02 & 4.2$\times 10^1$ & 18.6 & 2.46 & 1.39E-02 & 0.31 \\
2  & 1.88E+00 & 5.68E-01 & 1.20E-03 & 2.94E-02 & 7.90E-03 & 7.48E-01 & 1.52E-01 & 4.6$\times 10^1$ & 18.6 & 2.53 & 1.41E-02 & 0.34 \\
3  & 1.86E+00 & 5.76E-01 & 2.50E-03 & 2.81E-02 & 9.00E-03 & 6.95E-01 & 2.78E-01 & 4.3$\times 10^1$ & 17.8 & 2.62 & 1.46E-02 & 0.43 \\
4  & 1.84E+00 & 5.83E-01 & 3.70E-03 & 2.69E-02 & 9.90E-03 & 6.46E-01 & 3.74E-01 & 4.3$\times 10^1$& 17.0 & 2.69 & 1.53E-02 & 0.57 \\
5  & 1.82E+00 & 5.90E-01 & 4.50E-03 & 2.58E-02 & 1.07E-02 & 6.00E-01 & 4.21E-01 & 3.9$\times 10^1$ & 16.2 & 2.75 & 1.62E-02 & 0.75 \\
6  & 1.78E+00 & 5.96E-01 & 4.90E-03 & 2.48E-02 & 1.10E-02 & 5.67E-01 & 4.45E-01 & 3.6$\times 10^1$ & 15.3 & 2.80 & 1.72E-02 & 0.95 \\
\enddata \label{tab3}
\end{deluxetable}

\begin{deluxetable}{ccccccccccccc}
\rotate \tablecolumns{13} \tablewidth{0pc} \tablecaption{Evolution
of selected parameters for the Reimers ($\eta=0.4$) model at
$Z$=10$^{-4}$.} \tablehead{ \colhead{$n_{TP}^a$} &
\colhead{M$_{Tot}^{b}$} & \colhead{M$_H^{b}$} & \colhead{$\Delta
M_{TDU}^{b}$} & \colhead{$\Delta M_{CZ}^{b}$} & \colhead{$\Delta
M_{\rm H}^{b}$} & \colhead{$r$} & \colhead{$\lambda$} & \colhead{$n_c$}  &
\colhead{$\Delta t_{ip}^{c}$} & \colhead{$T_{MAX}^{d}$} &
\colhead{Z$_{surf}$} & \colhead{C/O}}\startdata
\cutinhead{\textbf{$Z$=0.0001  Y=0.245 Reimers $\eta=0.4$}}
-1 & 1.98E+00 & 6.34E-01 & 0.00E+00 & 1.83E-02 & 2.50E-03 & 9.19E-01 & 0.00E+00 &\nodata & 4.21 & 2.19 & 1.01E-04 & 0.25 \\
1  & 1.98E+00 & 6.41E-01 & 1.80E-03 & 1.96E-02 & 4.30E-03 & 7.97E-01 & 4.19E-01 & 4.6$\times 10^3$ & 9.50 & 2.55 & 3.36E-04 & 7.76 \\
5  & 1.93E+00 & 6.56E-01 & 7.60E-03 & 1.80E-02 & 1.01E-02 & 4.51E-01 & 7.52E-01 & 4.0$\times 10^3$ & 9.35 & 3.03 & 4.67E-03 & 35.0 \\
9  & 1.84E+00 & 6.70E-01 & 8.30E-03 & 1.63E-02 & 1.06E-02 & 3.38E-01 & 7.83E-01 & 3.7$\times 10^3$ & 7.60 & 3.22 & 1.18E-02 & 46.1 \\
14 & 1.75E+00 & 6.79E-01 & 7.00E-03 & 1.46E-02 & 9.60E-03 & 3.26E-01 & 7.29E-01 & 3.7$\times 10^3$ & 5.92 & 3.25 & 1.71E-02 & 51.5 \\
18 & 1.66E+00 & 6.88E-01 & 5.60E-03 & 1.29E-02 & 8.30E-03 & 3.39E-01 & 6.75E-01 & 3.6$\times 10^3$ & 4.79 & 3.25 & 2.18E-02 & 54.7 \\
25 & 1.50E+00 & 7.07E-01 & 3.20E-03 & 1.02E-02 & 6.50E-03 & 3.70E-01 & 4.92E-01 & 3.5$\times 10^3$ & 3.45 & 3.19 & 2.84E-02 & 57.7 \\
32 & 1.35E+00 & 7.31E-01 & 3.00E-04 & 7.74E-03 & 4.30E-03 & 4.28E-01 & 7.00E-02 & 3.0$\times 10^3$ & 2.19 & 3.08 & 3.14E-02 & 58.6 \\
\enddata \label{tab4}
\end{deluxetable}

\begin{deluxetable}{ccccccccccccc}
\rotate \tablecolumns{13} \tablewidth{0pc} \tablecaption{Evolution
of selected parameters for the low mixing length
($\alpha_{m.l.}=1.8$) model at $Z$=10$^{-4}$.} \tablehead{
\colhead{$n_{TP}^a$} & \colhead{$M_{Tot}^{b}$} & \colhead{$M_{\rm
H}^{b}$} & \colhead{$\Delta M_{TDU}^{b}$} & \colhead{$\Delta
M_{CZ}^{b}$} & \colhead{$\Delta M_{\rm H}^{b}$} & \colhead{$r$} &
\colhead{$\lambda$} & \colhead{$n_c$} & \colhead{$\Delta t_{ip}^{c}$} &
\colhead{$T_{MAX}^{d}$} & \colhead{Z$_{surf}$} &
\colhead{C/O}}\startdata \cutinhead{\textbf{$Z$=0.0001  Y=0.245
$\alpha_{m.l.}=1.8$}}
-1& 1.98E+00 & 6.35E-01 & 0.00E+00 & 1.81E-02 & 2.10E-03 & 9.52E-01 & 0.00E+00 & \nodata & 3.67   & 2.13 & 1.01E-04 & 0.25 \\
1 & 1.98E+00 & 6.39E-01 & 9.00E-04 & 1.97E-02 & 4.20E-03 & 8.15E-01 & 2.14E-01 & 4.6$\times 10^3$ & 8.94 & 2.50 & 1.63E-04 & 2.37 \\
2 & 1.97E+00 & 6.44E-01 & 2.60E-03 & 1.88E-02 & 5.80E-03 & 7.08E-01 & 4.48E-01 & 4.7$\times 10^3$ & 9.62 & 2.66 & 5.86E-04 & 13.0 \\
3 & 1.96E+00 & 6.49E-01 & 3.95E-03 & 1.81E-02 & 7.10E-03 & 6.24E-01 & 5.56E-01 & 5.1$\times 10^3$ & 8.70 & 2.77 & 1.30E-03 & 22.0 \\
4 & 1.95E+00 & 6.53E-01 & 4.70E-03 & 1.76E-02 & 8.05E-03 & 5.54E-01 & 5.84E-01 & 4.9$\times 10^3$ & 8.19 & 2.86 & 2.20E-03 & 27.3 \\
5 & 1.94E+00 & 6.57E-01 & 5.40E-03 & 1.71E-02 & 8.60E-03 & 5.10E-01 & 6.28E-01 & 4.6$\times 10^3$ & 7.80 & 2.92 & 3.21E-03 & 30.7 \\
6 & 1.93E+00 & 6.60E-01 & 5.70E-03 & 1.66E-02 & 8.90E-03 & 4.72E-01 & 6.40E-01 & 4.6$\times 10^3$ & 7.44 & 2.98 & 4.28E-03 & 33.3 \\
7 & 1.92E+00 & 6.63E-01 & 5.90E-03 & 1.61E-02 & 9.10E-03 & 4.44E-01 & 6.48E-01 & 4.6$\times 10^3$ & 7.09 & 3.02 & 5.38E-03 & 35.6 \\
\enddata \label{tab5}
\end{deluxetable}

\begin{deluxetable}{c|ccc|ccc} \tablecolumns{7} \tablewidth{0pc} \tablecaption{Evolution
of the $^{13}$C pockets for $Z$=0.0138 and $Z$=0.0001~. We report
the mass of the H-exhausted core ($M_{\rm H}$), the mass of the
effective $^{13}$C in the pocket ($\Sigma ^{13}$C$_{\rm eff}$) and
the mass extension of the pocket ($\Delta M_{pocket}$).} \tablehead{
\colhead{$N.$} & \colhead{$M_{\rm H}$$^{a}$} & \colhead{$\Sigma
^{13}$C$_{\rm eff}$$^{a}$} & \colhead{$\Delta M_{pocket}^{b}$} &
\colhead{$M_{\rm H}^{a}$} & \colhead{$\Sigma ^{13}$C$_{\rm
eff}$$^{a}$} & \colhead{$\Delta M_{pocket}^{b}$}} \startdata
 & \multicolumn{3}{|c|}{\textbf{$Z$=0.0138}}& \multicolumn{3}{c}{\textbf{$Z$=0.0001}}\\
1 & 0.5600  & 8.19E-06 & 5.96 & 0.6405 & 6.35E-06 & 5.79 \\
2 & 0.5677  & 7.81E-06 & 6.04 & 0.6450 & 6.84E-06 & 5.44 \\
3 & 0.5755  & 7.38E-06 & 5.54 & 0.6491 & 6.06E-06 & 4.58 \\
4 & 0.5829  & 6.41E-06 & 5.40 & 0.6529 & 5.33E-06 & 4.45 \\
5 & 0.5899  & 5.56E-06 & 5.09 & 0.6563 & 4.70E-06 & 4.35 \\
6 & 0.5964  & 4.77E-06 & 4.33 & 0.6593 & 4.12E-06 & 3.96 \\
7 & 0.6026  & 4.10E-06 & 4.01 & 0.6621 & 3.79E-06 & 3.79 \\
8 & 0.6087  & 3.35E-06 & 3.35 & 0.6648 & 3.44E-06 & 3.74 \\
9 & 0.6146  & 2.73E-06 & 2.81 & 0.6674 & 3.17E-06 & 3.29 \\
10 & 0.6205 & 2.00E-06 & 1.99 & 0.6698 & 2.81E-06 & 2.82 \\
11 & 0.6264 & 4.95E-07 & 0.71 & 0.6722 & 2.54E-06 & 2.60 \\
12 &   \nodata     &     \nodata     &   \nodata   & 0.6745 & 2.27E-06 & 2.51 \\
13 &   \nodata    &      \nodata    &    \nodata  & 0.6770 & 1.81E-06 & 1.88 \\
14 &   \nodata     &     \nodata     &    \nodata  & 0.6796 & 1.13E-06 & 1.31 \\
15 &   \nodata     &     \nodata     &   \nodata   & 0.6829 & 5.96E-07 & 0.73 \\
\cline{1-7}
\textbf{TOT.} & & \textbf{5.28E-05} &  & & \textbf{5.50E-05} & \\
\\
\multicolumn{7}{l}{$^{a)}$ M${_\odot}$}\\
\multicolumn{7}{l}{$^{b)}$ $10^{-4}$ M${_\odot}$}\\
\enddata
\label{tab6}
\end{deluxetable}

\begin{deluxetable}{ccccc}
\tablecolumns{5} \tablewidth{0pc} \tablecaption{Elemental surface
composition after selected TDUs of the $Z=0.0138$ model. Key
elements are tabulated.} \tablehead{ \colhead{[El/Fe]} &
\colhead{TDU 1} & \colhead{TDU 4}& \colhead{TDU 7}& \colhead{TDU
11}} \startdata
C   &   -2.00E-01   &   6.72E-02    &   3.65E-01       &   5.75E-01    \\
N   &   3.72E-01    &   3.71E-01    &   3.68E-01       &   3.65E-01    \\
O   &   -5.16E-04   &   6.71E-04    &   2.43E-03       &   1.36E-03    \\
F   &   -7.81E-03   &   7.03E-02    &   2.72E-01       &   4.88E-01    \\
Ne  &   -2.55E-03   &   2.77E-02    &   1.03E-01       &   2.02E-01    \\
Na  &   8.62E-02    &   9.87E-02    &   1.29E-01       &   1.72E-01    \\
Mg  &   -1.92E-07   &   6.82E-04    &   3.59E-03       &   1.06E-02    \\
Al  &   1.72E-04    &   1.76E-03    &   5.68E-03       &   1.06E-02    \\
Sr  &   1.27E-06    &   4.13E-01    &   8.80E-01       &   1.11E+00    \\
Y   &   -3.40E-07   &   4.05E-01    &   8.51E-01       &   1.08E+00    \\
Zr  &   1.75E-08    &   3.95E-01    &   8.51E-01       &   1.07E+00    \\
Ba  &   1.34E-06    &   3.95E-01    &   9.05E-01       &   1.10E+00    \\
La  &   3.04E-06    &   3.42E-01    &   8.35E-01       &   1.02E+00    \\
Ce  &   1.27E-06    &   2.87E-01    &   7.66E-01       &   9.51E-01    \\
Pr  &   3.04E-07    &   1.72E-01    &   5.41E-01       &   7.14E-01    \\
Nd  &   1.85E-06    &   1.90E-01    &   5.95E-01       &   7.69E-01    \\
Sm  &   3.19E-07    &   9.80E-02    &   3.83E-01       &   5.30E-01    \\
Eu  &   -1.83E-06   &   1.67E-02    &   8.93E-02       &   1.43E-01    \\
Pb  &   -1.19E-07   &   5.36E-02    &   3.61E-01       &   5.43E-01    \\
\enddata
\tablecomments{Table \ref{tab7} is published in its entirety in
the electronic edition of the {\it Astrophysical Journal}.  A
portion is shown here for guidance regarding its form and
content.} \label{tab7}
\end{deluxetable}

\newpage

\begin{deluxetable}{ccccc}
\tablecolumns{5} \tablewidth{0pc} \tablecaption{Elemental surface
composition after selected TDUs of the $Z=0.006$ model. Key
elements are tabulated.} \tablehead{ \colhead{[El/Fe]} &
\colhead{TDU 1} & \colhead{TDU 5}& \colhead{TDU 10}& \colhead{TDU
14}} \startdata

C   &   -9.26E-02    &   6.70E-01    &  1.03E+00   &   1.17E+00    \\
N   &   3.94E-01     &   3.90E-01    &  3.83E-01   &   3.97E-01    \\
O   &   -4.65E-04    &   1.50E-02    &  2.62E-02   &   2.90E-02    \\
F   &   -8.65E-03    &   3.87E-01    &  8.51E-01   &   1.06E+00    \\
Ne  &   4.94E-04     &   1.45E-01    &  4.09E-01   &   5.58E-01    \\
Na  &   1.14E-01     &   1.52E-01    &  2.53E-01   &   3.58E-01    \\
Mg  &   1.58E-04     &   6.57E-03    &  3.47E-02   &   6.77E-02    \\
Al  &   5.90E-04     &   8.78E-03    &  2.04E-02   &   2.72E-02    \\
Sr  &   4.65E-02     &   8.60E-01    &  1.22E+00   &   1.33E+00    \\
Y   &   5.73E-02     &   8.72E-01    &  1.24E+00   &   1.36E+00    \\
Zr  &   5.21E-02     &   9.18E-01    &  1.28E+00   &   1.39E+00    \\
Ba  &   1.06E-02     &   1.40E+00    &  1.78E+00   &   1.87E+00    \\
La  &   5.61E-03     &   1.43E+00    &  1.79E+00   &   1.88E+00    \\
Ce  &   2.97E-03     &   1.48E+00    &  1.83E+00   &   1.91E+00    \\
Pr  &   1.96E-03     &   1.23E+00    &  1.60E+00   &   1.69E+00    \\
Nd  &   1.45E-03     &   1.34E+00    &  1.68E+00   &   1.76E+00    \\
Sm  &   6.49E-04     &   1.09E+00    &  1.41E+00   &   1.49E+00    \\
Eu  &   -1.52E-04    &   4.73E-01    &  7.39E-01   &   8.04E-01    \\
Pb  &   3.05E-04     &   1.21E+00    &  1.54E+00   &   1.59E+00    \\
\enddata
\tablecomments{Table \ref{tab8} is published in its entirety in
the electronic edition of the {\it Astrophysical Journal}.  A
portion is shown here for guidance regarding its form and
content.} \label{tab8}
\end{deluxetable}

\newpage

\begin{deluxetable}{ccccc}
\tablecolumns{5} \tablewidth{0pc} \tablecaption{Elemental surface
composition after selected TDUs of the $Z=0.003$ model. Key
elements are tabulated.} \tablehead{ \colhead{[El/Fe]} &
\colhead{TDU 1} & \colhead{ TDU 6}& \colhead{ TDU 11}& \colhead{
TDU 16}} \startdata
C   &   -3.51E-02   &  1.06E+00  &  1.38E+00    &  1.54E+00   \\
N   &   4.16E-01    &  4.12E-01  &  4.10E-01    &  4.38E-01   \\
O   &   -1.42E-03   &  4.34E-02  &  7.14E-02    &  8.97E-02   \\
F   &   -1.81E-02   &  7.27E-01  &  1.19E+00    &  1.47E+00   \\
Ne  &   -8.52E-04   &  2.97E-01  &  6.36E-01    &  8.81E-01   \\
Na  &   1.37E-01    &  2.05E-01  &  3.61E-01    &  5.70E-01   \\
Mg  &   2.71E-04    &  1.72E-02  &  7.47E-02    &  1.70E-01   \\
Al  &   8.66E-04    &  1.81E-02  &  3.21E-02    &  4.41E-02   \\
Sr  &   4.20E-02    &  8.31E-01  &  1.12E+00    &  1.26E+00   \\
Y   &   5.65E-02    &  8.66E-01  &  1.17E+00    &  1.33E+00   \\
Zr  &   6.29E-02    &  9.17E-01  &  1.22E+00    &  1.38E+00   \\
Ba  &   3.10E-02    &  1.51E+00  &  1.86E+00    &  2.01E+00   \\
La  &   2.13E-02    &  1.56E+00  &  1.90E+00    &  2.05E+00   \\
Ce  &   1.34E-02    &  1.67E+00  &  2.01E+00    &  2.15E+00   \\
Pr  &   8.80E-03    &  1.46E+00  &  1.82E+00    &  1.98E+00   \\
Nd  &   8.21E-03    &  1.57E+00  &  1.90E+00    &  2.04E+00   \\
Sm  &   4.89E-03    &  1.34E+00  &  1.66E+00    &  1.80E+00   \\
Eu  &   6.43E-04    &  6.79E-01  &  9.67E-01    &  1.10E+00   \\
Pb  &   4.89E-04    &  2.02E+00  &  2.25E+00    &  2.32E+00   \\
\enddata
\tablecomments{Table \ref{tab9} is published in its entirety in
the electronic edition of the {\it Astrophysical Journal}.  A
portion is shown here for guidance regarding its form and
content.} \label{tab9}
\end{deluxetable}

\newpage

\begin{deluxetable}{ccccc}
\tablecolumns{5} \tablewidth{0pc} \tablecaption{Elemental surface
composition after selected TDUs of the $Z=0.001$ model. Key
elements are tabulated.} \tablehead{ \colhead{[El/Fe]} &
\colhead{TDU 1} & \colhead{ TDU 7}& \colhead{ TDU 12}& \colhead{
TDU 19}} \startdata
C   &   -2.13E-01   &  1.66E+00   &  1.95E+00  &   2.18E+00    \\
N   &   4.59E-01    &  4.59E-01   &  4.72E-01  &   5.56E-01    \\
O   &   -9.12E-03   &  1.59E-01   &  2.40E-01  &   3.22E-01    \\
F   &   -3.85E-02   &  1.40E+00   &  1.83E+00  &   2.22E+00    \\
Ne  &   -5.32E-03   &  7.21E-01   &  1.19E+00  &   1.60E+00    \\
Na  &   1.67E-01    &  4.06E-01   &  7.69E-01  &   1.28E+00    \\
Mg  &   9.40E-05    &  8.36E-02   &  3.27E-01  &   7.19E-01    \\
Al  &   2.91E-03    &  7.65E-02   &  1.12E-01  &   1.79E-01    \\
Sr  &   6.38E-06    &  7.94E-01   &  1.02E+00  &   1.26E+00    \\
Y   &   2.56E-06    &  8.47E-01   &  1.11E+00  &   1.37E+00    \\
Zr  &   2.56E-06    &  8.73E-01   &  1.14E+00  &   1.41E+00    \\
Ba  &   5.06E-06    &  1.39E+00   &  1.70E+00  &   1.99E+00    \\
La  &   2.56E-06    &  1.43E+00   &  1.74E+00  &   2.03E+00    \\
Ce  &   6.55E-07    &  1.54E+00   &  1.87E+00  &   2.15E+00    \\
Pr  &   -2.57E-06   &  1.38E+00   &  1.73E+00  &   2.02E+00    \\
Nd  &   2.56E-06    &  1.44E+00   &  1.75E+00  &   2.04E+00    \\
Sm  &   -4.37E-06   &  1.24E+00   &  1.57E+00  &   1.86E+00    \\
Eu  &   -2.54E-05   &  6.10E-01   &  9.05E-01  &   1.18E+00    \\
Pb  &   2.56E-06    &  2.54E+00   &  2.76E+00  &   2.87E+00    \\
\enddata
\tablecomments{Table \ref{tab10} is published in its entirety in
the electronic edition of the {\it Astrophysical Journal}.  A
portion is shown here for guidance regarding its form and
content.} \label{tab10}
\end{deluxetable}

\newpage

\begin{deluxetable}{cccccc}
\tablecolumns{5} \tablewidth{0pc} \tablecaption{Elemental surface
composition after selected TDUs of the $Z=0.0001$ model. Key
elements are tabulated.} \tablehead{ \colhead{[El/Fe]} &
\colhead{TDU 1} & \colhead{ TDU 5}& \colhead{TDU 10}& \colhead{
TDU 15}} \startdata
C   &   1.15E+00   &    2.41E+00  &    2.81E+00  &   3.00E+00   \\
N   &   4.93E-01   &    5.23E-01  &    6.17E-01  &   8.58E-01   \\
O   &   -9.15E-04  &    5.95E-01  &    8.76E-01  &   1.01E+00   \\
F   &   -8.31E-02  &    2.13E+00  &    2.62E+00  &   2.90E+00   \\
Ne  &   -3.48E-03  &    1.30E+00  &    1.91E+00  &   2.27E+00   \\
Na  &   2.22E-01   &    7.70E-01  &    1.40E+00  &   1.77E+00   \\
Mg  &   -2.92E-04  &    2.31E-01  &    8.59E-01  &   1.30E+00   \\
Al  &   7.02E-03   &    1.94E-01  &    4.37E-01  &   6.37E-01   \\
Sr  &   1.19E-04   &    5.17E-01  &    8.64E-01  &   1.01E+00   \\
Y   &   8.69E-06   &    5.66E-01  &    9.60E-01  &   1.12E+00   \\
Zr  &   9.89E-06   &    5.75E-01  &    9.92E-01  &   1.17E+00   \\
Ba  &   6.13E-05   &    9.21E-01  &    1.36E+00  &   1.55E+00   \\
La  &   -4.48E-06  &    9.37E-01  &    1.38E+00  &   1.57E+00   \\
Ce  &   2.41E-06   &    1.04E+00  &    1.49E+00  &   1.69E+00   \\
Pr  &   -1.07E-04  &    9.01E-01  &    1.35E+00  &   1.55E+00   \\
Nd  &   -4.98E-06  &    9.27E-01  &    1.37E+00  &   1.57E+00   \\
Sm  &   -1.09E-04  &    7.40E-01  &    1.20E+00  &   1.41E+00   \\
Eu  &   -4.44E-04  &    2.66E-01  &    5.87E-01  &   7.65E-01   \\
Pb  &   9.70E-06   &    2.37E+00  &    2.74E+00  &   2.88E+00   \\
\enddata
\tablecomments{Table \ref{tab11} is published in its entirety in
the electronic edition of the {\it Astrophysical Journal}.  A
portion is shown here for guidance regarding its form and
content.} \label{tab11}
\end{deluxetable}

\begin{deluxetable}{cccccc}
\tablecolumns{6} \tablewidth{0pc} \tablecaption{Final surface
composition of the 5 reference models at different $Z$. The final
s process indexes, the C/O and the $^{12}$C/$^{13}$C are given.}
\tablehead{ \colhead{} & \colhead{{\small $Z=1.38\times 10^{-2}$}}
& \colhead{{\small $Z=6.0\times 10^{-3}$}} & \colhead{{\small
$Z=3.0\times 10^{-3}$}} & \colhead{{\small $Z=1.0\times 10^{-3}$}}
& \colhead{{\small $Z=1.0\times 10^{-4}$}}} \startdata
$\rm[C/Fe] $& 0.57 & 1.17  & 1.54  & 2.17   &  3.00    \\
$\rm[N/Fe] $& 0.37 & 0.40  & 0.44  & 0.56   &  0.86    \\
$\rm[O/Fe] $& 0.00 & 0.03  & 0.09  & 0.32   &  1.01    \\
$\rm[F/Fe] $& 0.49 & 1.06  & 1.46  & 2.22   &  2.90    \\
$\rm[Ne/Fe]$ & 0.20 & 0.56  & 0.88  & 1.60   &  2.27    \\
$\rm[Na/Fe]$ & 0.17 & 0.36  & 0.57  & 1.28   &  1.77    \\
$\rm[Mg/Fe]$ & 0.01 & 0.07  & 0.17  & 0.72   &  1.30    \\
$\rm[Al/Fe]$ & 0.01 & 0.03  & 0.04  & 0.18   &  0.64    \\
\cline{1-6}
$\rm[ls/Fe]$ & 1.08 & 1.36  & 1.32  & 1.34   &  1.10    \\
$\rm[hs/Fe]$ & 0.87 & 1.78  & 2.00  & 2.01   &  1.56    \\
$\rm[Pb/Fe]$ & 0.54 & 1.59  & 2.32  & 2.87   &  2.88    \\
$\rm[hs/ls]$ & -0.21 & 0.42  & 0.68  & 0.67   & 0.46    \\
\cline{1-6}
C/O & 1.87 & 6.92  &  14.4 &  36.0  &  53.0    \\
C/N & 5.84 & 23.0  &  46.1 &  151.0  & 406.0    \\
$^{12}$C/$^{13}$C  & 1.42$\times 10^2$ & 5.75$\times 10^2$  & 1.37$\times 10^3$  &  5.20$\times 10^3$  &  1.72$\times 10^4$   \\
\enddata
\tablecomments{Table \ref{tab12} is published in its entirety in
the electronic edition of the {\it Astrophysical Journal}.  A
portion is shown here for guidance regarding its form and
content.} \label{tab12}
\end{deluxetable}

\begin{deluxetable}{cccccccc}
\tablecolumns{8} \tablewidth{0pc} \tablecaption{Selected surface
branching isotopic ratios at different $Z$. The tabulated ratios
relate to the branchings occurring at $^{85}$Kr, $^{86}$Rb,
$^{95}$Zr, $^{133}$Xe and $^{141}$Ce, respectively. Solar ratios
are also reported.} \tablehead{ \colhead{Isot. ratio} &
\colhead{$\odot$} & \colhead{{\small $Z=1.38\times 10^{-2}$}} &
\colhead{{\small $Z=6.0\times 10^{-3}$}} & \colhead{{\small
$Z=3.0\times 10^{-3}$}} & \colhead{{\small $Z=1.0\times 10^{-3}$}}
& \colhead{{\small $Z=1.0\times 10^{-4}$}}} \startdata
$^{86}$Kr/$^{84}$Kr& 0.31   &0.26  & 0.37   &  0.52   & 0.68 &  0.72    \\
$^{87}$Rb/$^{85}$Rb& 0.39   &0.40  & 0.58   &  0.94   & 1.32 &  1.30    \\
$^{96}$Zr/$^{94}$Zr& 0.16   &0.03  & 0.07   &  0.16   & 0.44 &  0.61    \\
$^{134}$Xe/$^{132}$Xe& 0.37 &0.07  & 0.06   &  0.13   & 0.47 &  0.71    \\
$^{142}$Ce/$^{140}$Ce& 0.13 &0.01  & 0.01   &  0.04   & 0.14 &  0.20    \\
\enddata
\label{tab13}
\end{deluxetable}

\begin{deluxetable}{cccccc}
\tablecolumns{6} \tablewidth{0pc} \tablecaption{Yields (in
M$_\odot$) of selected isotopes of the 5 reference models at
different $Z$.} \tablehead{ \colhead{Isotope} & \colhead{{\small
$Z=1.38\times 10^{-2}$}} & \colhead{{\small $Z=6.0\times
10^{-3}$}} & \colhead{{\small $Z=3.0\times 10^{-3}$}} &
\colhead{{\small $Z=1.0\times 10^{-3}$}} & \colhead{{\small
$Z=1.0\times 10^{-4}$}}} \startdata
$^{1}$H     &   -4.06E-02   &   -7.05E-02   &  -8.21E-02  &   -1.16E-01   &   -9.92E-02   \\
$^{2}$H     &   -3.77E-05   &   -3.79E-05   &  -3.86E-05  &   -5.42E-05   &   -5.39E-05   \\
$^{3}$He    &   3.77E-04    &   3.35E-04    &  3.03E-04   &   2.59E-04    &   1.83E-04    \\
$^{4}$He    &   3.23E-02    &   5.14E-02    &  6.31E-02   &   8.85E-02    &   7.99E-02    \\
$^{12}$C    &   5.84E-03    &   1.28E-02    &  1.64E-02   &   2.23E-02    &   1.71E-02    \\
$^{13}$C    &   5.16E-05    &   2.03E-05    &  1.01E-05   &   3.63E-06    &   6.97E-07    \\
$^{14}$N    &   1.31E-03    &   6.03E-04    &  3.45E-04   &   1.52E-04    &   3.40E-05    \\
$^{15}$N    &   -2.03E-06   &   -9.21E-07   &  -4.88E-07  &   -1.65E-07   &   -2.07E-08   \\
\enddata                                        
\label{tab14} \tablecomments{Table \ref{tab14} is published in its
entirety in the electronic edition of the {\it Astrophysical
Journal}.  A portion regarding selected isotopes is shown here for
guidance regarding its form and content.} \end{deluxetable}

\begin{deluxetable}{cccccc}
\tablecolumns{6} \tablewidth{0pc} \tablecaption{Final isotopic
surface abundances (by mass fraction) of selected short-lived
radioactive isotopes (with the corresponding stable isotopes) at
different $Z$.} \tablehead{ \colhead{Isotope} & \colhead{{\small
$Z=1.38\times 10^{-2}$}} & \colhead{{\small $Z=6.0\times
10^{-3}$}} & \colhead{{\small $Z=3.0\times 10^{-3}$}} &
\colhead{{\small $Z=1.0\times 10^{-3}$}} & \colhead{{\small
$Z=1.0\times 10^{-4}$}}} \startdata
$^{26}$Al   &    3.07E-07   &  1.41E-07     &   7.80E-08    &  7.37E-08     &  5.33E-08     \\
$^{27}$Al   &    6.14E-05   &  2.77E-05     &   1.43E-05    &  6.43E-06     &  1.98E-06     \\
$^{35}$Cl   &   3.72E-06    &   1.61E-06    &   7.97E-07    &  2.59E-07     &  2.78E-08    \\
$^{36}$Cl   &   1.70E-09    &   1.51E-09    &   9.29E-10    &  5.35E-10     &  3.50E-11    \\
$^{40}$Ca   &    6.26E-05   &  2.70E-05     &   1.34E-05    &  4.35E-06     &  4.78E-07     \\
$^{41}$Ca   &    5.13E-09   &  3.52E-09     &   1.90E-09    &  1.11E-09     &  7.11E-11     \\
$^{56}$Fe   &    1.20E-03   &  5.17E-04     &   2.56E-04    &  8.26E-05     &  8.35E-06     \\
$^{60}$Fe   &    4.78E-08   &  3.93E-08     &   5.70E-08    &  1.38E-07     &  2.04E-08     \\
$^{204}$Pb  &   2.42E-09    &  1.24E-08     &   1.58E-08    &  4.80E-09     &  1.50E-10     \\
$^{205}$Pb  &   1.58E-10    &  1.04E-09     &   1.55E-09    &  4.61E-10     &  1.31E-11     \\
\enddata
\label{tab15}
\end{deluxetable}

\begin{deluxetable}{cccc}
\tablecolumns{4} \tablewidth{0pc} \tablecaption{Final composition
of the three models at $Z$=$10^{-4}$ (Reference, Reimers and low
mixing length).}\tablehead{ \colhead{[El/Fe]} &
\colhead{Reference} & \colhead{Reimers} & \colhead{$\alpha$=1.8}}
\startdata
$\rm[C/Fe] $& 3.00  & 3.20 &  2.81   \\
$\rm[N/Fe] $& 0.86  & 1.15 &  0.68   \\
$\rm[O/Fe] $& 1.01  & 1.17 &  0.88   \\
$\rm[F/Fe] $& 2.90  & 3.27 &  2.66   \\
$\rm[Ne/Fe]$& 2.27  & 2.68 &  1.90   \\
$\rm[Na/Fe]$& 1.77  & 2.26 &  1.36   \\
$\rm[Mg/Fe]$& 1.30  & 1.82 &  0.84   \\
$\rm[Al/Fe]$& 0.64  & 0.97 &  0.47   \\
\cline{1-4}
$\rm[ls/Fe]$& 1.10  & 1.30 &  0.85   \\
$\rm[hs/Fe]$& 1.56  & 1.85 &  1.33   \\
$\rm[Pb/Fe]$& 2.88  & 3.00 &  2.78   \\
$\rm[hs/ls]$& 0.46  & 0.55 &  0.48   \\
\cline{1-4}
C/O &          53  & 52  &  46  \\
C/N &        406  & 325  &  388  \\
$^{12}$C/$^{13}$C  & 1.72$\times10^4$ & 1.35$\times10^4$ & 1.72$\times10^4$  \\
\enddata
\label{tab16}
\end{deluxetable}

\newpage

\clearpage

\newpage

\begin{deluxetable}{cccccccccccc}
\rotate \tablecolumns{12} \tablewidth{0pc} \tablecaption{Pulse by
pulse elemental surface composition of the $Z=0.0138$ model
(online only).} \tablehead{ \colhead{[El/Fe]} & \colhead{TDU 1} &
\colhead{TDU 2} & \colhead{TDU 3}& \colhead{TDU 4}& \colhead{TDU
5}& \colhead{TDU 6}& \colhead{TDU 7}& \colhead{TDU 8}&
\colhead{TDU 9}& \colhead{TDU 10}& \colhead{TDU 11}} \startdata

C   &   -2.00E-01   &   -1.58E-01   &   -5.55E-02   &   6.72E-02    &   1.84E-01    &   2.83E-01    &   3.65E-01    &   4.36E-01    &   4.98E-01    &   5.54E-01    &   5.75E-01    \\
N   &   3.72E-01    &   3.72E-01    &   3.72E-01    &   3.71E-01    &   3.71E-01    &   3.69E-01    &   3.68E-01    &   3.67E-01    &   3.66E-01    &   3.65E-01    &   3.65E-01    \\
O   &   -5.16E-04   &   -5.27E-04   &   -9.26E-05   &   6.71E-04    &   1.50E-03    &   2.13E-03    &   2.43E-03    &   2.44E-03    &   2.20E-03    &   1.73E-03    &   1.36E-03    \\
F   &   -7.81E-03   &   -2.14E-03   &   2.30E-02    &   7.03E-02    &   1.34E-01    &   2.04E-01    &   2.72E-01    &   3.38E-01    &   4.02E-01    &   4.63E-01    &   4.88E-01    \\
Ne  &   -2.55E-03   &   4.12E-04    &   1.05E-02    &   2.77E-02    &   5.05E-02    &   7.63E-02    &   1.03E-01    &   1.31E-01    &   1.60E-01    &   1.90E-01    &   2.02E-01    \\
Na  &   8.62E-02    &   8.78E-02    &   9.19E-02    &   9.87E-02    &   1.08E-01    &   1.18E-01    &   1.29E-01    &   1.40E-01    &   1.53E-01    &   1.66E-01    &   1.72E-01    \\
Mg  &   -1.92E-07   &   4.70E-05    &   2.60E-04    &   6.82E-04    &   1.36E-03    &   2.34E-03    &   3.59E-03    &   5.16E-03    &   7.10E-03    &   9.44E-03    &   1.06E-02    \\
Al  &   1.72E-04    &   4.57E-04    &   9.36E-04    &   1.76E-03    &   2.92E-03    &   4.26E-03    &   5.68E-03    &   7.10E-03    &   8.54E-03    &   9.99E-03    &   1.06E-02    \\
\enddata
\label{tab7_el}
\end{deluxetable}

\newpage

\begin{deluxetable}{ccccccccccccccc}
\rotate \tablecolumns{15} \tablewidth{0pc} \tablecaption{Pulse by
pulse elemental surface composition of the $Z=0.006$ model (online
only).} \tablehead{ \colhead{[El/Fe]} & \colhead{{\small TDU 1}} &
\colhead{{\small TDU 2}} & \colhead{{\small TDU 3}}&
\colhead{{\small TDU 4}}& \colhead{{\small TDU 5}}&
\colhead{{\small TDU 6}}& \colhead{{\small TDU 7}}&
\colhead{{\small TDU 8}}& \colhead{{\small TDU 9}}&
\colhead{{\small TDU 10}}& \colhead{{\small TDU 11}}&
\colhead{{\small TDU 12}}& \colhead{{\small TDU 13}}&
\colhead{{\small TDU 14}}} \startdata

C   &   {\tiny  -9.26E-02   }&{\tiny    1.47E-01    }&{\tiny    3.69E-01    }&{\tiny    5.41E-01    }&{\tiny    6.70E-01    }&{\tiny    7.71E-01    }&{\tiny    8.53E-01    }&{\tiny    9.22E-01    }&{\tiny    9.82E-01    }&{\tiny    1.03E+00    }&{\tiny    1.08E+00    }&{\tiny    1.13E+00    }&{\tiny    1.15E+00    }&{\tiny    1.17E+00    }   \\
N   &   {\tiny  3.94E-01    }&{\tiny    3.94E-01    }&{\tiny    3.93E-01    }&{\tiny    3.92E-01    }&{\tiny    3.90E-01    }&{\tiny    3.89E-01    }&{\tiny    3.87E-01    }&{\tiny    3.86E-01    }&{\tiny    3.84E-01    }&{\tiny    3.83E-01    }&{\tiny    3.82E-01    }&{\tiny    3.82E-01    }&{\tiny    3.86E-01    }&{\tiny    3.97E-01    }   \\
O   &   {\tiny  -4.65E-04   }&{\tiny    2.02E-03    }&{\tiny    6.21E-03    }&{\tiny    1.09E-02    }&{\tiny    1.50E-02    }&{\tiny    1.84E-02    }&{\tiny    2.10E-02    }&{\tiny    2.31E-02    }&{\tiny    2.48E-02    }&{\tiny    2.62E-02    }&{\tiny    2.74E-02    }&{\tiny    2.86E-02    }&{\tiny    2.90E-02    }&{\tiny    2.90E-02    }   \\
F   &   {\tiny  -8.65E-03   }&{\tiny    4.79E-02    }&{\tiny    1.51E-01    }&{\tiny    2.70E-01    }&{\tiny    3.87E-01    }&{\tiny    4.96E-01    }&{\tiny    5.95E-01    }&{\tiny    6.87E-01    }&{\tiny    7.73E-01    }&{\tiny    8.51E-01    }&{\tiny    9.24E-01    }&{\tiny    9.98E-01    }&{\tiny    1.03E+00    }&{\tiny    1.06E+00    }   \\
Ne  &   {\tiny  4.94E-04    }&{\tiny    1.82E-02    }&{\tiny    5.16E-02    }&{\tiny    9.57E-02    }&{\tiny    1.45E-01    }&{\tiny    1.97E-01    }&{\tiny    2.51E-01    }&{\tiny    3.04E-01    }&{\tiny    3.57E-01    }&{\tiny    4.09E-01    }&{\tiny    4.60E-01    }&{\tiny    5.13E-01    }&{\tiny    5.39E-01    }&{\tiny    5.58E-01    }   \\
\enddata
\label{tab8_el}
\end{deluxetable}

\newpage

\begin{deluxetable}{ccccccccccccccccc}
\rotate \tablecolumns{17} \tablewidth{0pc} \tablecaption{Pulse by
pulse elemental surface composition of the $Z=0.003$ model (online
only).} \tablehead{ \colhead{[El/Fe]} & \colhead{{\tiny TDU 1}} &
\colhead{{\tiny TDU 2}} & \colhead{{\tiny TDU 3}}& \colhead{{\tiny
TDU 4}}& \colhead{{\tiny TDU 5}}& \colhead{{\tiny TDU 6}}&
\colhead{{\tiny TDU 7}}& \colhead{{\tiny TDU 8}}& \colhead{{\tiny
TDU 9}}& \colhead{{\tiny TDU 10}}& \colhead{{\tiny TDU 11}}&
\colhead{{\tiny TDU 12}}& \colhead{{\tiny TDU 13}}&
\colhead{{\tiny TDU 14}}& \colhead{{\tiny TDU 15}}&
\colhead{{\tiny TDU 16}}} \startdata
C   &   {\tiny  -3.51E-02   }&{\tiny    3.41E-01    }&{\tiny    6.18E-01    }&{\tiny    8.12E-01    }&{\tiny    9.53E-01    }&{\tiny    1.06E+00    }&{\tiny    1.15E+00    }&{\tiny    1.22E+00    }&{\tiny    1.28E+00    }&{\tiny    1.33E+00    }&{\tiny    1.38E+00    }&{\tiny    1.42E+00    }&{\tiny    1.46E+00    }&{\tiny    1.50E+00    }&{\tiny    1.53E+00    }&{\tiny    1.54E+00    }   \\
N   &   {\tiny  4.16E-01    }&{\tiny    4.15E-01    }&{\tiny    4.15E-01    }&{\tiny    4.14E-01    }&{\tiny    4.13E-01    }&{\tiny    4.12E-01    }&{\tiny    4.11E-01    }&{\tiny    4.10E-01    }&{\tiny    4.10E-01    }&{\tiny    4.10E-01    }&{\tiny    4.10E-01    }&{\tiny    4.10E-01    }&{\tiny    4.11E-01    }&{\tiny    4.14E-01    }&{\tiny    4.20E-01    }&{\tiny    4.38E-01    }   \\
O   &   {\tiny  -1.42E-03   }&{\tiny    4.43E-03    }&{\tiny    1.41E-02    }&{\tiny    2.49E-02    }&{\tiny    3.49E-02    }&{\tiny    4.34E-02    }&{\tiny    5.05E-02    }&{\tiny    5.65E-02    }&{\tiny    6.19E-02    }&{\tiny    6.68E-02    }&{\tiny    7.14E-02    }&{\tiny    7.58E-02    }&{\tiny    8.01E-02    }&{\tiny    8.44E-02    }&{\tiny    8.87E-02    }&{\tiny    8.97E-02    }   \\
F   &   {\tiny  -1.81E-02   }&{\tiny    1.03E-01    }&{\tiny    2.82E-01    }&{\tiny    4.52E-01    }&{\tiny    5.99E-01    }&{\tiny    7.27E-01    }&{\tiny    8.39E-01    }&{\tiny    9.38E-01    }&{\tiny    1.03E+00    }&{\tiny    1.11E+00    }&{\tiny    1.19E+00    }&{\tiny    1.26E+00    }&{\tiny    1.32E+00    }&{\tiny    1.39E+00    }&{\tiny    1.45E+00    }&{\tiny    1.47E+00    }   \\
Ne  &   {\tiny  -8.52E-04   }&{\tiny    2.64E-02    }&{\tiny    7.96E-02    }&{\tiny    1.48E-01    }&{\tiny    2.23E-01    }&{\tiny    2.97E-01    }&{\tiny    3.70E-01    }&{\tiny    4.41E-01    }&{\tiny    5.09E-01    }&{\tiny    5.74E-01    }&{\tiny    6.36E-01    }&{\tiny    6.97E-01    }&{\tiny    7.54E-01    }&{\tiny    8.10E-01    }&{\tiny    8.63E-01    }&{\tiny    8.81E-01    }   \\
\enddata
\label{tab9_el}
\end{deluxetable}

\newpage

\begin{deluxetable}{ccccccccccccccccc}
\rotate \tablecolumns{17} \tablewidth{0pc} \tablecaption{Pulse by
pulse elemental surface composition of the $Z=0.001$ model (online
only).} \tablehead{ \colhead{[El/Fe]} & \colhead{{\tiny TDU 1}} &
\colhead{{\tiny TDU 2}} & \colhead{{\tiny TDU 3}}& \colhead{{\tiny
TDU 4}}& \colhead{{\tiny TDU 5}}& \colhead{{\tiny TDU 6}}&
\colhead{{\tiny TDU 7}}& \colhead{{\tiny TDU 8}}& \colhead{{\tiny
TDU 9}}& \colhead{{\tiny TDU 10}}& \colhead{{\tiny TDU 11}}&
\colhead{{\tiny TDU 12}}& \colhead{{\tiny TDU 13}}&
\colhead{{\tiny TDU 15}}& \colhead{{\tiny TDU 17}}&
\colhead{{\tiny TDU 19}}} \startdata
C   &   {\tiny  -2.13E-01   }&{\tiny    5.67E-01    }&{\tiny    9.88E-01    }&{\tiny    1.25E+00    }&{\tiny    1.43E+00    }&{\tiny    1.56E+00    }&{\tiny    1.66E+00    }&{\tiny    1.74E+00    }&{\tiny    1.80E+00    }&{\tiny    1.86E+00    }&{\tiny    1.91E+00    }&{\tiny    1.95E+00    }&{\tiny    1.99E+00    }&{\tiny    2.06E+00    }&{\tiny    2.13E+00    }&{\tiny    2.18E+00    }   \\
N   &   {\tiny  4.59E-01    }&{\tiny    4.59E-01    }&{\tiny    4.58E-01    }&{\tiny    4.58E-01    }&{\tiny    4.58E-01    }&{\tiny    4.59E-01    }&{\tiny    4.59E-01    }&{\tiny    4.61E-01    }&{\tiny    4.62E-01    }&{\tiny    4.65E-01    }&{\tiny    4.68E-01    }&{\tiny    4.72E-01    }&{\tiny    4.76E-01    }&{\tiny    4.88E-01    }&{\tiny    5.09E-01    }&{\tiny    5.56E-01    }   \\
O   &   {\tiny  -9.12E-03   }&{\tiny    4.42E-03    }&{\tiny    3.39E-02    }&{\tiny    7.04E-02    }&{\tiny    1.05E-01    }&{\tiny    1.35E-01    }&{\tiny    1.59E-01    }&{\tiny    1.79E-01    }&{\tiny    1.96E-01    }&{\tiny    2.12E-01    }&{\tiny    2.27E-01    }&{\tiny    2.40E-01    }&{\tiny    2.53E-01    }&{\tiny    2.78E-01    }&{\tiny    3.02E-01    }&{\tiny    3.22E-01    }   \\
F   &   {\tiny  -3.85E-02   }&{\tiny    3.32E-01    }&{\tiny    6.98E-01    }&{\tiny    9.50E-01    }&{\tiny    1.14E+00    }&{\tiny    1.28E+00    }&{\tiny    1.40E+00    }&{\tiny    1.51E+00    }&{\tiny    1.60E+00    }&{\tiny    1.68E+00    }&{\tiny    1.76E+00    }&{\tiny    1.83E+00    }&{\tiny    1.89E+00    }&{\tiny    2.02E+00    }&{\tiny    2.13E+00    }&{\tiny    2.22E+00    }   \\
Ne  &   {\tiny  -5.32E-03   }&{\tiny    5.01E-02    }&{\tiny    1.75E-01    }&{\tiny    3.21E-01    }&{\tiny    4.64E-01    }&{\tiny    5.98E-01    }&{\tiny    7.21E-01    }&{\tiny    8.32E-01    }&{\tiny    9.32E-01    }&{\tiny    1.03E+00    }&{\tiny    1.11E+00    }&{\tiny    1.19E+00    }&{\tiny    1.26E+00    }&{\tiny    1.39E+00    }&{\tiny    1.51E+00    }&{\tiny    1.60E+00    }   \\
\enddata
\label{tab10_el}
\end{deluxetable}

\newpage

\begin{deluxetable}{cccccccccccccccc}
\rotate \tablecolumns{16} \tablewidth{0pc} \tablecaption{Pulse by
pulse elemental surface composition of the $Z=0.0001$ model
(online only).} \tablehead{ \colhead{[El/Fe]} & \colhead{{\tiny
TDU 1}} & \colhead{{\tiny TDU 2}} & \colhead{{\tiny TDU 3}}&
\colhead{{\tiny TDU 4}}& \colhead{{\tiny TDU 5}}& \colhead{{\tiny
TDU 6}}& \colhead{{\tiny TDU 7}}& \colhead{{\tiny TDU 8}}&
\colhead{{\tiny TDU 9}}& \colhead{{\tiny TDU 10}}& \colhead{{\tiny
TDU 11}}& \colhead{{\tiny TDU 12}}& \colhead{{\tiny TDU 13}}&
\colhead{{\tiny TDU 14}}& \colhead{{\tiny TDU 15}}} \startdata
C   &   {\tiny  1.15E+00    }&{\tiny    1.72E+00    }&{\tiny    2.04E+00    }&{\tiny    2.26E+00    }&{\tiny    2.41E+00    }&{\tiny    2.52E+00    }&{\tiny    2.61E+00    }&{\tiny    2.69E+00    }&{\tiny    2.75E+00    }&{\tiny    2.81E+00    }&{\tiny    2.86E+00    }&{\tiny    2.90E+00    }&{\tiny    2.95E+00    }&{\tiny    2.99E+00    }&{\tiny    3.00E+00    }   \\
N   &   {\tiny  4.93E-01    }&{\tiny    4.95E-01    }&{\tiny    5.01E-01    }&{\tiny    5.11E-01    }&{\tiny    5.23E-01    }&{\tiny    5.39E-01    }&{\tiny    5.56E-01    }&{\tiny    5.75E-01    }&{\tiny    5.95E-01    }&{\tiny    6.17E-01    }&{\tiny    6.41E-01    }&{\tiny    6.68E-01    }&{\tiny    7.01E-01    }&{\tiny    7.51E-01    }&{\tiny    8.58E-01    }   \\
O   &   {\tiny  -9.15E-04   }&{\tiny    1.49E-01    }&{\tiny    3.30E-01    }&{\tiny    4.82E-01    }&{\tiny    5.95E-01    }&{\tiny    6.79E-01    }&{\tiny    7.43E-01    }&{\tiny    7.95E-01    }&{\tiny    8.38E-01    }&{\tiny    8.76E-01    }&{\tiny    9.09E-01    }&{\tiny    9.40E-01    }&{\tiny    9.71E-01    }&{\tiny    1.00E+00    }&{\tiny    1.01E+00    }   \\
F   &   {\tiny  -8.31E-02   }&{\tiny    1.16E+00    }&{\tiny    1.65E+00    }&{\tiny    1.94E+00    }&{\tiny    2.13E+00    }&{\tiny    2.27E+00    }&{\tiny    2.37E+00    }&{\tiny    2.47E+00    }&{\tiny    2.54E+00    }&{\tiny    2.62E+00    }&{\tiny    2.68E+00    }&{\tiny    2.75E+00    }&{\tiny    2.82E+00    }&{\tiny    2.88E+00    }&{\tiny    2.90E+00    }   \\
Ne  &   {\tiny  -3.48E-03   }&{\tiny    4.30E-01    }&{\tiny    8.19E-01    }&{\tiny    1.09E+00    }&{\tiny    1.30E+00    }&{\tiny    1.46E+00    }&{\tiny    1.60E+00    }&{\tiny    1.71E+00    }&{\tiny    1.82E+00    }&{\tiny    1.91E+00    }&{\tiny    2.00E+00    }&{\tiny    2.08E+00    }&{\tiny    2.17E+00    }&{\tiny    2.24E+00    }&{\tiny    2.27E+00    }   \\
\enddata
\label{tab11_el}
\end{deluxetable}

\newpage

\begin{deluxetable}{cccccc}
\tablecolumns{6} \tablewidth{0pc} \tablecaption{Final elemental
surface composition of the 5 reference models at different
metallicities (online only).} \tablehead{ \colhead{[El/Fe]} &
\colhead{{\small $Z=1.38\times 10^{-2}$}} & \colhead{{\small
$Z=6.0\times 10^{-3}$}} & \colhead{{\small $Z=3.0\times 10^{-3}$}}
& \colhead{{\small $Z=1.0\times 10^{-3}$}} & \colhead{{\small
$Z=1.0\times 10^{-4}$}}} \startdata

C   &   5.75E-01   &    1.17E+00     &    1.54E+00    &    2.18E+00     &     3.00E+00    \\
N   &   3.65E-01   &    3.97E-01     &    4.38E-01    &    5.56E-01     &     8.58E-01    \\
O   &   1.36E-03   &    2.90E-02     &    8.97E-02    &    3.22E-01     &     1.01E+00    \\
F   &   4.88E-01   &    1.06E+00     &    1.47E+00    &    2.22E+00     &     2.90E+00    \\
Ne  &   2.02E-01   &    5.58E-01     &    8.81E-01    &    1.60E+00     &     2.27E+00    \\
Na  &   1.72E-01   &    3.58E-01     &    5.70E-01    &    1.28E+00     &     1.77E+00    \\
Mg  &   1.06E-02   &    6.77E-02     &    1.70E-01    &    7.19E-01     &     1.30E+00    \\
Al  &   1.06E-02   &    2.72E-02     &    4.41E-02    &    1.79E-01     &     6.37E-01    \\
\enddata
\label{tab12_el}
\end{deluxetable}

\newpage

\begin{deluxetable}{cccccc}
\tablecolumns{6} \tablewidth{0pc} \tablecaption{Stellar yields of
the 5 reference models at different metallicities (online only).}
\tablehead{ \colhead{Isotope} & \colhead{{\small $Z=1.38\times
10^{-2}$}} & \colhead{{\small $Z=6.0\times 10^{-3}$}} &
\colhead{{\small $Z=3.0\times 10^{-3}$}} & \colhead{{\small
$Z=1.0\times 10^{-3}$}} & \colhead{{\small $Z=1.0\times
10^{-4}$}}}\startdata
$^{1}$H     &   -4.06E-02   &   -7.05E-02   &  -8.21E-02  &   -1.16E-01   &   -9.92E-02   \\
$^{2}$H     &   -3.77E-05   &   -3.79E-05   &  -3.86E-05  &   -5.42E-05   &   -5.39E-05   \\
$^{3}$He    &   3.77E-04    &   3.35E-04    &  3.03E-04   &   2.59E-04    &   1.83E-04    \\
$^{4}$He    &   3.23E-02    &   5.14E-02    &  6.31E-02   &   8.85E-02    &   7.99E-02    \\
$^{6}$Li    &   -9.09E-10   &   -3.87E-10   &  -1.97E-10  &   -6.38E-11   &   -6.80E-12   \\
$^{7}$Li    &   -1.29E-08   &   -5.43E-09   &  -2.70E-09  &   -7.86E-10   &   -1.60E-11   \\
$^{7}$Be    &   2.39E-17    &   6.06E-17    &  4.60E-16   &   1.47E-16    &   8.07E-17    \\
$^{9}$Be    &   -2.39E-10   &   -1.02E-10   &  -5.22E-11  &   -1.69E-11   &   -1.75E-12   \\
\enddata                                         
\label{tab14_el}                                
\end{deluxetable}

\end{document}